\documentclass[12pt]{book}
\usepackage{epsfig}
\usepackage{amsbsy}
\usepackage{amsfonts}
\usepackage{graphicx}
\def\bea{\begin{eqnarray}}
\def\eea{\end{eqnarray}}
\def\beqn{\begin{eqnarray}}
\def\eeqn{\end{eqnarray}}
\def\beq{\begin{equation}}
\def\eeq{\end{equation}}

\def\ra{\rightarrow}

\def\l{\left}
\def\r{\right}
\def\ie{{\it i.e.~}}
\def\eg{{\it e.g.~}}
\def\rhs{{\it r.h.s.~}}
\def\lhs{{\it l.h.s.~}}
\def\MS{\hbox{$\overline{\rm MS}$}} 
\def\Dslash{\not{\hbox{\kern-4pt $D$}}}
\def\pslash{\not{\hbox{\kern-4pt $p$}}}

\def\lan{\langle}
\def\ran{\rangle}

\def\tr{\,{\hbox{Tr}}\,}
\def\as{\alpha_s}

\def\nf{n_f}  
\def\tr{\,{\hbox{Tr}}\,}

\def\eps{\epsilon^{\mu\nu\rho\sigma}}
\def\lan{\langle}
\def\ran{\rangle}

\def\da{\partial_{\alpha}}
\def\db{\partial_{\beta}}

\def\ieight{\left(\frac{1}{1-x}\right)_+}  
  
\def\ionefive{\left(\frac{\log (1-x)}{1-x}\right)_+}

\def\pr{{\it Phys.~Rev.~}}
\def\prl{{\it Phys.~Rev.~Lett.~}}
\def\np{{\it Nucl.~Phys.~}}

\def\pl{{\it Phys.~Lett.~}}
\def\prep{{\it Phys.~Rep.~}}

\def\rmp{{\it Rev.~Mod.~Phys.~}}

\def\zp{{\it Zeit.~Phys.~}}

\def\epj{{\it Eur. Phys. J.}}

\def\vol#1{{\bf #1}}
\def\vyp#1#2#3{\vol{#1} (#2) #3}

\newcommand\sss{\scriptscriptstyle}  

\catcode`\@=11

\@addtoreset{equation}{section}
\def\theequation{\arabic{chapter}.\arabic{equation}}

\def\@chapapp{}
 
\def\section{\@startsection{section}{1}{\z@}{3.5ex plus 1ex minus
   .2ex}{2.3ex plus .2ex}{\large\bf}}

\def\thesection{\arabic{chapter}.\arabic{section}}

\def\theequation{\arabic{chapter}.\arabic{section}.\arabic{equation}}

\def\appendix{\setcounter{chapter}{0}
 \def\thechapter{\Alph{chapter}}
 \def\thesection{\Alph{chapter}.\arabic{section}}
 \def\theequation{\Alph{chapter}.\arabic{section}.\arabic{equation}}}

\newcommand{\tcaption}[2]{
  \refstepcounter{table}
             {\small Table~\thetable.~}{{\small\it #2}}
   }

\newcommand{\fcaption}[2]{
  \refstepcounter{figure}
             {\small Figure~\thefigure.~}{{\small\it #2}}
   }

\mark{{}{}}
\newbox\@intesta
\def\ps@plain{\let\@mkboth\@gobbletwo
\def\@oddfoot{\hbox{} \hfill \thepage \hfill \hbox{}}\def\@oddhead{}
\def\@evenhead{}\let\@evenfoot\@oddfoot}
\if@twoside
\def\ps@headings{\let\@mkboth\markboth%
\def\@oddfoot{\hbox{} \hfill \thepage \hfill \hbox{}}
\let\@evenfoot\@oddfoot
\def\@evenhead{\setbox\@intesta\hbox{\small\bf \leftmark}%
\underline{\makebox[\textwidth]{\rule[-0.5ex]{0pt}{0.5ex} \hfill
\footnotesize\sl\leftmark\hfill \hbox{}}}}

\def\@oddhead{\setbox\@intesta\hbox{\small\bf \rightmark}%
\underline{\makebox[\textwidth]{\rule[-0.5ex]{0pt}{0.5ex} \hfill
\footnotesize\sl\rightmark\hfill \hbox{}}}}%

\def\chaptermark##1{\markboth{\  ##1}{}}
\def\sectionmark##1{\markright{\ifnum \c@secnumdepth >\z@ \thesection -- %
\fi ##1}}}
\else
\def\ps@headings{\let\@mkboth\markboth
\def\@oddfoot{\hbox{} \hfill \thepage \hfill \hbox{}}
\def\@evenfoot{}
\def\@oddhead{\setbox\@intesta\hbox{\small\bf \rightmark}%
\underline{\makebox[\textwidth]{\rule[-0.5ex]{0pt}{0.5ex} \hfill
\footnotesize\sl\rightmark\hfill \hbox{}}}}%
\fi

\def\tableofcontents{\@restonecolfalse\if@twocolumn\@restonecoltrue\onecolumn
 \fi\chapter*{Contents\@mkboth{Contents}{Contents}}
  \@starttoc{toc}\if@restonecol\twocolumn\fi}
\def\l@part#1#2{\addpenalty{-\@highpenalty}
   \addvspace{2.25em plus 1pt}
   \begingroup   \@tempdima 3em \parindent \z@ \rightskip \@pnumwidth
     \parfillskip -\@pnumwidth {\large \bf \leavevmode #1\hfil
     \hbox to\@pnumwidth{\hss #2}}\par
     \nobreak
   \endgroup}

\def\thebibliography#1%
{\chapter*{References\@mkboth{References}{References}
\addcontentsline{toc}{chapter}{References}}
\list
  {[\arabic{enumi}]}{\settowidth\labelwidth{[#1]}\leftmargin\labelwidth
    \advance\leftmargin\labelsep
    \usecounter{enumi}}
    \def\newblock{\hskip .11em plus .33em minus -.07em}
    \sloppy
    \sfcode`\.=1000\relax}

\renewcommand{\baselinestretch}{1.0}

\begin{document}
\pagestyle{empty}
\begin{flushright}   
  
{\tt hep-ph/0207xxx}
\\GEF/TH-10-02
\end{flushright}   
  
\begin{center}   
\vspace*{0.5cm}  
{\Large \bf Aspects of QCD perturbative evolution}\\  
\vspace*{1.5cm}   
{\bf Andrea Piccione} \\
\vspace{0.6cm}  
{\it Dipartimento di Fisica, Universit\`a di Genova \\and\\ {}INFN,  
Sezione di Genova,\\  
Via Dodecaneso 33, I-16146 Genova, Italy}\\  
\vspace*{1.5cm}  
  
{\bf Abstract}

\end{center}  
\noindent  
This thesis is devoted to the study of some aspects of
perturbative QCD, and in particular to the development of high-precision
techniques for the extraction of physical parameters such as
structure functions, parton distributions, and the strong coupling from
the analysis of deep inelastic scattering data.
First, we will discuss scaling violations of 
singlet and nonsinglet truncated moments, and the use of
truncated momets to solve the Altarelli-Parisi equation.
Then we will suggest an approach based on neural networks to the
parametrization and interpolation of  experimental data, which 
retains information on experimental errors and correlations.
The method of truncated moments can be combined with the neural
network fit to extract various quantities of phenomenological
interest in a bias-free way.
As an example of such application, we will discuss the determination of
the strong coupling constant.

\vspace*{1cm}

\vfill  
\noindent  
  
\begin{flushleft} July 2002 \end{flushleft}   
\eject
\pagestyle{empty}

\begin{figure}[ht]
\begin{center}
\mbox{\includegraphics[width=0.15\textwidth,clip]{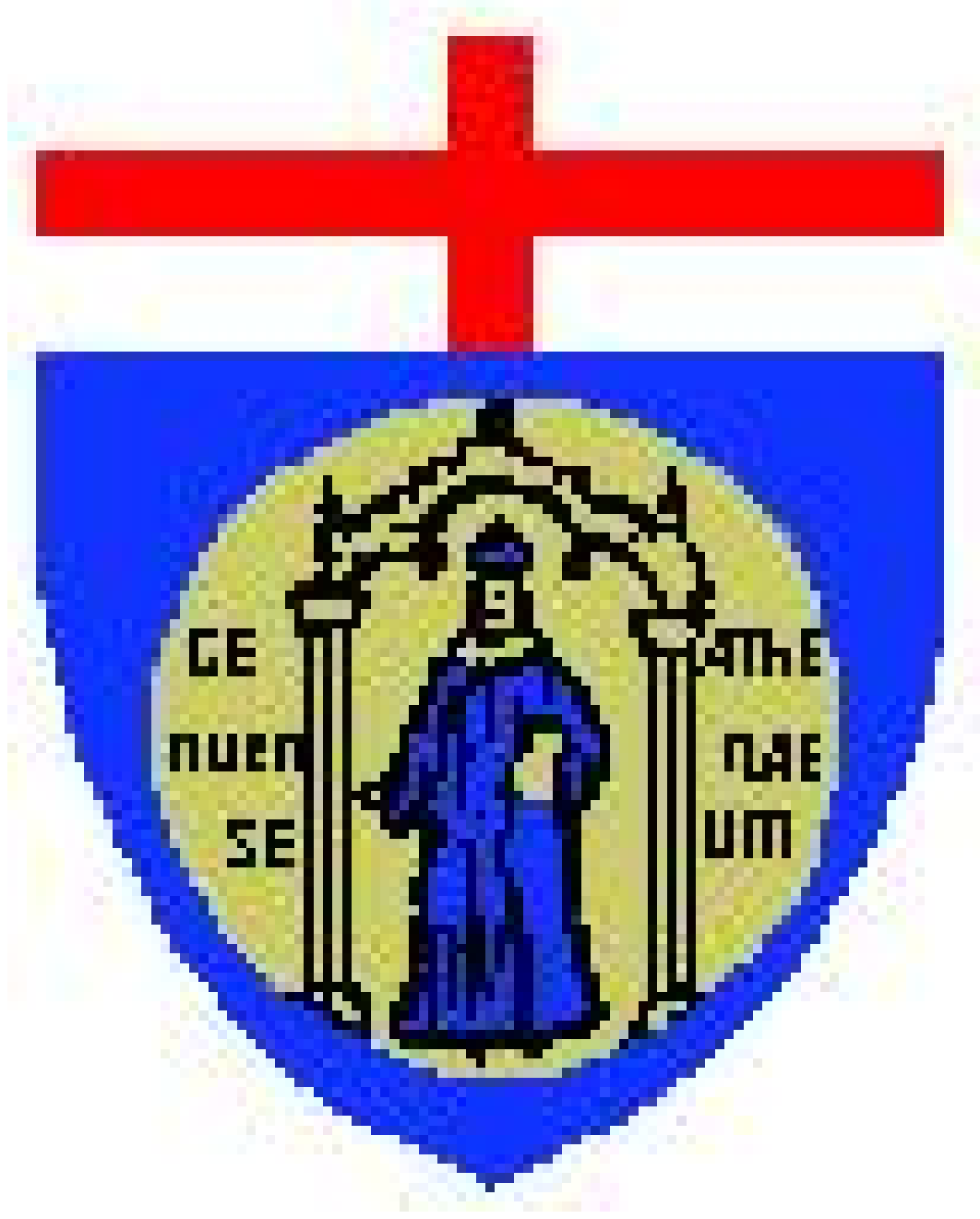}}
\end{center}
\end{figure}
\begin{center}
\large{\bf Universit\`a degli Studi di Genova}
\end{center}
\begin{center}
\normalsize Facolt\`a di Scienze Matematiche Fisiche e Naturali
\end{center}
\vspace{1 cm}

\begin{center}
Andrea Piccione
\end{center}
\vspace{3mm}
\begin{center}
\Large{\bf Aspects of QCD perturbative evolution}
\end{center}
\begin{center}
\large Tesi di Dottorato in Fisica 
\end{center}
\vspace{5mm}
\begin{table}[b]
\begin{center}
\begin{tabular}{|c|c|c|c|c|}
\multicolumn{1}{l}{\qquad}%
           &\multicolumn{1}{c}{\it Relatore:}%
           &\multicolumn{1}{c}{\qquad\qquad\qquad\qquad\qquad\qquad\qquad}%
           &\multicolumn{1}{c}{\it Relatore esterno:}%
           &\multicolumn{1}{r}{\qquad}\\
\multicolumn{1}{l}{\qquad}%
           &\multicolumn{1}{c}{Dr. G. Ridolfi}%
           &\multicolumn{1}{c}{}%
           &\multicolumn{1}{c}{Dr. S. Forte}%
           &\multicolumn{1}{r}{\qquad}\\
\end{tabular}
\end{center}
\end{table}
\begin{center}
Dicembre 2001
\end{center}
\newpage
\vspace*{\fill}
\newpage
\pagestyle{empty}
\pagenumbering{roman}
\begin{flushright}
{\it a Tania}
\end{flushright}
\vspace*{\fill}
\newpage

\pagenumbering{roman}
\pagestyle{headings}
\tableofcontents
\newpage
\setcounter{page}{1}
\pagenumbering{arabic}
\pagestyle{headings}

\chapter{Introduction}

Quantum Chromodynamics (QCD) is believed to be the theory of strong
interactions. QCD as a gauge theory of quarks and gluons is unique
among renormalizable theories in providing a basis for the parton
model within the principles of relativistic quantum field theory and
at present times it stands as a main building block of the ``Standard
Model'' of the fundamental interactions. 

The strength of electro-weak interaction is so
weak that perturbation theory is extremely reliable. Furthermore, the
leptons are at the same time the fields in the Lagrangian and the
particles in the detector. The case of QCD is different in both respects.
Perturbative methods are only applicable in those particular
domains of strong interactions physics where the asymptotic freedom
can actually be reached. Although there are several attempts to
describe non-perturbative effects of QCD, at present times we
have not yet a solution of QCD in the low-energy domain.
Also, QCD is a theory of quarks and gluons, while the real world is
made up of hadrons. Clearly, some model is needed to match one to the
other.

Essentially all physic aspects of present and future hadron 
colliders, from particles searching beyond the Standard Model to
electro-weak precision measu\-rements and study of heavy quarks, need
a detailed information on the hadronic initial states. 
Factorization is a central issue, as it allows to se\-pa\-ra\-te
perturbative and non-perturbative contributions. In particular, it
ensures that a non-perturbative quantity, as a structure function,
is process independent. As a consequence, we can extract structure
functions from, say, deep inelastic scattering processes, and use them as 
inputs for a hadron-hadron collision. Thus, we need a knowledge
as precise as possible on the inputs of hadron colliders that
we extract from other processes.

The present thesis is devoted to the study of some aspects of
perturbative QCD, and in particular to the development of high-precision
techniques for the extraction of phenomenological parameters such as
structure functions, parton distributions, and the strong coupling from
the analysis of deep inelastic scattering data. This analysis is
usually characterized by theoretical as\-sumptions on the phenomenological
parameters, such as the functional parametrizations of the structure
functions or the small-$x$ behavior of parton distributions.
These assumptions introduce potentially large biases, whose
size is very hard to assess. 

In this thesis we will develop methods to reduce the sources of
theoretical uncertainties. Specifically, we will first extend the method of
truncated moments from its original formulation for the non-singlet
parton distributions to all flavor combinations. Truncated
moments of parton distributions are defined by restricting the
integration range over the Bjorken variable to an experimentally
accessible subset $x_0<x<1$ of the allowed kinematic range $0<x<1$. As
a consequence this method provides a way to avoid theoretical biases
on the small-$x$ behavior of parton distributions. 

Special attention will be devoted to the numerical implementation of
the technique of truncated moments. We will write the evolution
equations for truncated moments in a particular form, which has the
advantage of increasing the efficiency of the method considerably.

A crucial ingredient of most phenomenological analyses is the
technique adopted to interpolate experimental data, and to reproduce
the corresponding errors and correlations. We will suggest an approach
to this problem based on neural networks.  We will show that with this
method it is possible to extract information from experimental data
without introducing a functional parametrization of structure
functions based on theoretical assumptions. Errors and correlations of
the data points will be determined by a Monte Carlo technique. We
will discuss results and details of the numerical implementation, based 
on a subset of the available experimental data for  
unpolarized deep inelastic scattering.

The method of truncated moments can be combined with the neural
network fit to extract various quantities of phenomenological
interest in a bias-free way.
Here, as an application, we will adopt these techniques to determine
the strong coupling constant. Other possible applications will be
sketched in the conclusions.
\begin{figure}[t]
\begin{center}
\mbox{\includegraphics[width=0.78\textwidth,clip]{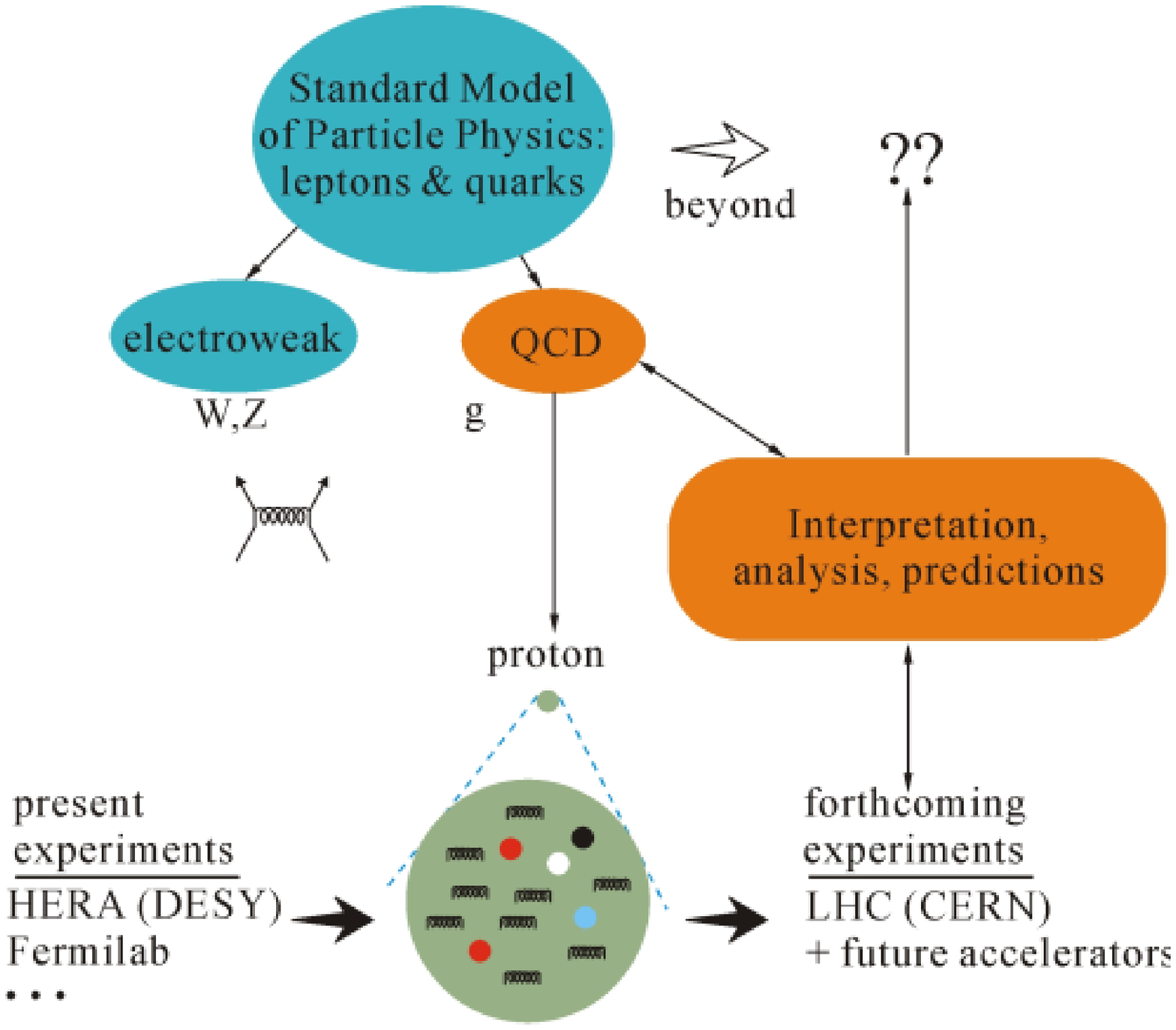}}
\end{center}
\begin{center}
\fcaption{}{General framework.}
\label{fig:sm}
\end{center}
\end{figure}

In Chapters 2 and 3 we will define our frameworks, \ie QCD and the
parton model. Besides all the general issues, such as the QCD Lagrangian,
the Feynman rules, running coupling constant, deep inelastic scattering
cross section and parton distributions evolution equations, we will
discuss in detail some technical issues, such as the matching of the
running coupling constant at quarks mass thresholds and the change of
factorization schemes, since they are essential ingredients of our
analysis.

In Chapter 4 we will introduce the method of truncated moments.
We will derive the relevant evolution equations and 
the corresponding solution.
We will discuss the numerical accuracy of the technique, and 
the way it can be implementeted for phenomenological applications.

Chapter 5 is a very brief introduction to neural networks and
it is mainly devoted to introduce the main algorithms and
some practical rules. In Chapter 6, we will first describe the main
features of the experimental data, and we will discuss under which conditions
we can reproduce them by a Monte Carlo technique. Then, we will
address to the neural network fit of data, and we will
give details on the behavior of neural networks, such as their
ability of finding an underlying law, and on the way we have
to train them.

As an application, the efforts of the previous Chapters will be used
in Chapter 7 to give a determination of the strong coupling constant
$\as$. The phenomenology of deep inelastic scattering will be
completed with the discussion of target mass corrections, the
renormalization scale dependence, and the corrections due to the
elastic contributions. The fitting procedure, as well as the estimation
of the theoretical errors, will be illustrated in detail. Finally, in
Chapter 8 we will summarize and discuss results and outlook.

\chapter{Basics of QCD}
\def\da{\partial_{\alpha}}
\def\db{\partial_{\beta}}
\def\eps{\epsilon^{\alpha\beta\gamma\delta}}
\def\fab{F_{\alpha\beta}}
\def\tfab{\widetilde{F}^{\alpha\beta}}
\def\ufab{F_{\alpha\beta}^A}
\def\dfab{F^{\alpha\beta}_A}
\def\vfgd{F_{\gamma\delta}}
\def\ta{t^A}
\def\Ta{T^a}
\def\ca{C_A}  
\def\cf{C_F}  

This chapter is a very basic and brief introduction on QCD. Lots of
reviews on this subject have been written (see
\eg \cite{repalt,kis,HM}), and we could not write about it better. 
Here we will limited ourselves to introduce the essential aspects of 
the QCD framework,
as the $SU(N)$ gauge invariance, the QCD Lagrangian and the Feynman rules.
We will pay more attention on the running coupling $\as$, as it is 
an essential tool for Chapter \ref{chapt:as}. For this purpose,
a very technical issue as matching conditions on quark mass thresholds
will be reviewed with details.

\section{$SU(N)$ gauge invariance}

\noindent 
We are interested in building a Lagrangian invariant under local phase
transformations related to a non-Abelian group. We consider the Lagrangian
\bea
{\cal L} = i \bar{\psi} \gamma_{\alpha} \partial^{\alpha} \psi
\label{lag1}
\eea
and the transformation
\bea
\psi (x) \ra U (x) \psi (x)
\eea
where
\bea
U (x) =  \exp[i g \omega^A (x) \ta]
\eea
in which the $\ta$ are the $N^2 -1$ traceless Hermitean matrices generating
$SU(N)$ and satisfying the relations
\bea
\l[T^A,T^B\r]=if^{ABC}T^C,~~
(T^A)_{BC}=-if^{ABC}\,.
\eea
By convention the normalization of the $SU(N)$ matrices is chosen to be
\bea
\tr t^A t^B = T_R \delta^{AB},~~~T_R=\frac{1}{2}\,.
\label{trt}
\eea
With this choice, the generators obey the following relations:
\bea
\sum_A t_{ab}^A t_{bc}^A &=& \cf\delta_{ac},~~\cf=\frac{N^2-1}{2N} 
\label{cf} \\
\tr T^C T^D &=& \sum_{A,B} f^{ABC}f^{ABD} = \ca\delta^{CD},~~\ca=N\,.
\label{ca}
\eea
As $U$ depends on $x$, the derivative term
$\partial_{\alpha} \psi$ no longer transforms as it should: indeed
\bea
\da \psi \ra \da U\psi = (\da U) \psi~+~U \da \psi
\neq U \da \psi.
\eea
We now look for a  generalization of the derivative which does not spoil the
invariance of ${\cal L}$. We define accordingly the {\it covariant derivative}
$D_{\alpha}$ by demanding that
\bea
D_{\alpha} \psi \ra U D_{\alpha} \psi
\eea
or in operator form
\bea
D_{\alpha} \ra D_{\alpha}'=U D_{\alpha} U_{\alpha}^{-1}.
\label{covdevtra}
\eea
Since $D_{\alpha}$ is to generalize $\da$, let us introduce the {\it ansatz}
\bea
D_{\alpha} = \da I + i g A_{\alpha} (x)
\label{covdivdef}
\eea
where $A_{\alpha} (x)$ is the $N \times N$ Hermitean matrix defined by
\bea
A_{\alpha} = A_{\alpha}^{A} \ta .
\eea
The transformation requirement (\ref{covdevtra}) implies that
\bea
A_{\alpha}' = U A_{\alpha} U^{-1}
               ~-~\frac{i}{g}U \da U^{-1}.
\eea
We have enlarged our Lagrangian in order to have $SU(N)$ local invariance,but
we have introduced $N^2 -1$ vector fields to built the covariant derivative.
In order to give these fields an existence on their own, we should include
their kinetic terms in a way that does not break the original local
symmetry. The Hermitean quantity
\bea
\fab \equiv -\frac{i}{g} \l[D_{\alpha},D_{\beta} \r]
\eea
will transform covariantly since $D_{\alpha}$ does, \ie
\bea
\fab (x) \ra U (x) \fab (x) U^{-1} (x).
\eea
Using the definition of the covariant derivative (\ref{covdivdef}), we obtain
\bea
\fab =  \da A_{\beta} - \db A_{\alpha} +
         i g \l[A_{\alpha},A_{\beta} \r]
\eea
that is the Yang-Mills generalization of the field strengths of
electromagnetism. The kinetic term is then given by
\bea
{\cal L}_{YM} = - \frac{1}{4} (F_{\alpha\beta} F^{\alpha\beta}) 
\eea
with the normalization of eq.~(\ref{trt}) for the $T$-matrices. ${\cal L}$
does not depend on the representation of the 
fermions and therefore stands on its own as a highly non-trivial theory.

If we consider the equation of motion without sources we have
\bea
D^{\alpha} \fab =0.
\eea
Just as in electrodynamics, there are plane waves solutions and they have
infinite energy (but finite energy density). However, unlike in Maxwell's
theory, they cannot be superimposed to produce finite energy solutions
because of the non-linear nature of this theory, unless they move in the
same direction. In addition the $\fab$ fields satisfy the kinematic (Bianchi)
constrains 
\bea
D_{\alpha} \tfab =0
\label{dualmot}
\eea
where
\bea
\tfab = \frac{1}{2} \eps \vfgd 
\eea
is the dual of $\fab$. We emphasize that (\ref{dualmot}) is not an equation 
of motion since it is trivially solved by expressing $\fab$ in terms of the
potentials.

There is even another way to construct a gauge and Lorentz invariant
kinetic term from $\fab$, that is
\bea
I = \tr ~\eps \fab \vfgd .
\eea
We did not consider this term as it can be expressed as a pure divergence
\bea
\eps \tr \fab \vfgd = 4 \partial_{\gamma} W^{\gamma}
\label{purediv}
\eea
with
\bea
W^{\gamma} = \eps \tr \l[ A_{\delta} \da A_{\beta} ~+~
          \frac{2ig}{3} A_{\delta}A_{\alpha}A_{\beta} \r].
\eea
This means that by taking $I$ as the kinetic Lagrangian, we could not
generate any equation of motion for the vector potential since it would only
affect the action at its end points.

Finally note that there is no gauge invariant way of including a
mass for the gauge boson. A term such as
\bea
m^2 A^{\alpha}A_{\alpha}\,,
\eea
is not gauge invariant. This is very similar to the Quantum Electrodynamics 
(QED) requirement of
a massless photon. 
On the other hand a mass term for the fermions given by
\bea
m\bar{\psi}\psi\,,
\eea
is gauge invariant.

\section{QCD Lagrangian}

\noindent 
QCD is an $SU(3)$ gauge invariant field theory.
The expression for the classical Lagrangian density is
\bea
\label{classlag}
{\cal L}_{\rm classical}=-\frac{1}{4}\ufab\dfab + 
\sum_{\rm flavors} \bar{q}_a\,(i\Dslash - m)_{ab}\,q_b\,,
\eea
where $\Dslash = \gamma_{\alpha}D^{\alpha}$, the metric is given by
$g^{\alpha\beta}=(1,-1,-1,-1)$ and $\hbar=c=1$.
These terms describe the interaction of spin-$\frac{1}{2}$ quarks of
mass $m$ and massless spin-1 gluons. $\ufab$ is the field strength
tensor derived from the gluon field $A_{\alpha}^A$\,
\bea
\label{gluonfield}
\ufab=\da A_{\beta}^A - \db A_{\alpha}^A -g\,f^{ABC}
A_{\alpha}^B A_{\beta}^C
\eea
and the indices $A,B,C$ run over the eight color degrees of freedom of
the gluon field. It is the third ``non-Abelian'' term on the \rhs of
eq.~(\ref{gluonfield}) which distinguishes QCD from 
QED, giving rise to triplet and quartic gluon
self-interactions and ultimately to the property of asymptotic
freedom. Note that each term in the Lagrangian has mass dimension
four, in order to give the correct dimensions for the action when
integrated over all space-time. It follows that the dimensions of the
fields $q_a$ and $A_{\alpha}^A$ are $\frac{3}{2}$ and 1, respectively.

The explicit sum in eq.~(\ref{classlag}) runs over the $\nf$ different
flavors of quarks, $g$ in eq.~(\ref{gluonfield}) is the coupling
constant which determines the strength of the interaction between
colored quanta, and $f^{ABC}$ $(A,B,C=1,\ldots,8)$ are the structure
constants of the $SU(3)$ color group. The quark fields $q_a$ are in
the triplet representation of the color group, $(a=1,2,3)$ and $D$ is
the covariant derivative (\ref{covdivdef}). Acting on triplet and
octet fields the covariant derivatives takes form
\bea
\label{covdev}
(D_{\alpha})_{ab}=\da \delta_{ab}+ig(t^C\,A_{\alpha}^C)_{ab}\,,~~~
(D_{\alpha})_{AB}=\da \delta_{AB}+ig(T^C\,A_{\alpha}^C)_{AB}\,,
\eea
where $t$ and $T$ are matrices in the fundamental and adjoint representations
of $SU(3)$ respectively.
A representation for the generators $t^A$ is provided by the eight Gell-Mann
matrices $\lambda^A$, which are Hermitean and trace-less,
\bea
t^A=\frac{1}{2}\lambda^A\,.
\eea
For the specific case of $SU(3)$, from eqs.~(\ref{cf}) and
(\ref{ca}) we have
\bea
\cf=\frac{4}{3}\,~~~\ca=3\,.
\eea

\section{Feynman rules}

\noindent
In may be shown (see \eg \cite{peskin,ramond}) that it is not possible
to define the propagator for the gluon field, without making a choice
of the gauge. As a consequence without a choice of gauge we can not
perform perturbation theory with the Lagrangian of
eq.~(\ref{classlag}). The choice
\bea
\label{gflag}
{\cal L}_{\rm gauge-fixing}=-\frac{1}{2\lambda}\l(\partial^{\alpha}
A_{\alpha}^A\r)^2\,,
\eea
fixes the class of {\it covariant gauges} with gauge parameter $\lambda$.
Two common choices are the Feynman gauge, $\lambda=1$, and the Landau
gauge, $\lambda=0$. In the following we will consider a general case.
In a non Abelian theory such as QCD the covariant gauge fixing term must
be supplemented by a ghost Lagrangian, which is given by
\bea
\label{ghostlag}
{\cal L}_{\rm ghost}=\da \eta^{A\dag}\l(D_{AB}^{\alpha}\eta^B\r)\,.
\eea
Here $\eta^A$ is a complex scalar field which obeys Fermi statistics.
The derivation of the form of the ghost Lagrangian is best provided
by the path integral formalism and the procedures of Faddeev and Popov.
The ghost field cancel unphysical degrees of freedom which would otherwise
propagate in covariant gauges.

Eqs.~(\ref{classlag}),(\ref{gflag}) and (\ref{ghostlag}) are sufficient
to derive the Feynman rules of the theory in a covariant gauge. The Feynman
rules are defined from the operator
\bea
S=i\int d^4x {\cal L}(x) 
\eea
which gives the phase of transition amplitude, rather than from the
Lagrangian density. The Lagrangian density can be separated into a free
piece ${\cal L}_0$, which normally contains all the terms bilinear in the
fields, and an interaction piece, ${\cal L}_I$, which contains 
all the rest:
\bea
S=S_0+S_I,
\nonumber
\eea
\bea
S_0=i\int d^4x {\cal L}_0(x),~~S_I=i\int d^4x {\cal L}_I(x)\,.
\eea
The practical recipe to determine the Feynman rules is that
the inverse propagator is derived from $-S_0$, whereas the Feynman rules
for the interacting parts of the theory which are treated as perturbations
are derived from $S_I$.

In order to understand this recipe, we will follow \cite{kis} and
compare the following two different approaches to the quantization
of a theory. For simplicity, we will take a theory which contains only a
complex scalar field $\phi$ and an action which contains only bilinear
terms, $S=\phi^* (K+K')\phi$. In the first approach, both $K$ and $K'$
are included in the free Lagrangian, $S_0=\phi^* (K+K')\phi$. Using the
above rule the propagator $\Delta$ for the $\phi$ field is given by
\bea
\Delta=\frac{-1}{K+K'}\,.
\eea
In the second approach $K$ is regarded as the free Lagrangian,
$S_0=\phi^*K\phi$, and $K'$ as the interaction Lagrangian,
$S_I=\phi^*K'\phi$. Now $S_I$ is included to all orders in
perturbation theory by inserting the interaction term an infinite number
of times:
\bea
\Delta &=& \frac{-1}{K}+\l(\frac{-1}{K}\r)K'\l(\frac{-1}{K}\r)+
\l(\frac{-1}{K}\r)K'\l(\frac{-1}{K}\r)K'\l(\frac{-1}{K}\r)+\dots
\nonumber \\
&=& \frac{-1}{K+K'}\,.
\eea
We observe that with the choice of signs described above the full propagator
of the $\phi$ field is the same in both approaches, demonstrating
the internal consistency of the recipe.

\begin{figure}[t]
\begin{center}
\includegraphics[width=0.75\textwidth]{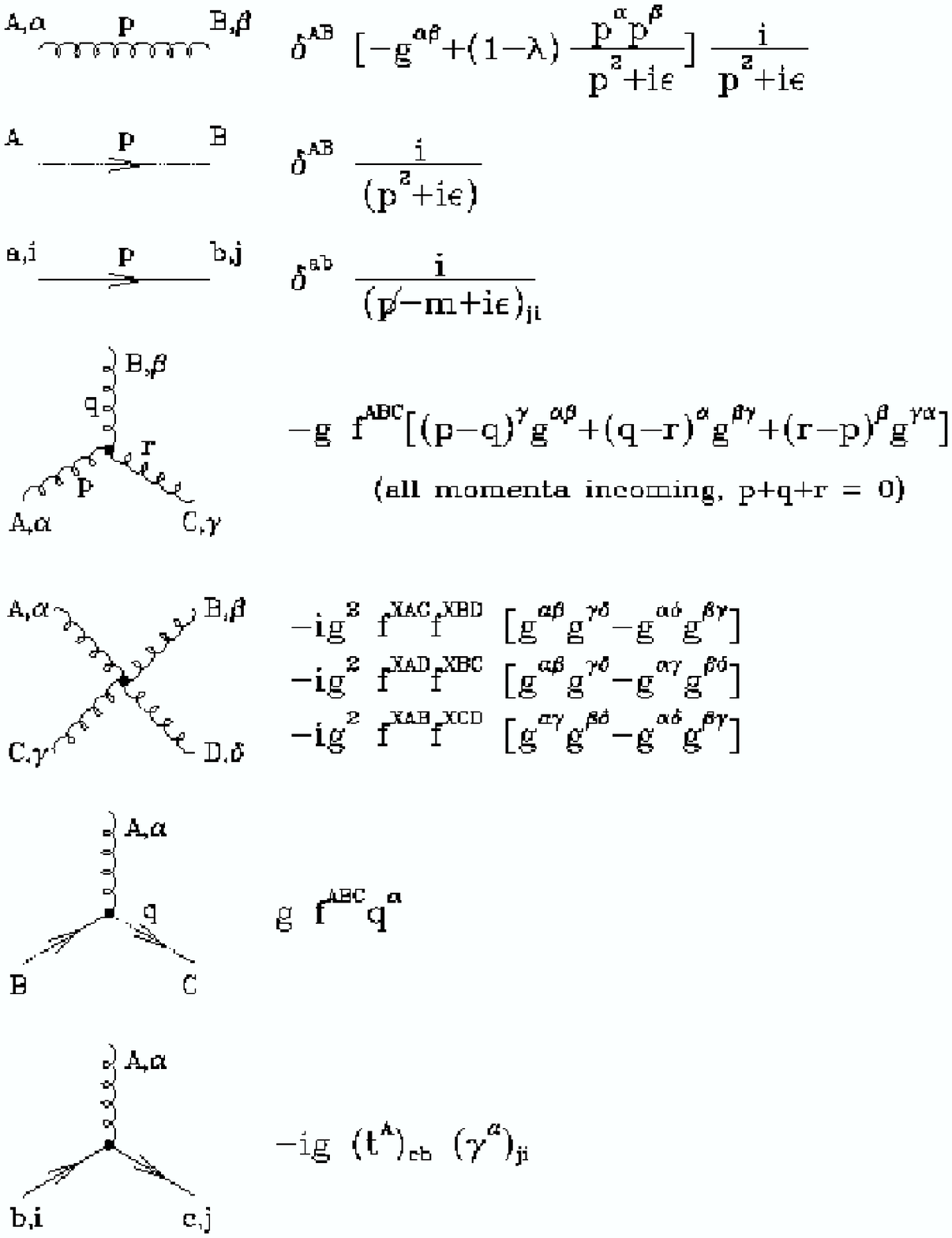}
\end{center}
\fcaption{}{Feynman rules for QCD in a covariant gauge for gluons
(curly lines), fermions (solid lines) and ghosts (dotted lines) 
as given in Ref.~\rm{\cite{kis}.}}
\label{fig:feynrules}
\end{figure}

The quark and the gluon propagators are obtained using the free piece
${\cal L}_0$ of the QCD Lagrangian given in eq.~(\ref{classlag}).
Thus, for example, the inverse fermion propagator in momentum space
can be obtained by making the identification
$\partial^{\alpha}=-ip^{\alpha}$ for an incoming field. In momentum
space the two point function of the quark field depends on a single
momentum $p$. We have
\bea
\Gamma^{(2)}_{ab}(p)=-i\delta_{ab}(\pslash -m)\,,
\eea
which is the inverse of the propagator given in Fig.~\ref{fig:feynrules}.
The $i\epsilon$ prescription  for the pole of the propagator is added
to preserve causality, exactly the same way as in QED. Similarly the inverse
propagator of the gluon field is given by
\bea
\Gamma^{(2)}_{\{AB,\alpha\beta\}}(p) = i\delta_{AB}\l[p^2g_{\alpha\beta}
-(1-\frac{1}{\lambda})p_{\alpha}p_{\beta}\r]\,.
\eea

It is straightforward to check that without the gauge fixing term this 
function would have no inverse. The result for the gluon propagator 
$\Delta$ is 
\bea
&& \Gamma^{(2)}_{\{AB,\alpha\beta\}}(p) 
\Delta^{(2)~\{BC,\beta\gamma\}}(p)
=\delta_A^Cg_{\alpha}^{\gamma} \\
&& \Delta^{(2)~\{BC,\beta\gamma\}}(p)=\delta_{BC}\frac{i}{p^2}
\l[-g_{\beta\gamma}+(1-\lambda)\frac{p_{\beta}p_{\gamma}}{p^2}\r]\,.
\eea
Replacing the derivatives with the appropriate momenta, eqs. 
(\ref{classlag}), (\ref{gflag}) and (\ref{ghostlag}) can be used to
derive all the rules in Fig.~\ref{fig:feynrules}.

\section{Running coupling}

\noindent
The QCD coupling constant is defined by
\bea
\as=\frac{g^2}{4\pi}\,,
\eea
that is the strong-interactions analogue of the fine-structure
constant. When we compute Feynman diagrams divergences arise 
and we need to regularize them. The most commonly used
regularization scheme is the  {\it Modified Mi\-ni\-mal 
Subtraction} (\MS). This involves continuing momentum
integrals from 4 to $4-2\epsilon$ dimensions, and
then subtracting off the resulting $1/\epsilon$ poles
and also $\log 4\pi - \gamma_E$  with $\gamma_E$ being the
Euler-Mascheroni constant. To preserve the dimensionless
nature of the coupling, a mass scale $\mu$ must also be introduced
and $g\ra\mu^{\epsilon}g$.
While in QED the coupling constant is defined in a natural way
by on-shell renormalization, 
in QCD, we would like to avoid discussing on-shell
quarks, since these are strongly interacting particles that are
significantly affected by non-perturbative forces. The use of an
arbitrary renormalization point $\mu$ allows us to avoid this
problem. We will define $\as$ by renormalization conditions imposed at
a large momentum scale $\mu$ where the coupling constant is small;
this value of $\as$ can then be used to predict the results of
scattering processes with any large momentum transfer.

However, the use of renormalization at a scale $\mu$ in a computation
invol\-ving momentum invariants of order $p^2$ involves some subtlety
when $p^2$ and $\mu^2$ are very different. In this circumstances,
Feynman diagrams with $n$ loops typically contain correction terms
proportional to $(\as\log(p^2/\mu^2))^n$. Fortunately, we can absorb
these corrections into the lowest order terms by using the
renormalization group to replace the fixed renormalized coupling with a
running coupling constant.

\subsection{The $\beta$ function}
\label{sect:beta}

The running of the coupling constant $\as$ is defined to satisfy 
the renormalization group equation (RGE)
\bea
\frac{\partial}{\partial t} \as (t)= \beta (t)\,,
~~~~t=\log\frac{Q^2}{\mu^2}\,,
\label{asren}
\eea
where $\mu$ is the renormalization scale and $Q$ the process energy.
In QCD, the $\beta$ function has the perturbative expansion
\bea
\beta (t) = -b \as^2 (1+b' \as+\dots)\,,
\label{beta}
\eea
with
\bea
b&=&\frac{11 C_A - 2 n_f}{12 \pi} = \frac{33 - 2 n_f}{12 \pi}\,,
\label{bbeta}\\ 
b'&=&\frac{17 C_A^2 - 5 C_A n_f -3 C_F n_f}{2\pi(11 C_A - 2 n_f)} 
= \frac{153-19n_f}{2\pi(33 - 2 n_f)}\,,
\label{bbeta1}
\eea
where $n_f$ is the number of active light flavors at the scale $Q^2$.

In the perturbative region ($\as$ small), we have
\bea
t &=& \int  \frac{dx}{\beta (x)} 
\\ \nonumber
&\simeq &
\frac{1}{b \as (Q^2)} + 
\frac{b'}{b}\log\frac{\as (Q^2)}{\as (\mu^2)}-C\,,
\label{asint}
\eea
where $C$ is the integration constant. It follows that
\bea
\frac{1}{\as (Q^2)} = bt - b' 
\log\frac{\as (Q^2)}{\as (\mu^2)}+bC.
\label{assol}
\eea
Neglecting the term $b'$, eq.~(\ref{assol}) can be written as
\bea
\frac{1}{\as (Q^2)} = bt+bC\,.
\label{aslla}
\eea
The choice of the integration constant is arbitrary, and it is
usually linked to the available scales of the energy.
Historically, the first choice was done in the '$60s$ where the
experimental energies were of the order of MeV. 
With this choice we can now write the integration constant as function 
of the normalization scale and of a parameter $\Lambda$ as
\bea
C=\log\frac{\mu^2}{\Lambda^2}\,.
\eea
With this definition $\Lambda$ represents the scale at which
the coupling would diverge, if extrapolated outside the perturbative domain.
Then, at the Leading Logarithmic Approximation (LLA) we have
\bea
\as (Q^2) = \frac{1}{b \log\frac{Q^2}{\Lambda^2}}.
\label{aslam}
\eea

We will now include the $b'$ term, too. In addition we will also let
the integration constant be dependent on a scale close to the present
experimental energies 
\bea
C=\frac{1}{b \as(M_Z^2)}\,,
\eea
where $M_Z$ is the mass of the $Z$ boson. 
Eq. (\ref{assol}) thus gives
\bea
\frac{1}{\as (Q^2)} = 
\frac{1}{\alpha_0 (Q^2)} - b' \log\frac{\as (Q^2)}{\as (M_Z^2)}
\eea
with $\alpha_0 (Q^2)$ satisfying eq. ({\ref{aslla}). Finally,
\bea
\as (Q^2) &=& \frac{\alpha_0 (Q^2)}{1 - \alpha_0 (Q^2) b' 
\log\frac{\as (Q^2)}{\as (M_Z^2)}}
\\ \nonumber 
&\simeq & \alpha_0 (Q^2)\Big[1 + \alpha_0 (Q^2)b' 
\log\frac{\alpha_0 (Q^2)}{\as (M_Z^2)}
+ {\cal O}\l(\alpha_0^2 (Q^2)\r)\Big]\,.
\eea
Since
\bea
\log\frac{\alpha_0 (Q^2)}{\as (M_Z^2)} = - \log (1+ \as (M_Z^2)\,b\, t)
\simeq - \log (1+ \alpha_0 (M_Z^2)\,b\, t)\,,
\eea
the expression we will use to fit $\as (M_Z^2)$ is given by
\bea
\as (Q^2) &=& \frac{\as (M_Z^2)}{1 + b\,\as (M_Z^2)\,\log \frac{Q^2}{M_Z^2}}
\label{asmz}
\\ \nonumber 
&\times&\l[1- \frac{b'\,\as (M_Z^2)}{1 + b\,\as (M_Z^2)\,\log \frac{Q^2}{M_Z^2}} 
\log \l(1+ b\,\as (M_Z^2)\, \log \frac{Q^2}{M_Z^2}\r)\r]\,.
\eea

For sake of completeness we give the Next-to-Logarithmic-Approximation (NLA) 
of $\as$ as a function of $\Lambda$ 
\bea
\as (Q^2) = 
\frac{1}{b\,\log \frac{Q^2}{\Lambda^2}}
\l[1 - \frac{b'}{b}\frac{\log \log \frac{Q^2}{\Lambda^2}}
{\log \frac{Q^2}{\Lambda^2}}\r].
\label{asnlo}
\eea
Finally note that $\Lambda$ depends on the
logarithmic approximation used. We have
\bea
\frac{\Lambda_1}{\Lambda_2}= \l( \frac{b}{b'}\r)^{\frac{b'}{2b}}.
\label{kislam} 
\eea
where $\Lambda_1$ and $\Lambda_2$ are defined at LLA and NLA
respectively.

\subsection{Asymptotic freedom and confinement}

From eq.~(\ref{asren}) and (\ref{beta}) we have that, if $b>0$,
\ie $n_f<16$ as follows from eq.~(\ref{bbeta}), the coupling constant 
tends to zero at a logarithmic rate as the momentum scale increases. Such
theories are called {\it asymptotically free}. In theories of this class,
the short distance behavior is completely solvable by Feynman diagram
methods. Though ultraviolet divergences appear in every order of
perturbation theory, from the renormalization group analysis
follows that the sum of
these divergences is completely harmless. If we interpret these
theories in terms of a bare coupling and a finite cutoff $\Lambda$,
eq.~(\ref{aslam}) indicates that there is a smooth limit in which
$\as$ tends to zero as $\Lambda$ tends to infinity.

To briefly describe the consequences of asymptotic freedom, we now
consider an example coming from electrodynamics.  The simplest phase
of the electrodynamics is the Coulomb phase. It is characterized by
massless photons which mediate a long range $1/R$ potential between
external sources.  When charged matter particles are present,
electrodynamics can be in another phase, the superconducting or the
Higgs phase. It is characterized by the condensation of a charged
field
\bea
\langle \Phi \rangle \neq 0.
\eea
This condensation creates a gap in the spectrum by making the photon
massive.  This phenomenon was first described in the context of
superconducti\-vity, where $\Phi$ is the Cooper pair. The condensation of
$\Phi$ makes electric currents superconducting. Its effect on magnetic
fields is known as the Meissner effect.  Magnetic fields cannot
penetrate the superconductor except in thin flux tubes.  Therefore,
when two magnetic monopoles (\eg the ends of a long magnet) are
inserted in a superconductor, the flux lines are not spread. Instead,
a thin flux tube is formed between them. The energy stored in the flux
tube is linear in its length and therefore the potential between two
external monopoles is linear (as opposed to the $1/R$ potential
outside the superconductor). Such a linear potential is known as a
{\it confining} potential. The same happens when we consider {\it
quark confinement}. As a result we have that if one
attempts to separate a color singlet state into colored components, \ie
to dissociate a meson into a quark and an antiquark, a tube of gauge
field forms between the two sources. In a non-Abelian gauge theory with
sufficiently strong coupling this tube has a fixed radius and energy
density, so the energy cost of separating color sources grows
proportionally to the separation. A force law of this type can
consistently be weak at short distances and strong at long distances,
accounting for the fact that isolated quarks are not observed.
An interesting description of confinement from a topological
point of view can be found in \cite{tH}.

\subsection{Thresholds at quark masses}
\label{sect:massth}

We now consider the case of a heavy quark with mass $m$ much greater
than the relevant scale $Q$. In this case it can be shown that the
effects of the heavy quark on cross sections are suppressed by inverse
powers of $m$ and can therefore be ignored for $m\ll Q$
\cite{AC75}. 

In the \MS~renormalization scheme $\as$ depends only on
the renormalization scale. As a consequence heavy
quarks contributions to the running of $\as$ must be taken into
account even when we compute a physical quantity at an energy lower
then the heavy quark mass scale. However, logarithms of large masses induced
by the renormalization group equations in the couplings are canceled
against other logarithms that appear in the calculation of physical
observables. This is obviously an inconvenient, since a lot of effort
must be invested in intermediate stages of a calculation to compute
terms that will cancel in physical quantities.

To remedy this problem the standard procedure has been the use of the
effective field theory language.  For example, in QCD with a heavy
quark and $\nf-1$ light quarks, one builds a theory with $\nf$ quarks
and an effective field theory with $\nf-1$ quarks. Around the
threshold of the heavy quark we require agreement of the two
theories. This gives a set of matching equations that relate the
couplings of the theory with $\nf$ quarks with the couplings of the
theory with $\nf-1$ quarks. This way, below the heavy quark threshold
one can work with the effective theory, but using effective
couplings. Then, by construction, decoupling is trivial. This
procedure is equivalent to other renormalization schemes and allows us
to correctly obtain the asymptotic value of the coupling constant.
The price one has to pay is that coupling constants might not be
continuous at thresholds.  However, the fact that one has to use
appropriate matching conditions in passing thresholds has been
frequently kept into account in the running of the QCD coupling constant by
just taking a continuous coupling constant across thresholds. Then,
the final results depend strongly on the exact scale one uses to
connect the couplings. To solve this ambiguity we can vary the
matching scale $\mu_{th}$ between $1/2$ and $2$ times the mass of the
heavy quark. It can be shown \cite{Rodrigo} that when appropriate
matching conditions are taken into account the final answer does not
depend on the exact $\mu_{th}$ used to connect the couplings.

As an example, we take eq. (\ref{assol}) at LLA and we require that
$\as (Q^2)$ should be continuous at $Q^2=\mu_{th}^2=k_{th}\, m^2$, 
with $m$ the heavy quark mass
\bea
\frac{1}{\alpha_+(Q^2)} - b_+\,t = \frac{1}{\alpha_-(Q^2)} - b_-\,t\,,~~~~
t=\log \frac{Q^2}{\mu_{th}^2}
\eea
where index $+$ and $-$ refers to variables above and below the threshold.
It follows that
\bea
\frac{1}{\alpha_+(Q^2)} = \frac{1}{\alpha_-(Q^2)} +(b_+ - b_-)\,t =
\frac{1 + \alpha_-(Q^2) (b_+ - b_-)\,t}{\alpha_-(Q^2)}\,.
\eea
Finally, we get
\bea
\alpha_+ (Q^2)&=& \frac{\alpha_-(Q^2)}{1 + \alpha_-(Q^2) (b_+ - b_-)\,t}=
\frac{\alpha_-(Q^2)}{1 - \frac{\alpha_-(Q^2)}{6\pi} 
\log \frac{Q^2}{\mu_{th}^2}}
\nonumber \\
&\simeq & \alpha_-(Q^2)\l(1+\frac{\alpha_-(Q^2)}{6\pi} 
\log \frac{Q^2}{\mu_{th}^2}\r)\,.
\label{thresh1}
\eea
The generalized version can be written as
\bea
\alpha_+ (Q^2)= \alpha_-(Q^2) + \sum_{k=1}^{\infty} 
C_k\l(\log \frac{Q^2}{\mu_{th}^2}\r)
\alpha_-^{k+1}(Q^2)
\eea
where
\bea
C_1(x) = \frac{1}{6\pi}\,x
\eea

Let now $\as (Q^2,n_f)$ be the solution of eq. (\ref{assol}) at LLA
with $\Lambda=\Lambda_5$ and fixed $n_f$ and let $\alpha_{s(n_f)}$ 
be the value of the coupling constant at a given $n_f$.
Eq. (\ref{thresh1}) can be written as (see \eg \cite{roberts})
\bea
\alpha_{s(6)}^{-1}(Q^2) &=& \as^{-1} (Q^2,6) + C_{65}\,, \nonumber \\
\alpha_{s(5)}(Q^2) &=& \as (Q^2,5)\,, \nonumber \\
\alpha_{s(4)}^{-1}(Q^2) &=& \as^{-1} (Q^2,4) + C_{45}\,, \nonumber \\
\alpha_{s(3)}^{-1}(Q^2) &=& \as^{-1} (Q^2,3) + C_{34} + C_{45}\,,
\label{ast1}
\eea
where
\bea
C_{65} &=& \as^{-1} (m_t^2,5)-\as^{-1} (m_t^2,6)\,, \nonumber \\ 
\label{ast2}
C_{45} &=& \as^{-1} (m_b^2,5)-\as^{-1} (m_b^2,4)\,, \\ \nonumber
C_{34} &=& \as^{-1} (m_c^2,4)-\as^{-1} (m_c^2,3)\,,
\eea
and we have set $\mu_{q\,th}=m_q$ for simplicity. As an example, we consider
\bea
&& \alpha_{s(5)}(Q^2)- \alpha_{s(4)}(Q^2) = \nonumber \\ 
&& = \alpha_{s(5)}(Q^2) -  
\frac{\as (Q^2,4)}{1 + \as (Q^2,4)\frac{\as (m_b^2,5)-\as (m_b^2,4)}
{\as (m_b^2,5) \as (m_b^2,4)}}
\label{thresh2}
\\ \nonumber 
&&\simeq \as (Q^2,5) - \as (Q^2,4) + 
\as (Q^2,4)\as (Q^2,5)\frac{\as (m_b^2,5)-\as (m_b^2,4)}
{\as (m_b^2,5) \as (m_b^2,4)}\, \\ \nonumber
&& \simeq  (b_5 - b_4) \as^2 (Q^2,4)\log \frac{Q^2}{m_b^2}\,.
\eea
This expression coincides with eq. (\ref{thresh1}) with 
$\alpha_{s(5)}(Q^2) = \alpha_+(Q^2)$ and 
$\alpha_{s(4)}(Q^2) =\alpha_-(Q^2)$. Note that eqs.~ 
(\ref{ast1}) and (\ref{ast2}) hold even at NLA.

To conclude we observe that, at each threshold also
$\Lambda_{n_f}$ changes. We now fix $\Lambda=\Lambda_5$ and we will
show how different values of $\Lambda$ for different $n_f$ are related
to each other. If we consider the NLA definition of $\Lambda$, we have
\bea
\Lambda = \frac{Q}{t/2} = Q \exp \l[-\frac{1}{2}\l(
\frac{1}{b \as (Q^2)} + \frac{b'}{b} \log b'\,\as (Q^2)\r)\r]\,.
\eea
If we require that $\as(Q^2)$ given by is
continuous at a threshold (we consider as an example $Q = m_b$) 
\bea
b_4\,\log \frac{m_b^2}{\Lambda_4^2}+
b_4'\log \log \frac{m_b^2}{\Lambda_4^2} =
b_5\,\log \frac{m_b^2}{\Lambda_5^2}+
b_5'\log \log \frac{m_b^2}{\Lambda_5^2}\,,
\eea
we find
\bea
\log\l[\Lambda_5^{-2 b_5} \Lambda_5^{2 b_4} m_b^{2(b_5-b_4)}\r] =
\log\l[\l(\log \frac{m_b^2}{\Lambda_5^2}\r)^{-b'_5}
\l(\log \frac{m_b^2}{\Lambda_4^2}\r)^{b'_4}\r]\,,
\eea
and
\bea
\Lambda_4^{2b_4} &=& \Lambda_5^{2b_5} m_b^{-2(b_5-b_4)}
 \l(\log \frac{m_b^2}{\Lambda_5^2}\r)^{-b'_5}
\l(\log \frac{m_b^2}{\Lambda_4^2}\r)^{b'_4}
\nonumber \\
&\simeq & \Lambda_5^{2b_5} m_b^{-2(b_5-b_4)}
\l(\log \frac{m_b^2}{\Lambda_5^2}\r)^{-b'_5+b'_4}\,;
\eea
it follows that \cite{marciano}
\bea
\Lambda_4 &=& \Lambda_5 \l(\frac{\Lambda_5}{m_b}\r)^{\frac{b_5-b_4}{b_4}}
\l[\log \frac{m_b^2}{\Lambda_5^2}\r]^{-\frac{b'_5-b'_4}{2 b_4}} 
\nonumber \\
&=& \Lambda_5 \l(\frac{m_b}{\Lambda_5}\r)^{\frac{2}{25}}
\l[\log \frac{m_b^2}{\Lambda_5^2}\r]^{\frac{963}{14375}}\,.
\eea

\chapter{QCD and the parton model}
\label{chapt:parton}
\def\ddt{\frac{d}{dt}}

\noindent
Deep inelastic scattering (DIS) has been the process that established
the first evidence of partons, and it is the traditional testing ground of
perturbative QCD. Nowadays, it is no longer the correctness of the
theory which come into question. Rather, the focus is shifted on the
precise determinations of the unknown parameters, and the development
of reliable computational techniques. The parameters to be determined
include not only the strong coupling $\as$, but also all quantities
determined from the non-perturbative low-energy dynamics, and that,
even though in principle computable, must be treated as
phenomenological input in the perturbative domain.

In this Chapter we will briefly discuss
deep inelastic scattering and the ``naive parton model''.
Then we will show how QCD modifies the simple Bjorken scaling property
of the parton model, and discuss how these scaling violations can be 
calculated in perturbation theory. We will also discuss issues 
useful for the following Chapters, as the factorization 
schemes.

\section{Deep inelastic scattering and the parton model}

\noindent
We now consider the scattering of a high-energy charged lepton, say, an
electron, off a hadron target as shown in Fig.~\ref{fig:dis}. If we
label the incoming and outgoing lepton four-momenta by $k^{\mu}$ and
$k'^{\mu}$ respectively, the momentum of the target hadron (assumed
henceforth to be a proton) by $p^{\mu}$ and the momentum transfer by
$q^{\mu}=k^{\mu}-k'^{\mu}$, then the standard deep inelastic variables
are defined by
\bea
Q^2 &=& -q^2 
\nonumber \\
M^2 &=& p^2
\nonumber \\
\nu &=& p\cdot q=M(E'-E)
\\
x &=& \frac{Q^2}{2\nu}=\frac{Q^2}{2M(E'-E)}
\nonumber \\
y &=& \frac{q\cdot p}{k\cdot p}=1-\frac{E'}{E}\,,
\nonumber 
\eea
where the energy variables refer to the target rest frame and $M$ is the 
proton mass. Henceforth we will consider only
charged lepton scattering, $lp\ra lX$, for $Q^2\ll M_Z^2$, where
the scattering is mediated by the exchange of a virtual photon.
A review on deep inelastic neutrino scattering can be found \eg in
Ref.~\cite{conrad}.

\begin{figure}[t]
\begin{center}
\mbox{\includegraphics[width=0.48\textwidth,clip]{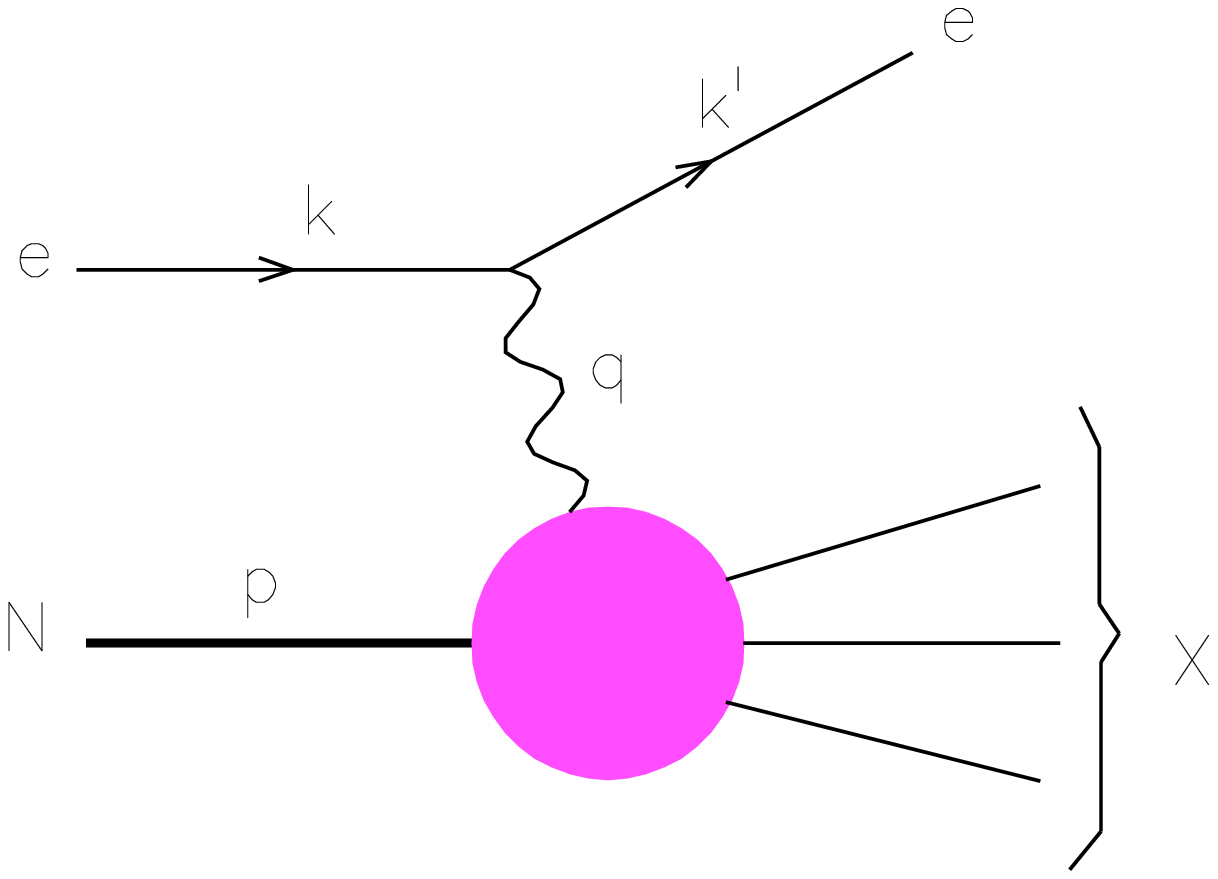}}\\
\fcaption{}{Deep inelastic charged lepton-proton scattering.}
\label{fig:dis}
\end{center}
\end{figure}

The DIS cross section is given by
\bea
\frac{d^2\sigma}{dxdy} &=& \frac{8\pi\alpha^2ME}{Q^4}
\l[\frac{1+(1-y)^2}{2}\,2xF_1 \r.
\nonumber \\
&+& \l. (1-y)(F_2-2xF_1) - \frac{M}{2E}xyF_2\r]\,,
\label{sigmaem}
\eea
where $F_i(x,Q^2)$ are the nucleon structure functions which carry the
information on the structure of the target as seen by the virtual photon.
The {\it Bjorken limit} is defined as $Q^2,\nu\ra\infty$ with $x$ fixed.
In this limit the structure functions are observed to obey an approximate
{\it scaling} law \cite{bjorken,fey}, \ie they depend only on the 
dimensionless variable $x$:
\bea
F_i(x,Q^2)\ra F_i(x)\,.
\eea
Bjorken scaling implies that the virtual photon scatters off {\it
point-like constituents}, since otherwise the dimensionless structure
functions would depend on the ratio $Q/Q_0$, with $1/Q_0$ some length
scale characterizing the size of constituents. The parton model
picture of deep inelastic scattering is most easily formulated in the
``infinite momentum frame'' in which the proton is moving very fast,
$p^{\mu}\approx (P,0,0,P)$ with $P\gg M$. In this frame, we can
consider a simple model where the photon scatters off a point-like
quark constituent which is moving parallel with the proton and
carrying a fraction $\xi$ of its momentum, \ie $p_q^{\mu}=\xi
p^{\mu}$. Neglecting the proton mass $M$, we can write
eq.~(\ref{sigmaem}) as
\bea
\frac{d^2\sigma}{dxdQ^2}=\frac{4\pi\alpha^2}{Q^4}
\l[[1+(1-y)^2]F_1+\frac{1-y}{x}\l(F_2 - 2xF_1\r)\r]\,.
\label{sigmaparton1}
\eea
In terms of the usual Mandelstam variables 
\bea
\hat{s}&=&(k+p_q)^2=\frac{\xi Q^2}{xy}\,,
\nonumber \\
\hat{t}&=&(k-k')^2=-Q^2\,,
\\ \nonumber
\hat{u}&=&(p_q-k')^2=\hat{s}(y-1)\,, 
\eea
the matrix element squared for the amplitude of the process
\bea
l(k)+q(p_q)\ra l(k')+q(p'_q)
\eea
is given by
\bea
\overline{\sum} \l|{\cal M}\r|^2 = 2e_q^2e^4
\frac{\hat{s}^2+\hat{u}^2}{\hat{t}^2}\,.
\eea
The notation $\overline{\sum}$ denotes the average (sum) over initial (final)
colors and spins. Using the standard result for massless $2\ra 2$ scattering,
\bea
\frac{d\hat{\sigma}}{d\hat{t}}=\frac{1}{16\pi\hat{s}^2}
\overline{\sum} \l|{\cal M}\r|^2\,,
\eea
and substituting for the kinematic variables in the matrix element squared,
gives
\bea
\frac{d\hat{\sigma}}{dQ^2}=\frac{2\pi\alpha^2e_q^2}{Q^4}[1+(1-y)^2]\,.
\eea
The mass-shell constraint for the outgoing quark,
\bea
p_q'=(p_q+q)^2=q^2+2p_q\cdot q=-2p\cdot q(x-\xi)=0\,,
\eea
implies $\xi=x$. By writing $\int_0^1dx\delta(x-\xi)=1$, we obtain the double
differential cross section for the quark scattering process:
\bea
\frac{d\hat{\sigma}}{dxdQ^2}=\frac{4\pi\alpha^2}{Q^4}[1+(1-y)^2]\frac{1}{2}
e_q^2\delta(x-\xi)\,.
\label{sigmaparton2}
\eea
By comparing eqs.~(\ref{sigmaparton1}) and (\ref{sigmaparton2}) 
we get that the structure functions in this simple model are
\bea
\hat{F}_2=xe_q^2\delta(x-\xi)=2x\hat{F}_1\,.
\eea
This result suggests that the structure function $F_2(x)$ ``probes'' a quark
constituent with momentum fraction $\xi=x$. 

The above ideas are incorporated in what is known as the ``naive
parton model'' \cite{fey}:
\begin{itemize}
\item $q(\xi)d\xi$ represents the probability that a quark $q$ carries a 
momentum fraction between $\xi$ and $\xi+d\xi$, where $0\leq\xi\leq 1$;
\item the virtual photon scatters incoherently off the quark constituents.
\end{itemize}
Thus the proton structure functions are obtained by weighting the
quark structure functions with the probability distribution $q(\xi)$,
\bea
F_2=2xF_1&=& \sum_{q,\bar{q}}\int_0^1d\xi\,q(\xi)xe_q^2\delta(x-\xi) 
\nonumber \\
&=& \sum_{q,\bar{q}} e_q^2 x q(x)\,.
\label{callangross}
\eea
The \lhs of eq.~(\ref{callangross}) is the {\it Callan-Gross relation} 
\cite{CG}) and it
is a direct consequence of the spin-$\frac{1}{2}$ property of the
quarks. Indeed, we observe that the two terms in the square
brackets on the \rhs of eq.~(\ref{sigmaparton1}) correspond to the
absorption of transversely ($F_1$) and longitudinally
($F_L\ra F_2-2xF_1$ in the Bjorken limit) polarized virtual photon. The
Callan-Gross relation follows from the fact that a spin-$\frac{1}{2}$
quark cannot absorb a longitudinally polarized vector boson. In
contrast, spin-0 (scalar) quarks cannot absorb transversely polarized
vector bosons and so would have $F_1=0$, \ie $F_L=F_2$. Structure
function measurements show that $F_L\ll F_2$, confirming that
spin-$\frac{1}{2}$ property of quarks. Note that in the QCD-improved
parton model $F_L$ in only non-zero at leading order in perturbation
theory, \ie $F_L=O(\as)$.

\section{Factorization}
\label{sect:fact}
\noindent
Factorization allows scattering amplitudes with incoming high-ener\-gy 
had\-rons to be written as a product of a hard scattering piece and a
reminder factor which carries the information on the physics at 
low-energies and momenta. The first term contains only high-energy and
momentum components, and, because of asymptotic freedom, it is calculable
in perturbation theory. The second term describes the non-perturbative
physics, and it is given by a single process-independent function
for each type of parton called the {\it parton distribution function} (PDF).  
Factorization is an essential tool as it ensures that
a parton distribution measured in one process can be used
in any other hard process. A detailed discussion of factorization is
beyond the aim of this thesis, hence only a brief description will be
given.

In the naive parton model, when terms down by powers of $1/Q^2$ are
neglected, the DIS cross section eq.~(\ref{sigmaparton1}) can be
obtained by the convolution of the distribution of a quark parton in the
target, with fraction $y$ of the target longitudinal momentum, with
the point-like cross section for quark-current scattering
\bea
d\sigma_{eN}=\sum_{q=1}^{n_f} q_{q}^{(0)} \otimes d\sigma_{eq}^{(0)} ,
\label{fftt}
\eea
where the index $0$ stands for quantities calculated without 
taking into account strong interactions and
the convolution with respect to $x$ is defined as
\bea
(f \otimes g) (x)=\int_{x}^{1}\frac{dy}{y} f(y)g\l(\frac{x}{y}\r)\,.
\eea
The qualitative observation that hadrons emerge with a transverse momentum
$k_T$ different from zero can be explained with the presence of gluon 
emission. In a parton model without gluons, all final-state jets
would be collinear with the virtual photon. Their hadron fragments 
will therefore be nearly collinear with the photon, that is, with a 
spread of $k_T$ of about 300 MeV as required by the uncertainty
principle for confined quarks. The data clearly establish an 
excess of large $k_T$ hadrons which are the fragments of the quark 
and gluon jets recoiling against one another.

We will now we consider QCD corrections to eq. (\ref{fftt}) at the
first order of the perturbative expansion in $\as$. In calculating these
partonic cross sections we encounter divergences. These divergences
are
\begin{itemize}
\item {\it collinear}, if a massless parton radiates a collinear massless
parton whose momentum is proportional to that of the emitting parton;
\item {\it soft}, if a parton emits or absorbs a massless parton with
zero energy. 
\end{itemize}
When we sum the virtual contributions to the cross section given by
diagrams in Fig.~\ref{fig:ir} a,b and the real contributions given
by diagrams in Fig.~\ref{fig:ir} c,d, we obtain that soft singularities
are canceled. We are then left with the only contribution of collinear
divergences that is proportional to
\bea
\frac{\as}{2\pi}\log\frac{Q^2}{\mu^2}\,,
\eea
where $\mu^2$ is an arbitrary scale.
\begin{figure}[t]
\begin{center}
\epsfig{width=0.48\textwidth,figure=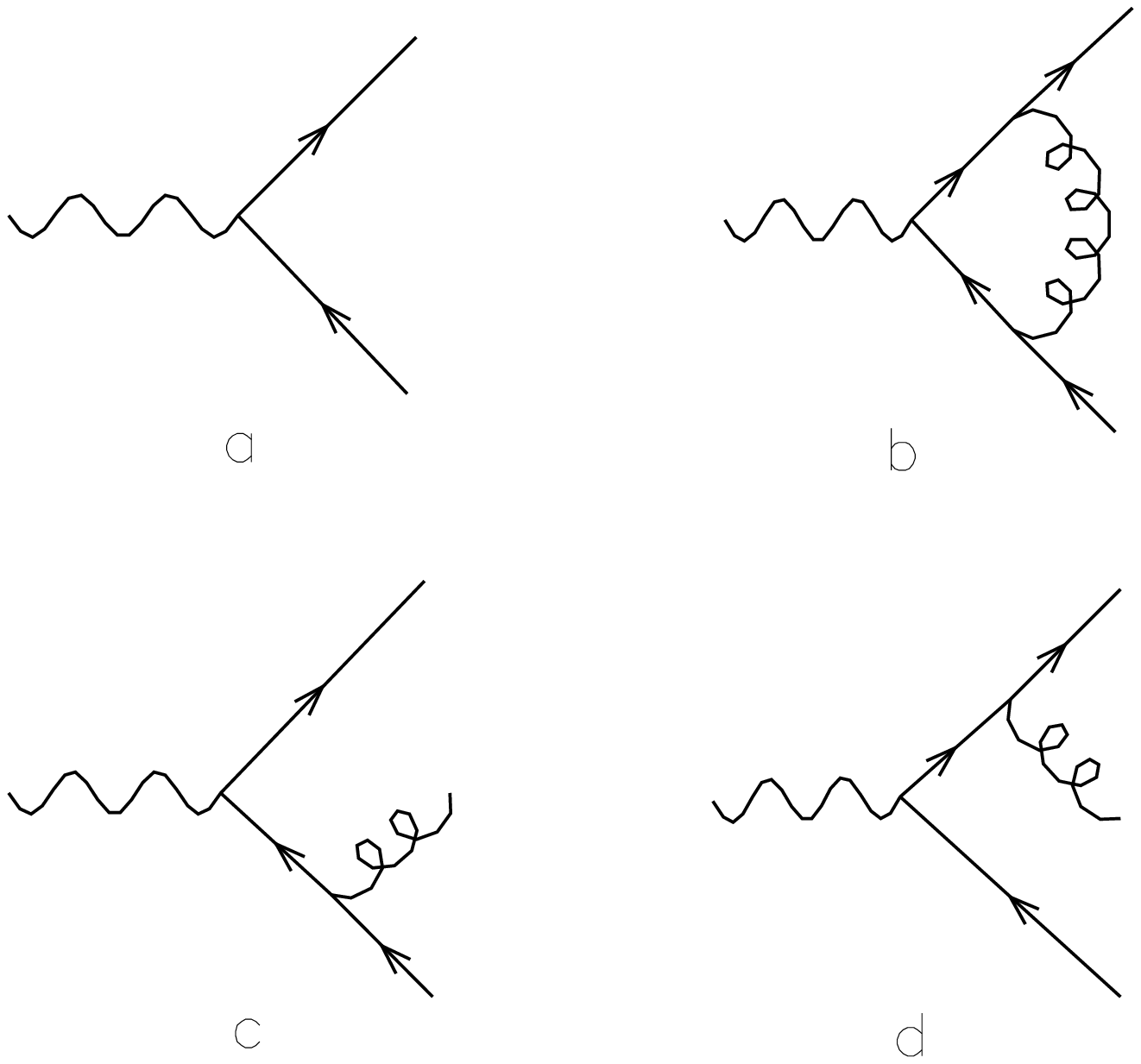}  
\end{center}
\fcaption{}{Diagrams giving corrections of order $\as$ to the
point-like quark-current cross sections.}
\label{fig:ir}
\end{figure}
If we take the initial parton, for instance, equal to a quark, the cross section can be
written as
\bea
d\sigma  &=& q^{(0)} \otimes \l[d\sigma_q^{(0)}+\frac{\as}{2\pi}
d\sigma_{q}^{(1)}\log
\frac{Q^2}{\mu^2}+\frac{\as}{2\pi}d\hat{\sigma}_{q}^{(1)}\r] 
\label{ridefdist1} \\ \nonumber
&+& g^{(0)} \otimes \l[\frac{\as}{2\pi}d\sigma_{g}^{(1)}\log\frac{Q^2}{\mu^2}+
\frac{\as}{2\pi}d\hat{\sigma}_{g}^{(1)}\r]
\eea
where we have neglected the sum over the flavors. It can be shown
by direct calculation that
\bea
d\sigma_{q}^{(1)} &=& P_{qq} \otimes d\sigma_q^{(0)}\,, \nonumber \\  
\label{fattotut}   \\
d\sigma_{g}^{(1)} &=& P_{qg} \otimes d\sigma_g^{(0)}\,. \nonumber
\eea
where $d\sigma_i^{(0)}$ are the Born cross sections and $P_{ij}$ are
process independent quantities that will be discussed in more detail
later. The cross sections $d\hat{\sigma}_{i}$ are regular.
Eq. (\ref{fattotut}) is very important as it allows to factorize the Born 
cross section $d\sigma^{(0)}$ and redefine the PDF including in this redefinition 
the contributions of the collinear divergences. We define
\bea
q(x,\,Q^2) &=& q^{(0)} \otimes \l[\delta (1-x)+\frac{\as}{2\pi}P_{qq}\log
\frac{Q^2}{\mu^2}\r]+g^{(0)}
\otimes \frac{\as}{2\pi}P_{qg}\log \frac{Q^2}{\mu^2}\nonumber \\
g(x,\,Q^2) &=& g^{(0)} (x)+O(\as)\,.
\label{redefq}
\eea
We can proceed in the same way when we have a gluon in the initial state.
Thus, we have
\bea
q(x,\,Q^2) &=& \l[\delta (1-x)+\frac{\as}{2\pi}P_{qq}\log
\frac{Q^2}{\mu^2}\r]\otimes q^{(0)} +
\frac{\as}{2\pi}P_{qg}\log \frac{Q^2}{\mu^2}\otimes g^{(0)}\,, \nonumber \\
\label{redefpdf}
\\ \nonumber
g(x,\,Q^2) &=& \frac{\as}{2\pi}P_{gq}\log \frac{Q^2}{\mu^2}\otimes q^{(0)} +
   \l[\delta (1-x)+\frac{\as}{2\pi}P_{gg}\log \frac{Q^2}{\mu^2}\r]
   \otimes g^{(0)}\,.
\eea
In a matrix form we have 
\bea
\left (\matrix{q^{(0)} \cr g^{(0)}}\right )=Z\otimes \left (
\matrix{q \cr g}\right )
\label{AP1}
\eea
with
\bea
Z=I-\frac{\as}{2\pi}\log \frac{Q^2}{\mu^2} \left (\matrix{P_{qq} & P_{qg}
\cr P_{gq}& P_{gg}}\right)+O(\as^2)\,,
\label{Zmatrix}
\eea
where $q^{(0)}$ and $g^{(0)}$ are the divergent scale independent
distributions, while $q$ and $g$ are the renormalized scale dependent
distributions.

We can picture the redefinition of PDFs as follows. 
As $Q^2$ is increased to $Q^2\sim
Q_0^2$ where $Q_0^2$ is some scale characterizing the process, say, 
the photon starts to ``see'' evidence for the point-like
valence quarks within the proton. If the quarks were non-interacting,
no further structure would be resolved increasing $Q^2$:
scaling would set in, and the naive parton model would be
satisfactory. However, QCD predicts that on increasing the resolution
($Q^2\gg Q_0^2$), we should see that each quark is itself surrounded by a
cloud of partons. The number of resolved partons which share the
proton's momentum increases with $Q^2$. 

The scale $\mu^2$ is not physical, being introduced arbitrarily in the
collinear divergences subtraction. Varying $\mu^2$ we have a
corresponding variation in the cross section at the order $\as$. As a
variation of $\as$ gives a contribution of order $\as^2$, these
correction are negligible. If we had the whole series in $\as$, all
the terms proportional to $\mu^2$ would cancel each other and we would
not have any dependence on $\mu^2$. Thus, the $\mu^2$ dependence is a
consequence of the truncation of the perturbative expansion. Usually,
we assign $\mu^2$ a value of the order of the process energy scale, as
to avoid large logarithm to appear. In the DIS case we choose $\mu^2
\simeq Q^2$.

\subsection{Coefficient functions}
 
With the redefinition of PDFs given in eq. (\ref{redefpdf}),
the cross section in eq. (\ref{ridefdist1}) can be written as
\bea
d\sigma  (x,\,Q^2)&=& q (x,\,Q^2) \otimes \l[d\sigma_q^{(0)}+
\frac{\as}{2\pi}d\hat{\sigma}_{q}^{(1)}\r] + 
g (x,\,Q^2) \otimes \frac{\as}{2\pi}d\hat{\sigma}_{g}^{(1)}
\nonumber \\
&=&  {\cal C}^q (x) \otimes q (x,\,Q^2)+{\cal C}^g (x)\otimes g (x,\,Q^2)\,,
\eea
where ${\cal C}^{q,g}$ are the coefficient functions that contain the finite
part of the partonic cross section.
The structure function $F_2$ is given by
\bea
F_2 (x,\,Q^2) &=&
x \int_x^1 \frac{dy}{y} \l\{
\sum_{q=1}^{\nf} e_q^2 \,q_q (y, Q^2) \l[\delta\l(\frac{x}{y}-1\r) 
+ \frac{\as}{2\pi} d\hat{\sigma}_{q}^{(1)}(x)\r]\r. 
\nonumber \\ 
&+&  2 \nf \,g (y, Q^2) \frac{\as}{2\pi}
d\hat{\sigma}_{g}^{(1)}(x) \Bigg\} 
\label{f2LT}
\eea
where we have used the fact that $d\sigma_q^{(0)} (x) = \delta (1-x)$.
In the \MS\ scheme the NLO contributions to the
coefficient functions are given by \cite{buras}
\bea 
\label{cfq}
d\hat{\sigma}_{q}^{(1)}(x) &=& C^q(x)
\\ \nonumber 
&=& \cf\Bigg[2\ionefive-\frac{3}{2}\ieight
-(1+x)\log(1-x)
\nonumber \\ \nonumber 
&&\phantom{aaa} -\frac{1+x^2}{1-x}\log x+3+2x
-\left(\frac{\pi^2}{3}+\frac{9}{2}\right)\delta(1-x) \Bigg]~,
\\
d\hat{\sigma}_{g}^{(1)}(x) &=& C^g(x)
\label{cfg}\\ \nonumber
&=&
T_R\Bigg[\left((1-x^2)+x^2\right)\log\frac{1-x}{x}-8x^2+8x-1\Bigg]\,.
\nonumber
\eea
The next-to-next-to-leading order (NNLO) contribution to coefficient functions
have been calculated in \cite{NNLO}. 
The coefficient functions are process dependent quantities and can be
computed from the Renormalization Group Equation approach to the
Operator Product Expansion of two currents. They describe the
deviation from the canonical behavior of free field theory.

\section{Evolution equations}

\subsection{Scaling violations and the Altarelli-Parisi \\~equations}
\label{sect:AP}

The redefinition of parton distributions given in eq.~(\ref{AP1}),
introduces a scale dependence, and the redefinition of parton
distribution means that we have substituted the bare quarks and gluons
with clouds of partons. This is right what it happens when we
renormalize a coupling constant. We will now derive the analogue of
the RGE for parton distribution. Thus we require the invariance of
eq.~(\ref{AP1}) from the scale $\mu^2$; we get
\bea
\frac{d}{dt} \left (\matrix{ q \cr g}\right )=\frac{\as}{2\pi}
\left (\matrix{P_{qq} & P_{qg} \cr P_{gq}& P_{gg}}\right) \otimes
\left (\matrix{ q \cr g }\right )+O(\as^2)
\label{AP}
\eea
where $t=\log \frac{Q^2}{\mu^2}$, which is the Altarelli-Parisi
equation for the unpolarized case \cite{AP}. We can interpreted this 
equation saying that a quark with momentum fraction $x$ can 
be produced by another quark with a larger momentum fraction $y$ which 
has emitted a gluon. 
The functions $P_{ij}$ are the Altarelli-Parisi splitting functions. They
are calculated perturbatively as power series in $\as$. At the leading-order
they have an attractive physical interpretation as the probabilities of
finding a parton of type $i$ in a parton of type $j$ with a fraction
$x$ of the longitudinal momentum of the parent parton and a transverse 
momentum squared much less than $\mu^2$. The interpretation as
probabilities implies that the splitting functions are positive definite
for $x<1$, and satisfy the sum rules
\bea
\int_{0}^{1} dx P_{qq}(x) &=& 0, \nonumber \\
\int_{0}^{1} dx x[P_{qq}(x)+P_{gq}(x)] &=& 0,  \\
\int_{0}^{1} dx x[2n_{f}P_{qg}(x)+P_{gg}(x)] &=& 0. \nonumber
\label{sumrulesP}
\eea
The leading-order \cite{AP} and the next-to-leading-order
\cite{Furmanski} contribution to the Altarelli-Parisi splitting
functions have been calculated. Here we show only the LO
contributions:
\bea
P_{qq}^{(0)}(x) &=& C_{F}\l[\frac{(1+x^{2})}{(1-x)_{+}}+\frac{3}{2}\delta (1-x)\r],
              \nonumber \\
P_{qg}^{(0)}(x) &=& T_{R}\l[x^{2}+(1-x)^{2}\r], \nonumber \\  \\
P_{gq}^{(0)}(x) &=& C_{F}\l[\frac{1+(1-x)^{2}}{x}\r], \nonumber \\
P_{gg}^{(0)}(x) &=& 2N \l[\frac{x}{(1-x)_{+}}+\frac{1-x}{x}+x(1-x)\r]+
               \delta (1-x)\frac{(11N-4n_{f}T_{R})}{6} \nonumber,
\eea
where $1/(1-x)_+$ means:
\begin{equation}
\int_{0}^{1} dx \frac{f(x)}{(1-x)_+}=\int_{0}^{1}\frac{f(x)-f(1)}{1-x}.
\end{equation}

In the general case in the Altarelli-Parisi equations we have a
$(2n_{f}+1)$-dimension matrix in the space of quarks and gluons.
However, not all parton distributions evolve independently and we can
restrict ourselves to two case. We define the non-singlet and the
singlet parton distributions as
\bea
q^{(NS)}(x,Q^2) &=& \sum_{i\neq j=1}^{\nf}(q_{i}(x,Q^2)-q_{j}(x,Q^2)),  
\label{nsdef}     \\
\Sigma (x,Q^2) &=& \sum_{i=1}^{\nf}(q_{i}(x,Q^2)+\bar{q}_{i}(x,Q^2)),
\label{singdef}
\eea
where in the non-singlet distribution definition $q_j$ may coincide 
with $\bar{q}_i$. Thus the analysis of the evolution is simplified.
As the gluon emission is flavor independent, in the difference between
two distributions the gluonic terms cancel and the evolution of
the non-singlet term is independent from the gluon term. We have
\bea
\frac{d}{dt}q^{(NS)}=\frac{\as(t)}{2\pi}\l[P_{qq} \otimes q^{(NS)}\r].
\label{nsAPunp}
\eea
In the singlet case we have a mixing between quarks and gluon in the
evolution and the Altarelli-Parisi equation is given by
\bea
\frac{d}{dt} \left (\matrix{\Sigma \cr g}\right)
=\frac{\as (t)}{2\pi}
\left (\matrix{P_{qq} & 2\nf P_{qg}
\cr P_{gq} & P_{gg}} \right) \otimes
\left (\matrix {\Sigma \cr g} \right )\,.
\label{sapunp}
\eea
Note that the parton evolution is causal, \ie known the parton
distributions at an initial scale $Q_0^2$, we know their values at an
arbitrary value $Q^2$. This provides to be an essential tool to fit
experimental data.

\subsection{Solution of the Altarelli-Parisi equations}
\label{sect:APsolution}
 
There are different techniques to solve the Altarelli-Parisi
equations. One can solve them by performing numerically the
convolution integrals starting from input distributions obtained from
data. This is particularly convenient if a simultaneous solution of
all $2 n_f +1$ parton distributions is required, given a set of
starting functions $q(x, Q_0^2)$ at some reference scale $Q_0^2$. This
approach is adopted \eg in Monte Carlo simulations for parton
branching processes \cite{kis}.
A second way is the analytical solution of the evolution equations 
in the Mellin moment space, where the convolution products turn into
ordinary ones. Mellin moments are defined by
\bea
f_n = \int_0^1 dx\,x^{n-1} f(x)\,.
\eea

In this section, we find  the general solution of the equation  
\beq  
\label{APi}  
\frac{d}{d\tau}\,q = C\,q\,,  
\eeq  
where $q$ is a vector with $M$ components, and $C$ is a generic $M \times M$  
matrix. The usual QCD evolution equations are special cases of this  
equation in which $M\le2$.  
We will assume that $C$ has a perturbative expansion in powers  
of a parameter $a(\tau)$:  
\beq  
\label{Cexp}  
C=C_0+a(\tau)C_1+\ldots~,  
\eeq  
where
\beq  
\frac{d a(\tau)}{d\tau}=  
-b_0\,a\,\left(1+b_1\,a\,+\ldots\right)\,,  
\label{rge}  
\eeq  
and $b_0=2\pi\,b$ and $b_1=2\pi\,b_1$, 
with $b$ and $b'$ given in eqs.~(\ref{bbeta}) and (\ref{bbeta1}).
For QCD applications  
\beq  
a=\frac{\as}{2\pi};\;\;\;\tau=\frac{1}{2\pi}\int_{t_0}^tdt' \as(t')~,  
\label{tau}
\eeq  
with $t=\log(Q^2/\Lambda_{\sss QCD}^2)$. 
  
The solution of eq.~(\ref{APi}) can be obtained perturbatively. Expanding  
$q$ to order $a$,  
\beq  
q=q^{(0)}+a\,q^{(1)}\;,  
\eeq  
we find  
\beqn  
\label{LO}  
&&\frac{d}{d\tau}\,q^{(0)}=C_0\,q^{(0)}~,  
\\  
\label{NLO}  
&&\frac{d}{d\tau}\,q^{(1)} = (C_0+b_0)\,q^{(1)}+C_1 q^{(0)}\,.  
\eeqn  
The solutions of eqs.~(\ref{LO}) and (\ref{NLO}) are  
\beqn  
q^{(0)}(\tau)&=&R^{-1}e^{\gamma\tau}R\,q^{(0)}(0)~,  
\\  
q^{(1)}(\tau)  
&=&R^{-1}e^{(\gamma+b_0)\tau}R\,q^{(1)}(0)
\\ \nonumber
&+&R^{-1}e^{(\gamma+b_0)\tau}\int_0^\tau d\sigma\, e^{-(\gamma+b_0)\sigma}  
{\hat C}_1 e^{\gamma\sigma}\,R\,q^{(0)}(0)\;,  
\eeqn  
where the matrix $R$ diagonalizes $C_0$,  
\beq  
RC_0R^{-1}={\rm diag}(\gamma_1,\ldots,\gamma_M)\equiv \gamma~,  
\eeq  
and   
\beq  
{\hat C}_1=RC_1R^{-1}\,.  
\eeq  
  
Collecting these results, and noting that $a\exp(b_0\tau)=a(0)$, up to  
terms of order $a^2$, we can write the solution as  
\beq  
\label{q1ter}  
q(\tau)\equiv U(C,\tau)q(0)=R^{-1}\left[e^{\gamma\tau}  
+ae^{(\gamma+b_0)\tau}\int_0^\tau d\sigma\, e^{-(\gamma+b_0)\sigma}  
{\hat C}_1 e^{\gamma\sigma}\right]R\,q(0)\,,  
\eeq  
with the initial condition  
\beq  
q(0)=q^{(0)}(0)+a(0)q^{(1)}(0)\,.  
\eeq  
The explicit expression of $U(C,\tau)$ is  
\beq  
U_{ij}(C,\tau)= R_{im}^{-1} \left[ \delta_{mn} e^{\gamma^n \tau}  
+ a(\tau)\, {\hat C}_1^{mn}\,\frac  
{e^{\gamma^n\tau}-e^{(\gamma^m+b_0)\tau}}  
{\gamma^n - \gamma^m - b_0}  \right] R_{nj}~,  
\eeq  
which, expanded to next-to-leading order reduces to  
\beqn  
U_{ij}(C,\tau)&=& R_{im}^{-1}\l\{\delta_{mn}  
\left(\frac{a(0)}{a(\tau)}\right)^{\gamma_n/b_0}  
+\frac{{\hat C}_1^{mn}-b_1\gamma_n\delta_{mn}}{\gamma^m-\gamma^n+b_0} \r.  
\\
&\times &  \l. \left[  
a(0)\left(\frac{a(0)}{a(\tau)}\right)^{\gamma_m/b_0}  
 -a(\tau)\left(\frac{a(0)}{a(\tau)}\right)^{\gamma_n/b_0}   
\right] \r\}  
R_{nj}~.  
\nonumber 
\label{singletsol}  
\eeqn  
In the case of standard QCD evolution equations the matrix $C_0$ is at  
most $2\times2$ and is easily diagonalized. 

\section{Factorization schemes}
\label{regschemes}

\noindent
The way we subtract divergences is arbitrary. We may add a finite term
of order $\as$ to $Z$ defined in eq.~(\ref{Zmatrix}), obtaining an
infinity of choices, all equivalent to each other as it happens for
ultraviolet renormalization schemes.
Once we fix the finite term, a collinear subtraction scheme is defined.

If we sum over the number of flavors eq. (\ref{AP1}) in the
Mellin space, we can write the column vector as 
(here the index $n$ is omitted to simplify the notation)
\bea
q^{(0)} = \l(\matrix{\Sigma^{(0)} \cr g^{(0)}}\r).
\eea
As the matrix $Z$ is not uniquely defined up to term of order $\as$,
we can consider two transformations:
\bea
&& q^{(0)} = Z q = Z' q' \\
&& q' = W q\,,~~~~W = I + \frac{\as}{2\pi}\, E\,
\label{W}
\eea
where $Z$ and $Z'$ are $2\times2$ matrices. As $q_0$ is independent of the 
factorization scale $\mu$ we get
\bea
\ddt q^{(0)} =0.
\eea
From the Altarelli-Parisi evolution equations we obtain
\bea
\ddt Z q + Z \ddt q =0 &\ra& \ddt q = \frac{\as}{2\pi}\, P
=- Z^{-1} \ddt Z q \\
\ddt Z' q' + Z' \ddt q' =0 &\ra& \ddt q' =\frac{\as}{2\pi}\, P'
= - Z'^{-1} \ddt Z' q'
\eea
If we perform a dimensional 
regularization, the $Z$-matrices can be written as
\bea
Z &=& I + \frac{\as}{2\pi} \frac{1}{\epsilon} {\cal Z} + O(\as^2) \\
Z' &=& I + \frac{\as}{2\pi} \frac{1}{\epsilon} {\cal Z}'+ O(\as^2) =
ZW^{-1} + O(\as^2)\,.  
\eea
We have
\bea
\ddt q' = - W\,Z^{-1}\l[\l(\ddt Z\r)W^{-1} + Z \l(\ddt W^{-1}\r)\r]q',
\eea
and
\bea
\frac{\as}{2\pi}\,P'= \frac{\as}{2\pi}\,W P W^{-1} 
- W \ddt W^{-1}\,, 
\eea
with
\bea
\ddt W^{-1} = \frac{\as^2}{2\pi}\,b\,E + O(\as^2).
\eea
We obtain
\bea
P' = W P W^{-1} - \as\,b E + O(\as^2). 
\eea
The substitution of $W$ with its expansion (\ref{W}) yields
\bea
P' &=& \l(I + \frac{\as}{2\pi}\,E\r) P \l(I-\frac{\as}{2\pi}\,E\r) 
- b \as\,E \\ \nonumber
&=&  P + \frac{\as}{2\pi}\, [E,P] - \frac{\as}{2\pi}\,b_0\,E\,.
\eea
Expanding $P$ in powers of $\as$ at NLO and comparing terms proportional to
same powers of $\as$, we finally obtain
\bea
P'^{(0)} &=& P^{(0)} \\
P'^{(1)} &=& P^{(1)} + [E,P^{(0)}] - b_0 E
\label{p0p1}
\eea

\subsection{The unpolarized case: \MS $\ra$ DIS}
\label{sect:DISscheme}

We will apply the results of the previous Section to the particular
case in which the starting scheme is the \MS\ and the final scheme is the
DIS. We will write the $(n-1)^{th}$ moment of $F_2$ in both the schemes.
Since $F_2$ is independent of the factorization scheme, comparing
the two expression we will find a transformation between the schemes.

In the \MS\ scheme, the Mellin moment of eq.~(\ref{f2LT}) with the 
definition of non-singlet and singlet PDFs given in eqs.~(\ref{nsdef})
and (\ref{singdef}), yields
\bea
\l(\frac{F_2}{x}\r)_n = {\cal C}^{q}_n q^{NS}_n +
\langle e^2\rangle({\cal C}^q_n \Sigma_n + 2 n_f {\cal C}^g_n g_n)\,,
\eea
where
\bea
{\cal C}^{q}_n  &=& I + \frac{\as}{2\pi}\, C^{q}_n  
\label{cfmsbar}\nonumber \\ \\ \nonumber
{\cal C}^g_n &=& \frac{\as}{2\pi} C^g_n\,.
\eea
are the Mellin moments of the coefficient functions eqs.~(\ref{cfq}) and
(\ref{cfg}) and $\langle e^2 \rangle=\sum_{i=q,{\bar
q}}e^2_i/(2\nf)$ . In a partonic scheme \cite{aem} we impose that the quark
distribution is identified with $F_2$ (as it is in the parton model at
LO) at any order in $\as$
\bea
\l(\frac{F_2}{x}\r)_n =\sum_{q=1}^{nf} e_q^2 q_{qn}' = q_n'^{NS}+
\langle e^2\rangle\Sigma_n'.
\label{pdfsc}
\eea
If we take
\bea
E=\left[ \matrix{E_{qq}(x) &      2 n_f  E_{qg}(x) \cr   
                   E_{gq}(x) & E_{gg}(x)}\right]\,,
\eea
comparing the two expression of $F_2$, we get
\bea
&&\l(I + \frac{\as}{2\pi}\,E_{NS}\r) q^{NS}_n + 
\l(I + \frac{\as}{2\pi}\,E_{qq}\r)\Sigma_n
+ 2 n_f \frac{\as}{2\pi}\,E_{qg} g_n = \nonumber \\
&&\l(I + \frac{\as}{2\pi}\,C^{q}_n\r) q^{NS}_n + 
\l(I + \frac{\as}{2\pi}\,C^{q}_n\r)\Sigma_n
+ 2 n_f \frac{\as}{2\pi}\,C^{g}_n g_n\,. 
\eea
It follows
\beqn  
E=\left[ \matrix{C^{q}_n &      2 n_f  C^{g}_n \cr   
                   E_{gq} & E_{gg}}\right]\,.  
\eeqn
and 
\beqn
E_{NS} =  C^q_n\,.
\eeqn
The two other matrix elements of $E$ specify completely the scheme.
Here we will fix the scheme by assuming that all moments satisfy the  
relations between the parton--scheme gluon and the \MS\ quark and  
gluons imposed by momentum conservation on second  
moments~\cite{DFLM,catani}\footnote{
This condition is verified by the second moments of the NLO splitting 
functions in the \MS\ scheme and fix their behavior at $x=1$}. 
This is the prescription used in common parton  
sets, and usually referred to as DIS scheme. For $n=2$ we have
\bea 
&& \int_0^1 dx\,x \l[P_{qq}'^{(1)}(x) + P_{gq}'^{(1)}(x)\r] =
\nonumber \\ &&
\int_0^1 dx\,x \l[-b_0 (E_{qq} + E_{gq}) - E_{gq}(2 n_fP_{qg}^{(0)}(x) +
P_{gg}^{(0)}(x)) + P_{gq}^{(0)}(x) (E_{gg} \r.+ 
\nonumber \\ &&
\l. 
2 n_f E_{qg}) + E_{gq} P_{qq}^{(0)}(x) - E_{qq} P_{gq}^{(0)}(x)
+ P_{qq}^{(1)}(x) + P_{gq}^{(1)}(x) \r]= 0\,,
\nonumber \\ &&
\\ && 
\int_0^1 dx\,x \l[2 n_f P_{qg}'^{(1)}(x) + P_{gg}'^{(1)}(x)\r] =
\nonumber \\ &&
\int_0^1 dx\,x \l[-b_0 (2 n_f E_{qg} + E_{gg}) - 2 n_f E_{qg}(P_{qq}^{(0)}(x) +
P_{gq}^{(0)}(x)) + \r.
\nonumber \\ && 
2 n_f E_{qg} P_{gg}^{(0)}(x) - 2 n_f E_{gg}P_{qg}^{(0)}(x) +
2 n_f P_{qg}^{(0)}(x)(E_{gq}  + E_{qq})
\nonumber \\ && \nonumber
\l.
+ 2 n_f P_{qg}^{(1)}(x) + P_{gg}^{(1)}(x) \r]= 0\,,
\eea 
From eq.~(\ref{sumrulesP}) we get
\bea
E_{gq} = - E_{qq} = -C_q^n ~~~ E_{gg} = - 2 n_f E_{qg} = - 2 n_f C_g^n\,.
\eea 
Thus, generalizing for all $n$, we have
\beqn E^n=\left[ \matrix{C^q_n & 2 n_f C^g_n \cr -
    C^q_n & - 2 n_f C^g_n}\right]\,. 
\label{matrixe}
\eea

\chapter{Truncated Moments of parton distributions}
\def\eq#1{eq.~(\ref{#1})} 

In the previous Chapter we have introduced the Altarelli-Parisi
equations which describe 
the evolution of parton distributions. As we have seen,
we can solve the evolution equations analytically by taking their
Mellin transform, which turns convolution products into ordinary ones,
and therefore the $x$-space integro-differential equation into a set
of independent ordinary first order differential equations. Usually,
a parametrization of the parton distributions is assigned at some initial
scale, and the parameters are then determined by fitting to data the
evolved distributions. Mellin moments of structure functions, however,
cannot be measured even indirectly, since they are defined as
integrals over the whole range $0\leq x\leq 1$, and thus require
knowledge of the structure functions for arbitrarily small $x$, {\it
i.e.} arbitrarily large energy $W^2=\frac{Q^2(1-x)}{x}$.

We can solve this problem using the Altarelli-Parisi equation
to evolve parton distributions directly: the scale dependence of any parton  
distribution at $x_0$ is then determined by knowledge of parton  
distributions for all $x > x_0$, {\it i.e.}, parton evolution is  
causal. In fact, through a judicious choice of factorization  
scheme~\cite{aem,catani} all parton distributions can be identified with  
physical observables, and it is then possible to use the  
Altarelli-Parisi equations to express the scaling violations of  
structure functions entirely in terms of physically observable  
quantities. It is, however, hard to measure local scaling violations  
of structure functions in all the relevant processes: in practice, a  
detailed comparison with the data requires the solution of the  
evolution equations.

The frequently-adopted input on the $x$ dependence of the parton
distributions at the initial scale is for example~\cite{pdfparref}
\bea
q(x,Q_0^2)=a_0\,x^{a_1}\,(1-x)^{a_2}\,P(x;,a_3,\ldots)\,,
\nonumber
\eea
where $Q_0^2$ is a reference scale. The parameter $a_1$ is associated
with the small-$x$ behavior while $a_2$ is associated with the
large-$x$ valence counting rules.  The term $P(x;,a_3,\ldots)$ is a
suitably chosen smooth function, depending on one or more parameters,
that adds more flexibility to the parton distributions
parametrization. It has however become increasingly clear that in
practice this procedure introduces a potentially large theoretical
bias, whose size is very hard to assess~\cite{pdfrev}. In Ref.~\cite{pdfer}
it was proposed to adopt a functional method to keep this theoretical
error under control. Another suitable way to minimize the bias introduced
by the parton distributions parametrization is to project the parton
distributions on an optimized basis of orthogonal functions. Different methods
have been suggested with suitable families of orthogonal polynomials
({\it e.g.}~Bernstein \cite{ynd}, Jacobi \cite{par} or Laguerre polynomials
\cite{fur}) as basis of function.
A different approach has been suggested in Ref.~\cite{FM}, which
makes use of truncated moments of parton distributions. 

In this Chapter the technique of truncated moments
will be extended from its original
formulation for the non-singlet parton distributions to all flavor
combinations, and we will show its numerical implementation.

In the following we will introduce the relevant
evolution equations for truncated moments, and the corresponding
solution. We will assess the accuracy of the method, and discuss
techniques for phenomenological applications \cite{fmpr}. 
In the last Section, we
will show how a numerical technique to increase the numerical
efficiency of the method will provide to be a useful tool to solve
the Altarelli-Parisi equation \cite{mine}.

\section{Evolution equations for truncated moments \\~and their solutions}  

\noindent  
In Sect.~\ref{sect:APsolution} it has been described the solution 
of the Altarelli-Parisi equations eq.~(\ref{AP}) by taking ordinary 
Mellin moments. Here, we are interested in  
the evolution of truncated moments, defined for a generic function  
$f(x)$ by  
\beq   
f_n(x_0) = \int_{x_0}^1 dx x^{n-1} f(x)\,.    
\eeq   
One finds immediately that the truncated moments of $q(x,Q^2)$ 
obey the equation  
\beq   
\frac{d}{d\tau} q_n(x_0, Q^2) = 
\int_{x_0}^1 dy y^{n - 1} G_n  
\left(\frac{x_0}{y}\right) q(y, Q^2) \, ,  
\label{apsinglet}  
\eeq  
with $\tau$ given by eq.~(\ref{tau}) and where 
\beq  
G_n(x) = \int_x^1 d z z^{n - 1} P(z)  
\label{kern}  
\eeq  
is perturbatively calculable as a power series in $\alpha_s$.  
  
Expanding $G_n(x_0/y)$ in powers of $y$ around $y=1$,  
\beq  
G_n \left(\frac{x_0}{y}\right) = \sum_{p=0}^\infty  
\frac{g_p^n (x_0)}{p!} (y - 1)^p\,;  
\;\;\;\;  
g_p^n (x_0)=\left[\frac{\partial^p}{\partial y^p}   
G_n  \left(\frac{x_0}{y}\right) \right]_{y = 1} \, ,  
\label{gntaylor}  
\eeq  
one obtains  
\beq  
\frac{d}{d\tau}q_n(x_0, Q^2)=  
\sum_{p=0}^{\infty}\sum_{k=0}^p \frac{(-1)^{k+p}   
g_p^n (x_0)}{k! (p - k)!}\, q_{n + k}(x_0, Q^2)\,.  
\label{finsyst}  
\eeq  
The key step in the derivation of \eq{finsyst} is the term-by-term  
integration of the series expansion. This is allowed, despite the fact  
that the radius of convergence of the series in \eq{gntaylor} is $1 -  
x_0$, because the singularity of $G_n(x_0/y)$ at $y = x_0$ is  
integrable (this can be proved~\cite{Lebe} using the Lebesgue  
definition of the integral).  One can then express each power of  
$(y-1)$ using the binomial expansion, which leads to \eq{finsyst}.  
  
Equation~(\ref{finsyst}) expresses the fact that, while full moments of  
parton distributions evolve independently of each other, truncated  
moments obey a system of coupled evolution equations. In particular,  
the evolution of the $n^{th}$ moment is determined by all the moments  
$q_j$, with $j\ge n$.  In practice, the expansion in \eq{gntaylor},  
because of its convergence, can be truncated to a finite order $p =  
M$. The error associated with this procedure will be discussed in  
Sect.~\ref{accuracy}. In this case, \eq{finsyst} can be rewritten as  
\beq  
\frac{d}{d\tau} q_n(x_0, Q^2) = 
\sum_{k=0}^M c^{(M)}_{n k} (x_0) \, q_{n + k}(x_0, Q^2) \, ,  
\label{finsyst2}  
\eeq  
where  
\beq  
c^{(M)}_{nk}(x_0) = \sum_{p=k}^M \frac{(-1)^{p+k}   
g_{p}^n (x_0)}{k! (p-k)!}\,.  
\label{cMnkx0}
\eeq  
To solve the system of equations (\ref{finsyst2}), it is necessary to  
include a decreasing number of terms ($M$, $M-1$, and so on) in the  
evolution equations for higher moments ($n+1$, $n+2$, \dots), obtaining  
$M+1$ equations for the $M+1$ truncated moments $\{q_n, \ldots, q_{n +  
M}\}$.  We will see in the next Section that this approximation is  
fully justified.  In this case, the coupled system of evolution  
equations takes the form  
\beq  
\frac{d}{d\tau} \,q_k = \sum_{l=n}^{n+M} C_{kl}\, q_l \, ;   
\;\;\;\; n \leq k \leq n+M \,,  
\label{finsyst3}  
\eeq  
where  $C$ is now a triangular matrix:  
\beq  
\left\{  
\begin{array}{cccc}  
C_{k l} & = & c_{k,l - k}^{(M - k + n)} & (l \geq k)~~, \\  
C_{k l} & = & 0 &(l < k)~~.  
\end{array}  
\right.  
\label{matr0}  
\eeq  
  
In the non-singlet case, discussed in Ref.~\cite{FM}, the matrix   
elements $C_{k l}$ are just numbers, and the matrix $C$ in  
eq.~(\ref{matr0}) is triangular, which makes it easy to solve  
\eq{finsyst3} perturbatively. In the singlet case, each entry  
$C_{kl}$ is a $2 \times 2$ matrix. As a consequence, the matrix $C$,  
which is given in terms of partial moments of the evolution kernels,  
is no longer triangular, but has non-vanishing $2 \times 2$ blocks  
along the diagonal. 

Here, we will show how to solve the singlet case that easily reduce to
the non-singlet case. By writing the perturbative expansion of $C$,
we get  
\beq  
C = C_0 + a C_1 + \ldots \, = (A_0 + B_0) + a ( A_1 + B_1) + \ldots,  
\eeq  
where $A = A_0 + a A_1$ is block-diagonal, with $2 \times 2$ blocks on its   
diagonal,  
\beq  
A_{kl} = C_{kk} \delta_{kl} \, ,  
\eeq  
while $B = B_0 + a B_1$, considered as a matrix of $2\times 2$ blocks,  
is upper-triangular with vanishing diagonal entries. Now one can  
define a matrix $S$ that diagonalizes $A_0$,  
\beq  
S A_0 S^{-1} = \mbox{diag} (\gamma_1, \ldots, \gamma_{2M}) \, .  
\eeq  
Clearly, $S$ is $\tau$-independent, block-diagonal, and easily  
computed. Equation~(\ref{finsyst3}) can then be rewritten as  
\beq  
\frac{d}{d\tau}\,{\tilde q} = T \,{\tilde q}\,,  
\label{finsyst4}  
\eeq  
where ${\tilde q} = S \, q$ and $T = S C S^{-1}$.    
  
The new evolution matrix $T$ is triangular at leading order (with the  
same eigenvalues as $A_0$).  This is enough to solve the evolution  
equation to next-to-leading order.
The general solution has been worked out in detail in 
Sect.~\ref{sect:APsolution}; the result is  
\beq  
{\tilde q}(\tau) = U(T, \tau) \, {\tilde q} (0) \, ,  
\label{qevol}  
\eeq  
where  
\beqn  
\label{singletsoltext}  
U_{ij}(T, \tau) & =  & R_{ik}^{-1} \left\{ \delta_{kl}  
\left(\frac{a(0)}{a(\tau)}\right)^{\gamma_l/b_0} \right. \\  
& + & \left. \frac{ {\hat T_1}^{kl} - b_1 \gamma_l \delta_{kl}}  
{\gamma_k - \gamma_l + b_0} \left[  
a(0) \left( \frac{a(0)}{a(\tau)} \right)^{\gamma_k/b_0}  
- a(\tau) \left( \frac{a(0)}{a(\tau)} \right)^{\gamma_l/b_0}   
\right] \right\} R_{lj} \, . \nonumber  
\eeqn  
In eqs.~(\ref{qevol}) and (\ref{singletsoltext}), $T$ is expanded as  
$T=T_0+a T_1$; $R$ is the matrix which diagonalizes $T_0$, $R T_0  
R^{-1}=\mbox{diag}(\gamma_1,\dots\gamma_{2M})$; finally, $\hat T_1 = R T_1  
R^{-1}$.  
  
The matrix $R$ can be computed recursively, using the technique applied  
in Refs.~\cite{FM,fmpr} and proved in Appendix~A.  One finds  
\beqn  
& & R_{i j} = \frac{1}{\gamma_i - \gamma_j}  
              \sum_{p = i}^{j - 1} R_{i p} \, T_0^{p j}~,  
\label{trir}\\  
& & R^{-1}_{i j}=\frac{1}{\gamma_j - \gamma_i}  
                 \sum_{p = i + 1}^j T_0^{i p}\,R^{-1}_{p j},  
\label{invtrir}  
\eeqn  
which, together with the conditions $R_{ii}=1$ and $R_{ij}=0$ when  
$i>j$, determine the matrix $R$ completely.  
  
The general solution for the parton distributions is then  
\beq  
\label{generalsolution}  
q(\tau) = U(C, \tau) \, q(0) \, ,  
\eeq  
where  
\beq  
U(C, \tau) = S^{-1} U(T,\tau) S \, .  
\eeq  
We have calculated the splitting functions and partial moment integrals 
which should be  used in eq.~(\ref{singletsoltext}) in order 
compute this solution explicitly, and they  
are listed in Appendices~C and D of Ref.~\cite{fmpr}.  
  
For the sake of completeness, we describe a different method to solve  
\eq{finsyst2}. It is immediate to check that the matrix  
\beq  
\label{U}  
U (C, \tau) = I + \sum_{n = 1}^\infty \int_0^\tau d\tau_1 \ldots   
\int_0^{\tau_{n - 1}} d \tau_n C(\tau_1) \ldots C(\tau_n)  
\eeq  
obeys the differential equation  
\beq  
\frac{d}{d \tau} \, U(C, \tau) = C U(C, \tau)~,  
\eeq  
with the initial condition $U(C, 0) = I$. In general, \eq{U} is not  
very useful, since it involves an infinite sum.  In the present case,  
however, the infinite sum collapses to a finite sum.  To see this,  
consider again the decomposition $C = A + B$, where $A$ is  
block-diagonal and $B$ is upper-triangular.  It is easy to prove that  
\beq  
\label{newsol}  
U(C, \tau) = U(A, \tau) U({\tilde B}, \tau) \, ,  
\eeq  
where  
\beq  
{\tilde B} = U^{-1}(A, \tau) B U(A, \tau) \, .  
\label{tilb}  
\eeq  
Since $A$ is block diagonal, $U(A,\tau)$ is also block-diagonal, and  
it can be computed perturbatively using the procedure described in  
Sect. 3.3.2. Furthermore, once $U(A)$ is known, the upper-triangular  
matrix $\tilde B$ can be computed through eq.~(\ref{tilb}). Now one  
can use the fact that upper-triangular matrices have the property that  
their $M^{th}$ power vanishes. Hence, the solution can be expressed as  
the finite sum  
\beq  
U({\tilde B},\tau) = I + \sum_{n = 1}^{M - 1} \int_0^\tau d \tau_1   
\ldots \int_0^{\tau_{n - 1}} d \tau_n  
{\tilde B}(\tau_1) \ldots {\tilde B}(\tau_n) \, ,  
\eeq  
and from the knowledge of $U(\tilde B)$ and $U(A)$ one can determine  
the solution to the evolution equations explicitly.  
  
\section{Numerical methods and their accuracy}  
\label{accuracy}  
\def\R{{\cal R}}  
  
\noindent
In this Section we will assess the accuracy of our method when the  
series of contributions to the right-hand side of the evolution  
equation (\ref{finsyst}) is approximated by retaining a finite number  
$M$ of terms. The loss of accuracy due to this truncation is the price  
to pay for eliminating the dependence on parton parametrizations and  
extrapolations in the unmeasured region. However, unlike the latter  
uncertainties, which are difficult to estimate, the truncation  
uncertainty can be simply assessed by studying the convergence of the  
series.  A reasonable goal, suitable for state-of-the-art  
phenomenology, is to reproduce the evolution equations to about $5\%$  
accuracy: indeed, we expect the uncertainties related to the  
parametrization of parton distributions in the conventional approach  
to be somewhat larger ($\sim 10\%$)\footnote{Notice that this is {\it  
not} the uncertainty associated with evolution of a {\it given}  
parametrization with, say, an $x$-space code; rather, it is the  
uncertainty associated with the {\it choice} of the parametrization,  
and with the bias it introduces in the shape of the  
distribution.}. Notice that there is no obstacle to achieve a higher  
level of precision when necessary, by simply including more terms in  
the relevant expansions.  To this level of accuracy it is enough to  
study the behavior of the leading order contribution to the evolution  
equation: indeed, next-to-leading corrections to the anomalous  
dimension are themselves of order $10\%$. We have verified explicitly  
that the inclusion of the next-to-leading corrections does not affect  
our conclusions.  
  
We can compare the exact evolution equation~(\ref{finsyst}) with its  
approximate form, eq.~(\ref{finsyst2}), by defining the percentage error  
on the right-hand side of the evolution equations for the quark  
non-singlet, singlet and gluon:  
\bea  
\R_{n,M}^{\sss NS} = \frac{1}{{\cal N}_{\sss NS}}  
\int_{x_0}^1 dy~y^{n-1} \l[G^{\sss NS}_n \l(\frac{x_0}{y}\right)   
- \sum_{k=0}^M y^k  c^{\sss NS}_{nk} \right] q^{\sss NS}(y,Q^2)\,,
\nonumber \\  
\label{rns}
\eea
\bea
&&\R_{n,M}^\Sigma = \frac{1}{{\cal N}_\Sigma}  
\int_{x_0}^1 dy~y^{n-1} \l\{\l[G^{qq}_n \l(\frac{x_0}{y}\right)   
- \sum_{k=0}^M y^k  c^{qq}_{nk} \right] \Sigma(y,Q^2) \right.   
\\ \nonumber
&&\phantom{aaaaaaaaaaaaaaaaaaa}  
+\l.\l[G^{qg}_n\l(\frac{x_0}{y}\right) - \sum_{k=0}^M y^k  c^{qg}_{nk}\right]   
g(y,Q^2)\right\},  
\\  
&&\R_{n,M}^g = \frac{1}{{\cal N}_g}  
\int_{x_0}^1 dy~y^{n-1} \l\{\l[G^{gq}_n \l(\frac{x_0}{y}\right)   
- \sum_{k=0}^M y^k  c^{gq}_{nk} \right] \Sigma(y,Q^2) \right.   
\\ \nonumber
&&\phantom{aaaaaaaaaaaaaaaaaaa}  
+\l. \l[G^{gg}_n\l(\frac{x_0}{y}\right) - \sum_{k=0}^M y^k  c^{gg}_{nk}\right]   
g(y,Q^2)\right\},  
\eea  
where ${\cal N}_{{\sss NS},\Sigma,g}$ are the exact right-hand sides  
of the evolution equation~(\ref{finsyst}).  We study the dependence of  
the percentage error on the value of $M$ for typical values of the  
cutoff $x_0$ and for representative choices of test parton  
distributions. In particular, we parametrize parton distributions as  
\beq  
q(x,Q^2) = a_0 x^{-a_1} (1-x)^{a_2} \, .  
\label{distr}  
\eeq  
We begin by choosing, as a representative case, $a_2=4$ and  
$a_1=1$ for the singlet distributions and $a_1=0$ for the  
non-singlet. The non-singlet is assumed to behave qualitatively as  
$q_{NS}\sim x g\sim x\Sigma$, in accordance with the behavior of the  
respective splitting functions. Furthermore, the normalization factors  
$a_0$ for the singlet and gluon are fixed by requiring that the second  
moments of $\Sigma(x,Q^2)$ and $g(x,Q^2)$ are in the ratio $0.6/0.4$,  
which is the approximate relative size of the quark and gluon momentum  
fractions at a scale of a few GeV$^2$.  We will then show that  
changing the values of $a_1$ and $a_2$ within a physically  
reasonable range does not affect the qualitative features of our  
results.  
  
The accuracy of the truncation of the evolution equation is determined  
by the convergence of the expansion in \eq{gntaylor}. Because this  
expansion is centered at $y=1$, and diverges at $y=x_0$, the small $y$  
region of the integration range in \eq{apsinglet} is poorly reproduced  
by the expansion.  Hence, even though the series in \eq{finsyst}  
converges, as discussed in Sect.~4.1, the convergence will be slower for  
low moments, which receive a larger contribution from the region of  
integration $y\sim x_0$. In fact, for low enough values of $n$, the  
convolution integral on the right-hand side of the evolution equation  
(\ref{apsinglet}) does not exist: this happens for the same value for  
which the full moment of the structure function does not exist, {\it  
i.e.} $n\le1$ in the unpolarized singlet and $n\le0$ in the  
unpolarized non-singlet and in the polarized case.  Therefore, we  
concentrate on the lowest existing integer moments of unpolarized  
distributions, {\it i.e.}  the cases $n=2,3$ for the singlet  
distributions, and correspondingly $n=1,2$ for the non-singlet, which  
are the cases in which the accuracy of the truncation will be worse.  
  
The values of $\R_{n,M}^{{\sss NS},\Sigma,g}$, computed at leading  
order with $x_0=0.1$, are shown in Table~\ref{rhs01}.  
\begin{table}[t]  
\begin{center}  
\begin{tabular}{rrrrrrr} 
\multicolumn{7}{c}{$x_0=0.1$}\\ \hline  
M&  $\R_{1,M}^{\sss NS}$ & $\R_{2,M}^\Sigma$ & $\R_{2,M}^g$   
& $\R_{2,M}^{\sss NS}$ & $\R_{3,M}^\Sigma$ & $\R_{3,M}^g$ \\  
\hline  
 5  & 0.63 &  0.43  & 0.55  &  0.16  &  0.12  &  0.16   \\ 
10  & 0.49 &  0.36  & 0.38  &  0.13  &  0.10  &  0.12   \\ 
20  & 0.34 &  0.27  & 0.26  &  0.10  &  0.08  &  0.08   \\ 
40  & 0.20 &  0.17  & 0.17  &  0.06  &  0.05  &  0.05   \\ 
70  & 0.12 &  0.10  & 0.10  &  0.04  &  0.03  &  0.03   \\ 
100 & 0.09 &  0.07  & 0.07  &  0.03  &  0.02  &  0.02   \\ 
150 & 0.06 &  0.05  & 0.05  &  0.02  &  0.01  &  0.01   \\ 
\hline       
\end{tabular}  
\end{center}  
\tcaption{}  
{Values of $\R_{n,M}^{{\sss NS},\Sigma,g}$ for $x_0=0.1$  
and different values of $n$ and $M$.}  
\label{rhs01}  
\begin{center}
\begin{tabular}{rrrrrrr} 
\multicolumn{7}{c}{$x_0=0.01$}\\ \hline
M&$\R_{1,M}^{NS}$ &$\R_{2,M}^\Sigma$ &$\R_{2,M}^g$ 
&$\R_{2,M}^{NS}$ &$\R_{3,M}^\Sigma$ &$\R_{3,M}^g$ \\
\hline
 5 &  0.80  &   0.12 & -42.81 & 0.0050  & 0.0024 & 0.0080\\ 
10 &  0.71  &   0.12 & -34.67 & 0.0047  & 0.0024 & 0.0071\\ 
15 &  0.64  &   0.11 & -29.23 & 0.0044  & 0.0024 & 0.0064\\ 
20 &  0.59  &   0.11 & -25.28 & 0.0042  & 0.0023 & 0.0059 \\ 
\hline     
\end{tabular}
\end{center}
\tcaption{}{Values of $\R_{N,M}^\Sigma$ and $\R_{N,M}^g$ for $x_0=0.01$
and different values of $N$ and $M$.}
\label{rhs001}
\end{table}

The table shows that non-singlet moments of order $n$ behave as singlet  
moments of order $n - 1$. This is a consequence of the fact that, as  
discussed above, the convergence of the expansion is determined by the  
singularity of the integrand $G_n\left({x_0/y}\right) q(y) $ of  
\eq{apsinglet} as $y\to x_0$; near $y = x_0$, the function  
$G_n\left({x_0/y}\right)$ is well approximated by the singular  
contribution $\log\left(1-{x_0/y}\right)$, while parton distributions  
carry an extra power of $y^{-1}$ in the singlet case in comparison to  
the non-singlet.  We also observe in Table~\ref{rhs01} that, as expected, the  
convergence is slower for the lowest moments, and rapidly improves as  
the order of the moment increases.  This rapid improvement is a  
consequence of the fact that the convergence of the expansion of  
$G(x_0/y)$ is only slow in the immediate vicinity of the point  
$y=x_0$, and the contribution of this region to the $n^{th}$ moment is  
suppressed by a factor of $x_0^{n-1}$.  Due to this fast improvement,  
the approximation introduced by including one less term in the  
expansion as the order of the moment is increased by one, which is  
necessary to obtain the closed system of evolution equations  
(\ref{finsyst3}), is certainly justified.  

The 5\% accuracy goal which we set to ourselves requires the inclusion  
of more than 100 terms for the lowest moment, but only about 40 terms  
for the next-to-lowest. The computation of series with such a large  
number of contributions does not present any problem, since the  
splitting functions are known and their truncated moments are easily  
determined numerically. The implications of this requirement for  
phenomenology will be discussed in the next Section.  
We can now study the dependence of these results on the value of the  
truncation point $x_0$ by plotting the exact and approximate  
right-hand side of the evolution equations as a function of $x_0$,  
as shown in Fig.~\ref{fig:RHS}.  

The figures show that the case $x_0 = 0.1$ studied in Table~\ref{rhs01} is a  
generic one between the two limiting (and physically uninteresting)  
cases $x_0 = 0$ and $x_0 = 1$, where the approximation is exact. In  
fact, with this particular choice of parton distributions, $x_0 = 0.1$  
is essentially a worst case and the error estimates of Table~\ref{rhs01} 
are therefore conservative.  

An interesting feature of these plots is the presence of zeroes of the  
lowest moment evolution at $x_0 = 0$ in the non-singlet and around  
$x_0 \approx 10^{-2} $ in the gluon case. The physical origin of these  
zeroes is clear. At leading order, the first non-singlet full moment  
does not evolve.  On the other hand, the second gluon full moment  
grows with $Q^2$, while higher gluon full moments decrease, \ie
the gluon distribution decreases at large $x$; this implies that the  
second truncated moment of the gluon must decrease for a high enough  
value of the cutoff $x_0$, while it must increase for very small  
$x_0$; its derivative is thus bound to vanish at some intermediate  
point. Of course, the phenomenology of scaling violations (such as a  
determination of $\alpha_s$) cannot be performed at or close to these  
zeroes, where there is no evolution. From the point of view of a  
truncated moment analysis, this means that the value of $x_0$ should  
be chosen with care in order to avoid these regions.  

In Table \ref{rhs001} we show for sake of completeness 
values of ${\cal R}_{n,M}^{{\sss NS},\Sigma,g}$ for $x_0=0.01$.
The large percentage errors on the first moment of the gluon
are due to its zero value as it has been discussed above.
\begin{figure}[t!]  
\begin{center}  
\epsfig{figure=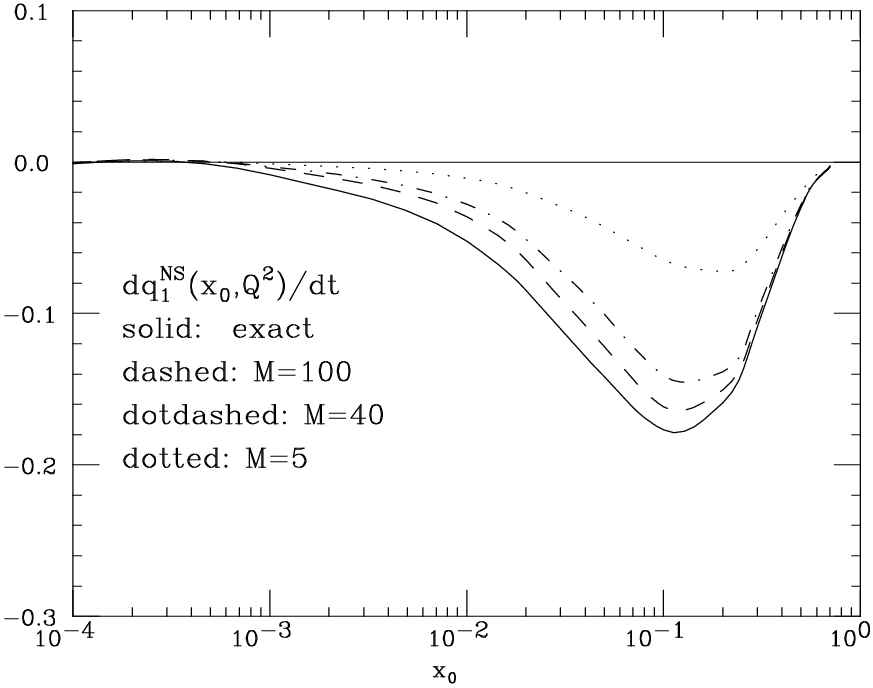,width=0.42\textwidth}  
\epsfig{figure=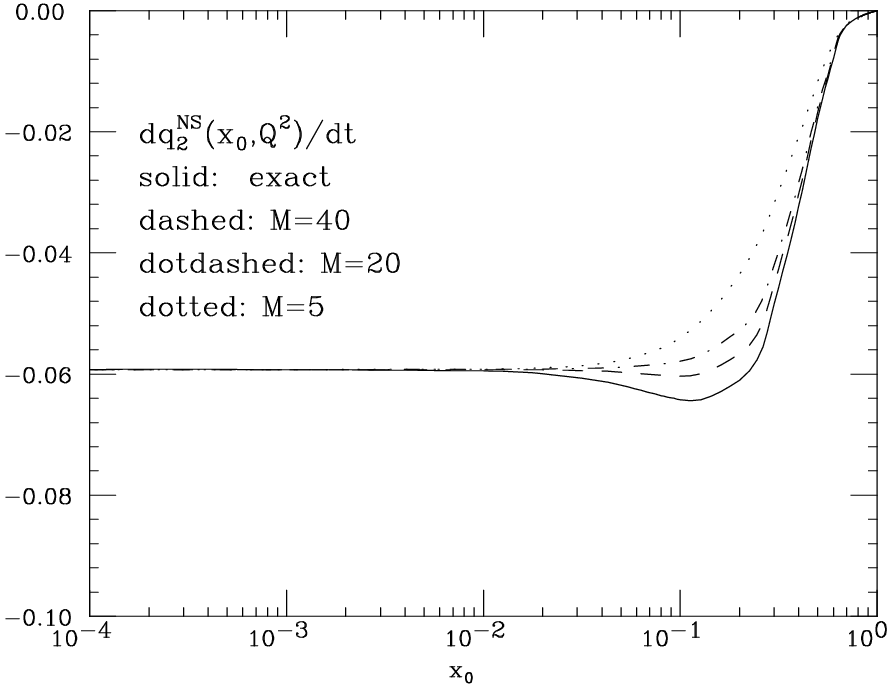,width=0.42\textwidth}  
\epsfig{figure=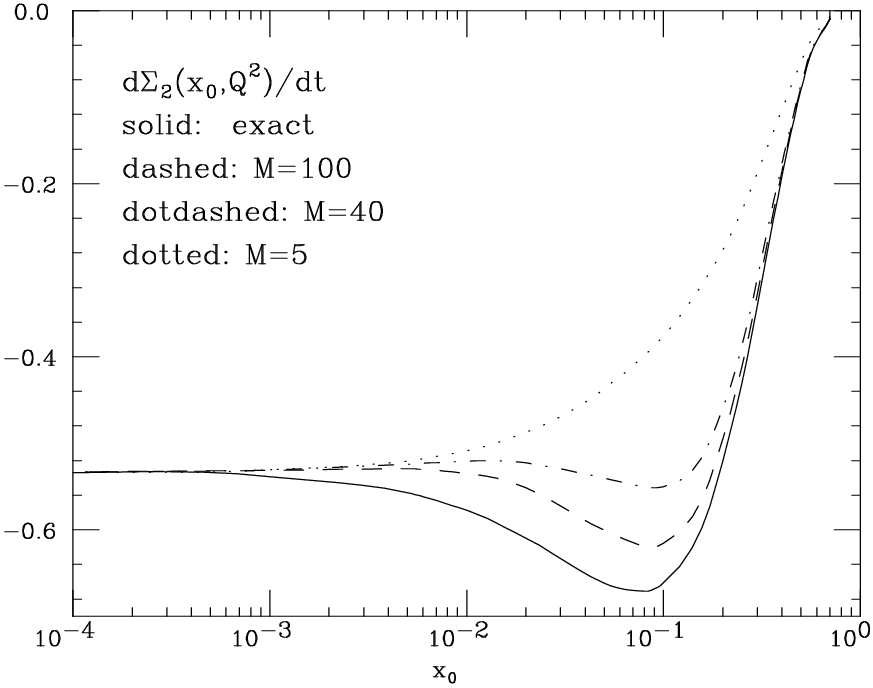,width=0.42\textwidth}  
\epsfig{figure=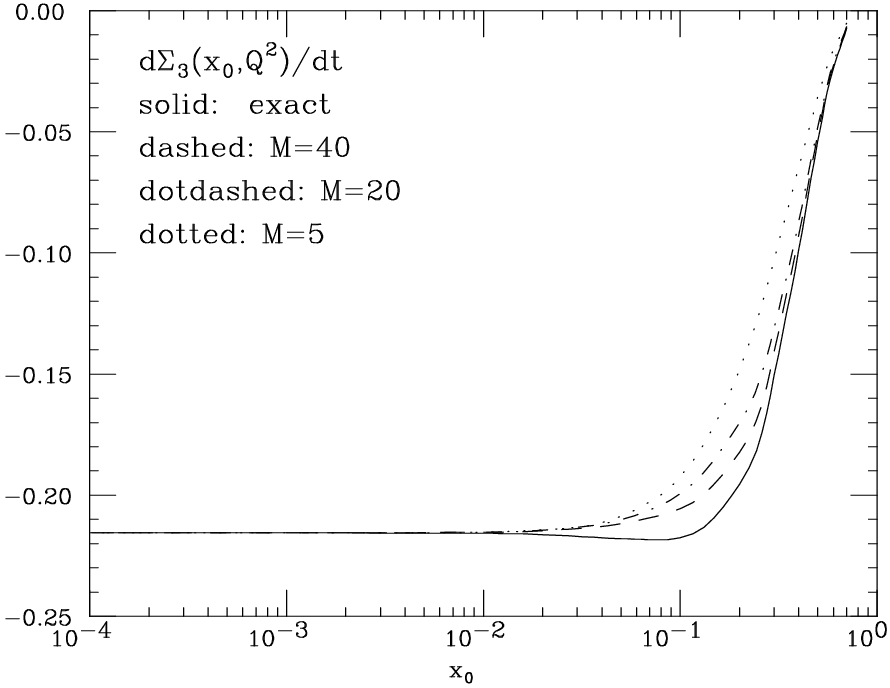,width=0.42\textwidth}  
\epsfig{figure=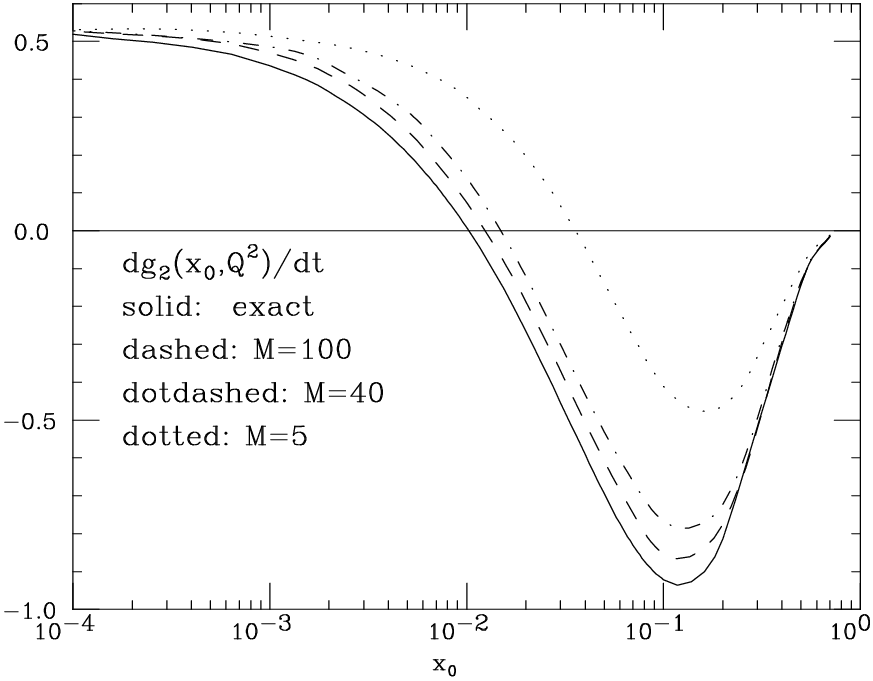,width=0.42\textwidth}  
\epsfig{figure=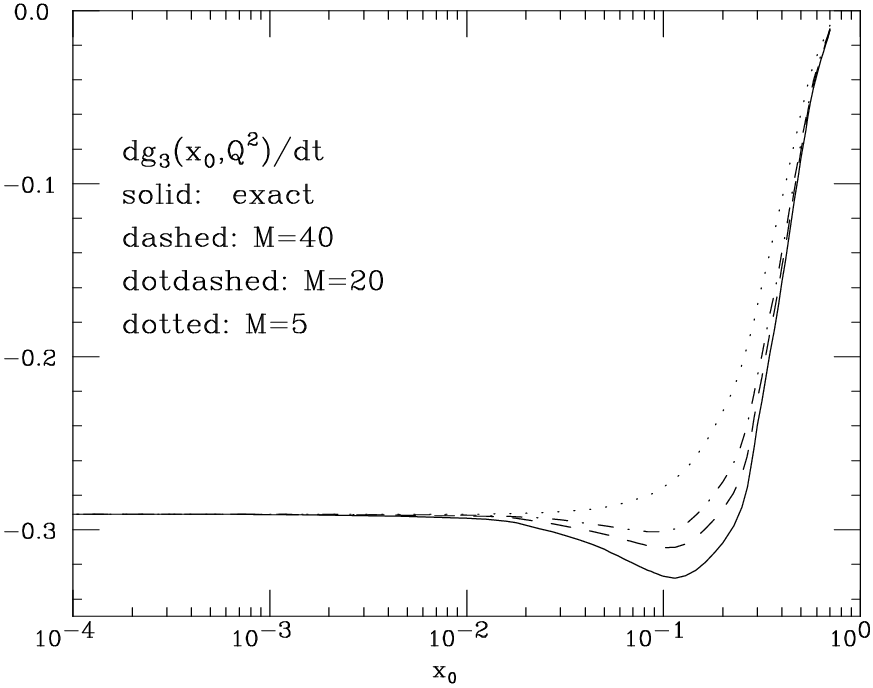,width=0.42\textwidth}  
\end{center}  
\fcaption{}{Right-hand sides of the evolution equations for the  
first and second truncated moments of the non-singlet distribution, and  
for the second and third moments of singlet distributions. The overall  
scale is set by $\alpha_s(2\hbox{GeV}^2)$.}  
\label{fig:RHS}  
\end{figure}  

Finally, in Table~\ref{confev01} we study the dependence of our  
results on the form of the parton distributions, by varying the  
parameters $a_1$ and $a_2$ within a reasonable range.  
\begin{table}[t]  
\begin{center}  
\begin{tabular}{rrrrrr} 
\multicolumn{6}{c}{$n=2$, $x_0=0.1$}\\ \hline  
$a_1$ & $a_2$ &$\R_{n,20}^\Sigma$   
              &$\R_{n,70}^\Sigma$ &$\R_{n,20}^g$ &$\R_{n,70}^g$ \\  
\hline  
1.5 & 2.0 & 0.26 & 0.10 & 0.27 &  0.11\\  
1.0 & 2.0 & 0.20 & 0.07 & 0.23 &  0.09\\  
0.5 & 2.0 & 0.14 & 0.05 & 0.18 & 0.07\\  
1.5 & 4.0 & 0.32 & 0.12 & 0.30 &  0.12\\  
1.0 & 4.0 & 0.27 & 0.10 & 0.26 &  0.10\\  
0.5 & 4.0 & 0.22 & 0.08 & 0.22 &  0.09\\  
1.5 & 6.0 & 0.36 & 0.14 & 0.32 &  0.14\\  
1.0 & 6.0 & 0.32 & 0.12 & 0.29 &  0.12\\  
0.5 & 6.0 & 0.27 & 0.10 & 0.26 &  0.10\\  
\hline  
\multicolumn{6}{c}{$n=3$, $x_0=0.1$}\\ 
\hline  
1.5 & 2.0 & 0.07 & 0.03 & 0.07 &  0.03\\  
1.0 & 2.0 & 0.04 & 0.02 & 0.04 &  0.02\\  
0.5 & 2.0 & 0.02 & 0.01 & 0.02 &  0.01\\  
1.5 & 4.0 & 0.12 & 0.05 & 0.11 &  0.05\\  
1.0 & 4.0 & 0.08 & 0.03 & 0.08 &  0.03\\  
0.5 & 4.0 & 0.05 & 0.02 & 0.05 &  0.02\\  
1.5 & 6.0 & 0.16 & 0.07 & 0.15 &  0.06\\  
1.0 & 6.0 & 0.12 & 0.05 & 0.12 &  0.05\\  
0.5 & 6.0 & 0.09 & 0.03 & 0.08 & 0.03  \\  
\hline  
\end{tabular}  
\end{center}  
\tcaption{}{Values of $\R_{n,M}^\Sigma$ and $\R_{n,M}^g$ for $x_0=0.1$  
and different choices of parton distribution parameters.}  
\label{confev01}  
\end{table}  
Of course, parton distributions which are more concentrated at small  
$y$ give rise to slower convergence. However, we can safely conclude  
that the effect of varying the shape of parton distributions is  
generally rather small. We have also verified that varying the  
relative normalization of the quark and gluon distributions has a  
negligible effect on the convergence of the series, even though it may  
change by a moderate amount the position of the zeroes in gluon  
evolution discussed above.  
  
\section{Techniques for phenomenological applications}  
\label{pheno}  
  
\noindent
So far we have discussed scaling violations of parton distributions.  
In a generic factorization scheme, the measured structure functions  
are convolutions of parton distributions and coefficient functions.  
When taking moments, convolutions turn into ordinary products and  
moments of coefficient functions are identified with Wilson  
coefficients. In the present case, however, as shown in  
eqs.~(\ref{apsinglet}-\ref{finsyst}), truncated moments turn  
convolutions into products of triangular matrices. Hence, in a generic  
factorization scheme, truncated moments of parton distributions are  
related to truncated moments of structure functions by a further  
triangular matrix of truncated moments of coefficient functions.  
  
This complication can be avoided by working in a parton  
scheme~\cite{aem}, where the quark distribution is identified with the  
structure function $F_2$. This still does not fix the factorization  
scheme completely in the gluon sector. One way of fixing it is to use  
a ``physical'' scheme, where all parton distributions are identified  
with physical observables~\cite{catani}. This may eventually prove the  
most convenient choice for the sake of precision phenomenology, once  
accurate data on all the relevant physical observables are  
available. At present, however, the gluon distribution is mostly  
determined from scaling violations of $F_2$, so within the parton  
family of schemes the choice of gluon factorization is immaterial.  
Here we will fix the scheme by assuming that all moments satisfy the  
relations between the parton--scheme gluon and the \MS\ quark and  
gluons imposed by momentum conservation on second  
moments~\cite{DFLM}. This is the prescription used in common parton  
sets, and usually referred to as DIS scheme. The explicit expressions of  
Altarelli-Parisi kernels in the DIS scheme that we have evaluated 
are given in Appendix D of Ref.~\cite{fmpr}.  
  
With this prescription, the phenomenology of scaling violations can be  
studied by computing a sufficiently large  
number of truncated moments of structure  
functions, so as  to guarantee the required accuracy. If the aim is, for  
instance, a determination of $\alpha_s$ from non-singlet scaling  
violations, all we need is a large enough number of truncated moments  
of the non-singlet structure function.  Once an interpolation of the  
data in the measured region is available, the determination of such  
truncated moments is straightforward. This interpolation can be  
performed in an unbiased way using neural networks, as we will
see in the following.   
One may wonder, however, whether the need to use the values  
of very high moments would not be a problem. Indeed, very high moments  
depend strongly on the behavior of the structure function at large  
$y$, which is experimentally known very poorly. Furthermore, it seems  
contradictory that scaling violations of the lowest moments   
should be most dependent on the structure function  
at large $y$.  
  
This dependence is only apparent, however. Indeed, \eq{finsyst}  
for, say, the non-singlet first moment can be rewritten as  
\beq  
{d\over d\tau} q_1(x_0,Q^2) = \sum_{p=0}^\infty   
{g_p^1 (x_0)  \over p!} {\hat q}_1^p(x_0,Q^2)~,  
\label{hatmom1}  
\eeq  
where  
\beq  
{\hat q}_k^p(x_0,Q^2) = \int_{x_0}^1 dy y^{k-1} (y-1)^p q(y,Q^2)~.  
\label{hatmom2}  
\eeq  
The need to include high orders in the expansion in \eq{finsyst2} is  
due to the slow convergence of the series in \eq{hatmom1}, in turn  
determined by the fact that $G_n(x_0/y)$ diverges logarithmically at  
$y = x_0$. Correspondingly, the right-hand side of the evolution  
equation depends significantly on $\hat q_1^p$ for large values of  
$p$, which signal a sensitivity to the value of $q(y,Q^2)$ in the  
neighborhood of the point $y = x_0$.  The dependence on high truncated  
moments $q_n$ is introduced when $\hat q_1^p$ is re-expressed in terms  
of $q_n$, by expanding the binomial series for $(y-1)^p$. Since this  
re-expansion is exact, it cannot introduce a dependence on the large  
$y$ region which is not there in the original expression.  The high  
orders of the expansion do instead introduce a significant dependence  
on the value of the structure function in the neighborhood of $x_0$,  
which can be kept under control provided $x_0$ is not too small,  
{\it i.e.} well into the measured region. There is therefore no obstacle  
even in practice in performing an accurate determination of $\alpha_s$  
from scaling violations of truncated moments.  
  
Let us now consider a second typical application of our method, namely  
the determination of truncated moments of the gluon distribution. In  
particular, the physically interesting case is the lowest integer  
moment, {\it i.e.} the momentum fraction in the unpolarized case or the spin  
fraction in the polarized case. The need to include a large number of  
terms in the expansion of the evolution equations seems to imply the  
need to introduce an equally large number of parameters, one for each  
gluon truncated moment. This would be problematic since it is appears  
unrealistic to fit a very large number of parameters of the gluon from  
currently available data on scaling violations.  We may, however, take  
advantage of the fact that the dependence on high order truncated  
moments is fictitious, as we have just seen, and it rather indicates  
an enhanced sensitivity to the value of $q(y,Q^2)$ as $y\to x_0$.  This  
suggests that a natural set of parameters to describe the gluon  
distribution should include the first several truncated moments, as  
well as further information on the behavior of the distribution around  
the truncation point $x_0$, such as the value of the distribution (and  
possibly of some of its derivatives) at the point $x_0$.  
  
To understand how such a parametrization might work, notice that if  
$q(y,Q^2)$ is regular around $y=x_0$, then it is easy to prove that  
\beq  
\lim_{p\to \infty} {\int_{x_0}^1 d y (y - 1)^p q(y,Q^2) \over  
q(x_0,Q^2) \int_{x_0}^1 d y (y - 1)^p } = 1~,  
\eeq  
by Taylor expanding $q(y)$ about $y=x_0$. We may therefore approximate  
the series  
which appears on the {\it r.h.s.} of \eq{hatmom1} by  
\bea  
S(x_0, n_0) &=& \sum_{p = 0}^{n_0 - 1} \frac{g_p^1 (x_0)}{p!}   
\int_{x_0}^1 d y (y - 1)^p q(y,Q^2) \nonumber \\  
&+& \sum_{p = n_0}^\infty   
\frac{g_p^1 (x_0)}{p!} q(x_0,Q^2) \int_{x_0}^1 d y (y - 1)^p. 
\label{appro} 
\eea  
Equation~(\ref{appro}) describes the evolution of the first truncated  
moment of $q(y,Q^2)$ in terms of the first $n_0$ truncated moments and of  
the value of $q(y)$ at the truncation point $x_0$.  Of course, the  
approximation gets better if $n_0$ increases. It is easy to check that  
when $x_0=0.1$ the accuracy is already better than 10\% when $n_0\sim  
7$.  This means that a parametrization of the distribution in terms of  
less than ten parameters is fully adequate. It is easy to convince  
oneself that this estimate is reliable, and essentially independent of  
the shape of the distribution $q(y,Q^2)$.  In fact, because slow  
convergence arises due to the logarithmic singularity in $G_n(x_0/y)$,  
we can estimate the error of the approximation in eq.~(\ref{appro}) by  
replacing the functions $g_p^1 (x_0)/p!$ with the coefficients of the  
Taylor expansion of $\log(1 - x_0/y)$ in powers of $y - 1$, which   
we may denote by $\hat{g}_p^1 (x_0)/p!$. The error is then  
\beqn  
&&\left|\sum_{p = n_0}^\infty \frac{\hat{g}_p^1 (x_0)}{p!}  
\int_{x_0}^1 d y (y - 1)^p \left( q(y,Q^2) - q(x_0,Q^2) \right)  
\right|   
\label{bound} \\   
&&
\leq  
\int_{x_0}^1 d y \left| \log (1 - x_0/y) -  
\sum_{p = 0}^{n_0 - 1} \frac{\hat{g}_p^1 (x_0)}{p!} (y - 1)^p \right|  
\left| q(y,Q^2) - q(x_0,Q^2) \right|~.  
\nonumber  
\eeqn  
The expression inside the first absolute value on the {\it r.h.s.} of  
eq.~(\ref{bound}) is just the error made in approximating the  
logarithm with its Taylor expansion around $y = 1$; thus, it is a  
slowly decreasing function of $n_0$, it is integrable, and the  
integral receives the largest contribution from the region $y \sim  
x_0$; the second absolute value, on the other hand, is a bounded  
function of $y$ in the range $x_0 \leq y \leq 1$, which vanishes as $y  
\to x_0$ for any choice of $q(y,Q^2)$. These two facts combine to limit  
the size of the error.  One can check directly that, choosing for  
example $q(y,Q^2)=(1-y)^4$ as in the previous Section, the accuracy is  
better than 10\% with $n_0\sim 10$ and $x_0=0.1$, in agreement with  
the previous estimate. One may also verify that, as expected, changing  
the shape of $q(y,Q^2)$ does not significantly affect the result.

\section{Solving the Altarelli-Parisi equation with \\~truncated moments}

\noindent
The number $M$ of truncated moments needed to achieve a precision on
the evolution of the lowest moment comparable to that of other
techniques is rather large ($M\sim 150$), and in some cases it may
lead to problems in the numerical implementation of the method, even
if in practice, it is sufficient to parametrize the parton
distributions using the first few (between 7 and 10) truncated
moments, plus the value of the parton distributions at $x=x_0$.
In this Section we present a different way to improve the numerical
efficiency of the method of truncated moments. 
We begin by studying the unpolarized non-singlet case at leading order, 
leaving at the end the extension to next-to-leading order.

We now integrate by parts the \rhs of eq.~(\ref{apsinglet}). We get
\bea
\label{rhs}
&&\int_{x_0}^1 dy\,y^{n-1} q(y,Q^2) G_n \l(\frac{x_0}{y}\r) 
\\
&& =\l[\widetilde{G}_n (x_0,y) y^{n-1} q(y,Q^2)\r]_{x_0}^1 -
\int_{x_0}^1 dy \,\widetilde{G}_n (x_0,y) \frac{d}{dy}
\l(y^{n-1} q(y,Q^2)\r) \nonumber
\eea
where 
\bea
\widetilde{G}_n (x_0,y) = \int_{x_0}^y \,dz\,G_n \l(\frac{x_0}{z}\r)
\label{gt1}
\eea
(the lower integration bound is irrelevant here; it has been chosen
equal to $x_0$ for later convenience). Using the definition of 
$\widetilde{G}_n(x_0,y)$ and eq.~(\ref{kern}), we get 
\bea
\widetilde{G}_n (x_0,y)
&=&\int_{x_0}^y dz\int_{x_0/z}^1dx\,x^{n-1}P(x)
=\int_{x_0/y}^1dx\,x^{n-1}P(x)\int_{x_0/x}^y dz
\nonumber \\
&=&yG_n \l(\frac{x_0}{y}\r) - x_0G_{n-1} \l(\frac{x_0}{y}\r)\,.
\eea
By taking the Taylor expansion of $\widetilde{G}_n (x_0,y)$ 
around $y=1$, we obtain 
\bea
&&\frac{d}{d\tau}q_n(x_0,Q^2) =
\l[\widetilde{G}_n (x_0,y) y^{n-1} q(y,Q^2)\r]_{x_0}^1
\\ \nonumber
&&- \sum_{p=0}^{\infty}\frac{\widetilde{g}_n^p (x_0)}{p!}
\int_{x_0}^1 dy (y-1)^p \frac{d}{dy} \l(y^{n-1} q(y,Q^2)\r)
\label{tayexp1}
\eea
where
\bea
\widetilde{g}_n^p (x_0) = 
\l[\frac{d^p}{dy^p} \widetilde{G}_n (x_0,y)\r]_{y=1} = 
\l[\frac{d^{p-1}}{dy^{p-1}} \frac{d}{dy}\widetilde{G}_n
(x_0,y)\r]_{y=1} = g_n^{p-1} (x_0)\,.
\label{dertaycoeff}
\eea
The functions $\widetilde{G}_n(x_0,y)$ are regular in the whole
interval $[x_0,1]$. In fact, the $G_n(x_0/y)$ are regular for all
values of $y$ except $y=x_0$, as they contain singular terms
proportional to $\log(1-x_0/y)$. However, these terms are integrable,
and independent of $n$. Thus, $\widetilde{G}_n (x_0,y)$ is regular in
the limit $y\ra x_0$  and tends to zero.  Furthermore, we observe
that the Taylor coefficient of order $p$ of $\widetilde{G}_n(x_0,y)$
is equal to that of $G_n(x_0/y)$, times a factor $1/p$ (see
eq.~(\ref{dertaycoeff})). For this reason, the convergence of the
expansion of $\widetilde{G}_n(x_0,y)$ is faster than that of
$G_n(x_0/y)$.

Integrating by parts the second term of the \rhs of eq.~(\ref{tayexp1}), 
we have:
\bea
\frac{d}{d\tau}q_n(x_0,Q^2) &=&
\l[\widetilde{G}_n (x_0,y) y^{n-1} q(y,Q^2)\r]_{x_0}^1 \\
&-& \l[\sum_{p=0}^{\infty}\frac{\widetilde{g}_n^p (x_0)}{p!} 
(y-1)^p y^{n-1} q(y,Q^2)\r]_{x_0}^1 \nonumber \\ \nonumber
&+& \sum_{p=1}^{\infty} \frac{g_n^{p-1} (x_0)}{(p-1)!}
\int_{x_0}^1 dy\,y^{n-1} (y-1)^{p-1} \,q(y,Q^2)
\eea
Truncating the series, expanding the binomial $(y-1)^{p-1}$
with $q(1,Q^2)=0$ (this is our only assumption on the behavior 
of the parton distributions), we get
\bea
\frac{d}{d\tau}q_n(x_0,Q^2) &=& 
x_0^{n-1} q(x_0,Q^2)\,\sum_{p=0}^{M} 
\frac{\widetilde{g}_n^p (x_0)}{p!} (x_0-1)^p \\ \nonumber 
&+& \sum_{k=0}^{M-1} c_{nk}^{(M-1)}(x_0) q_{n+k} (x_0,Q^2)
\eea
where $c^{(M)}_{nk}(x_0)$ are defined as in eq.~(\ref{cMnkx0}).
Defining the triangular matrix as in eq.~(\ref{matr0})
we can finally write the truncated evolution equation as
\bea
&&\frac{d}{d\tau}q_n(x_0,Q^2) = 
\nonumber \\ \label{eqev1}
&&x_0^{n-1} q(x_0,Q^2)\,\sum_{p=0}^{M} 
\frac{\widetilde{g}_n^p (x_0)}{p!} (x_0-1)^p 
+ \sum_{l=1}^{M} C_{nl} q_{l} (x_0,Q^2)\,.
\eea
Notice that the first term in the \rhs of eq.~(\ref{eqev1}) vanishes
in the limit $M\ra\infty$ and the original expression given in
eq.~(\ref{finsyst3}) with $n_0=1$ is recovered. 
However, for finite values of $M$
this term must be taken into account (in a sense, it is the price we
have to pay for the better convergence of the expansion after the
integration by parts).  This term poses special problems because it
depends on the value of the parton distributions at $x=x_0$. In the
following we will show how to obtain an approximated expression of
$q(x_0,Q^2)$ in terms of a finite number $N$ (not necessarily equal to
$M$) of truncated moments.  The evolution equation (\ref{eqev1}) will
then be solved with a technique similar to that presented in
Sect.~4.1.

We begin by taking the Taylor expansion of $q(x,Q^2)$
around $x=y_0$:
\bea
q(x,Q^2) = \sum_{k=1}^{\infty} \eta_k (Q^2)(x-y_0)^{k-1}\,,
\label{pdftaylor}
\eea
The initial point of the expansion, $y_0$, must be carefully chosen.
Parton distributions pa\-ra\-me\-tri\-zed as in eq.~(\ref{distr}) are
non-analytical in $x=1$ when the exponent $a_2$ is not an integer; and
even in that case, an essential singularity in $x=1$ is generated by
perturbative evolution.  One should therefore choose
$y_0\leq(1+x_0)/2$, so that the expansion (\ref{pdftaylor}) is
convergent everywhere in $[x_0,1)$.  The series will not be convergent
in $x=1$, no matter what $y_0$ is; however, the singularity in $x=1$
is integrable, and the term-by-term integration is allowed using the
Lebesgue definition of the integral as we did in Sect.~4.1.
We have therefore
\bea
\label{qjx0}
q_j (x_0,Q^2) = \int_{x_0}^1\,dx\, x^{j-1} q(x,Q^2) 
= \sum_{k=1}^{\infty} \beta_{jk}(x_0,y_0)\,\eta_k(Q^2)\,,
\eea
where
\bea
\label{betadef}
\beta_{jk}(x_0,y_0) = \int_{x_0}^1\,dx\, x^{j-1} (x-y_0)^{k-1}\,.
\eea
Our task is now to find a way of inverting eq.~(\ref{qjx0}), in order to
express the coefficients $\eta_k(Q^2)$ in terms of the truncated moments
$q_j (x_0,Q^2)$. This can be done in the following way.
Define a matrix $\widetilde{\beta}^{-1}$ by
\bea
\widetilde{\beta}_{kj}^{-1} = 
\left\{  
\begin{array}{cc}  
\l(\beta^{(N)}\r)^{-1}_{kj} & k,j \leq N \\
0 & \rm{otherwise}
\end{array}  
\right.  
\eea
where $\beta^{(N)}$ is the $N\times N$ upper left 
sub-matrix of $\beta$. For example, in the case $N=2$ the matrix
$\widetilde{\beta}^{-1}$ is such that
\bea
\widetilde{\beta}^{-1} \cdot \beta
= \l(\matrix{1 & 0 & \frac{\beta_{13}\beta_{22}-\beta_{23}\beta_{12}}{\det\beta^{(2)}} & \dots \cr
             0 & 1 & \frac{\beta_{23}\beta_{11} - \beta_{13}\beta_{21}}{\det\beta^{(2)}} & \dots \cr
             0 & 0 & 0                                     & \dots \cr
             \vdots     & \vdots & \vdots                  & \ddots
 }\r)
\eea
Multiplying eq.~(\ref{qjx0}) by $\widetilde{\beta}^{-1}$ on the right,
we obtain
\bea
\sum_{j=1}^{N} \widetilde{\beta}_{ij}^{-1}\,q_j (x_0,Q^2) =
\sum_{k=1}^{\infty} \sum_{j=1}^{N} \widetilde{\beta}_{ij}^{-1}\,
\beta_{jk}\,\eta_k(Q^2)\,,
\eea
where for simplicity we have not shown the dependence on $x_0,y_0$.
Using the definition of $\widetilde{\beta}^{-1}$ we get
\bea
\sum_{j=1}^{N} \widetilde{\beta}_{ij}^{-1}\,q_j (x_0,Q^2) \equiv
\eta_i(Q^2) +
\sum_{k=N+1}^{\infty} \sum_{j=1}^{N} \widetilde{\beta}_{ij}^{-1}\,
\beta_{jk}\,\eta_k(Q^2) 
\label{etadef}
\eea
for $i\leq N$.
Substituting eq.~(\ref{etadef}) in
eq.~(\ref{pdftaylor}) gives
\bea
q(x_0,Q^2) =
\sum_{k=1}^{N} \sum_{j=1}^{N} \widetilde{\beta}^{-1}_{kj}\,
q_j (x_0,Q^2)\, (x_0-y_0)^{k-1} +R(x_0,y_0,Q^2)\,,
\label{pdfapprox}
\eea
where
\bea
\label{errapprox}
R(x_0,y_0,Q^2)&=&\sum_{k=N+1}^{\infty} \eta_k(Q^2) \Big[(x_0-y_0)^{k-1}
\\ \nonumber 
&-&\
\sum_{i=1}^{N} (x_0-y_0)^{i-1}\sum_{j=1}^{N} 
\widetilde{\beta}^{-1}_{ij}\,
\beta_{jk} \Big] \,.
\eea
We have thus obtained an approximate expression of $q(x_0,Q^2)$ as a
function of the first $N$ truncated moments of $q$,
eq.~(\ref{pdfapprox}); the quantity $R$ in eq.~(\ref{errapprox})
represents the error on this reconstruction. The quantity in square
brackets in eq.~(\ref{errapprox}) is independent of the parton
distributions, and can be computed for any $N$ and $k$ starting from
the coefficients $\beta_{ij}$, given by eq.~(\ref{betadef}). 
The analytic expression of this quantity is very complicated. We have
checked that, for $y_0=(1+x_0)/2$, it decreases as
$[(x_0-1)/2]^{k-1}$, for any value of $N$. Therefore,
$R(x_0,y_0,Q^2)$ vanishes, for $N\to\infty$, at least as fast as the
remainder of order $N$ of the Taylor expansion in
eq.~(\ref{pdftaylor}).

In order to assess the accuracy of our approximation, we have
computed the percentage error given by ratio 
$|R/q(x_0,Q^2)|$ for some representative choices of
the parton density, namely $q(x,Q^2)=(1-x)^{a_2}$ with
$a_2=2.5,3.5,4.5$.  We have fixed $x_0=0.1$ and $y_0=(1+x_0)/2$.  The
results are shown in Table~\ref{tabpdf}. We see that an excellent
approximation is achieved already with $N=5$, independently of the
value of the large-$x$ exponent $a_2$.  The accuracy increases with
increasing $N$; however, it should be noted that a numerical
evaluation of the matrix $\widetilde{\beta}^{-1}$ requires a numerical
precision which also rapidly increases with $N$. Therefore, for a
practical implementation of the method, $N$ cannot be very large. We
see from Table~\ref{tabpdf} that for $5\leq N\leq 10$ the accuracy is
already better than $10^{-3}$ in the cases we have studied.
\begin{table}[t]  
\begin{center}  
\begin{tabular}{cccc} 
\hline
\multicolumn{4}{c}{$x_0=0.1$}\\ \hline  
N & $a_2=2.5$ & $a_2=3.5$ & $a_2=4.5$ \\
\hline
5  & $3.3\times 10^{-4}$ & $3.2\times 10^{-4}$ & $1.4\times 10^{-3}~$ \\ 
10 & $3.8\times 10^{-6}$ & $5.4\times 10^{-7}$ & $1.4\times 10^{-7}~$ \\ 
15 & $3.2\times 10^{-7}$ & $1.8\times 10^{-8}$ & $1.8\times 10^{-9}$ \\
\hline
\end{tabular}
\end{center}
\tcaption{}{Precision in the reconstruction of
$q(x_0,Q^2)=(1-x_0)^{a_2}$ in terms of a finite number $N$ of
truncated moments, for different values of $N$ and for three different
choices of $a_2$.}
\label{tabpdf}
\end{table}
We conclude that
\bea
q(x_0,Q^2) \simeq \sum_{j=1}^{N} \l[\sum_{k=1}^{N} 
\widetilde{\beta}^{-1}_{kj}(x_0,y_0)\,(x_0-y_0)^{k-1}\r]q_j (x_0,Q^2)
\label{reconpdf}
\eea
to an accuracy of about $10^{-3}$ for $N=5$, 
independent of the detailed shape of $q(x,Q^2)$, and rapidly increasing
with $N$.

We are now ready to re-write the original evolution equation
(\ref{eqev1}) using our result eq.~(\ref{reconpdf}). We get
\bea
\frac{d}{d\tau}q_n(x_0,Q^2) =
 \sum_{l=1}^{M} C_{nl}\,q_{l} (x_0,Q^2)
+\sum_{l=1}^{M} D_{nl}^{(N)}\,q_{l} (x_0,Q^2)\,,
\label{eqev2}
\eea
where $C_{nl}$ is defined in eq.~(\ref{matr0}), and
\bea
D_{nl}^{(N)} =x_0^{n-1} 
\l[\sum_{p=0}^{M} \frac{\widetilde{g}_n^p (x_0)}{p!} (x_0-1)^p\r]\,   
\l[\sum_{k=1}^N \widetilde{\beta}^{-1}_{kl}(x_0,y_0)\,(x_0-y_0)^{k-1}\r]\,.
\nonumber \\
\label{Dnl}  
\eea

We now turn to a test of the accuracy of the evolution equation. We will also
compare the accuracy achieved with the method presented in this Section, 
and that of Sect.~4.1. The original evolution equation (\ref{apsinglet})
and its truncated version, eq.~(\ref{finsyst2}), can be
written schematically as
\bea
\frac{d}{d\tau}q_n(x_0,Q^2) = S_n \qquad \qquad \qquad ~~
\frac{d}{d\tau}q_n(x_0,Q^2) = S_n^{(M)}
\eea
respectively. Therefore, the quantity
\bea
{\cal R}_{n,M}^a = 1-\frac{S_n^{(M)}}{S_n}
\eea
is the same measure of the accuracy of the method given by 
eq.~(\ref{rns}).
Similarly, we write eqs.~(\ref{eqev2}) in the form
\bea
\frac{d}{d\tau}q_n(x_0,Q^2) = S_n^{(M-1)} + T_n^{(M,N)}\,,
\eea
and define
\bea
{\cal R}_{n,M,N}^b=1-\frac{S_n^{(M-1)}+T_n^{(M,N)}}{S_n}
\eea
to test the error of the method presented above.
The values of ${\cal R}_{n,M}^{a,b}$, computed
at leading order with $x_0=0.1$ for $n=1$ and $n=2$,
are shown in Table~\ref{tabrhs} for
different values of $M$ and $N=6$.
\begin{table}[t]  
\begin{center}  
\begin{tabular}{rrrrr} \hline 
\multicolumn{5}{c}{$x_0=0.1$}\\ \hline  
$M$  & ${\cal R}_{1,M}^a$ & ${\cal R}_{1,M,6}^b$
&  ${\cal R}_{2,M}^a$ & ${\cal R}_{2,M,6}^b$ \\
\hline
5  & 0.62 & 0.14 & 0.14 & 0.020 \\
10 & 0.48 & 0.07 & 0.12 & 0.016  \\
20 & 0.33 & 0.03 & 0.09 & 0.009  \\
40 & 0.20 & 0.01 & 0.05 & 0.004  \\
\hline
\end{tabular}
\end{center}
\tcaption{}{Comparison between percentage errors for the first and the second 
truncated moment at LO with $N=6$, $x_0=0.1$, $y_0=(1+x_0)/2$
and $q(x,Q^2)=(1-x)^{3.5}$.}
\label{tabrhs}
\end{table}
We observe that the error ${\cal R}_{n,M,N}^b$ computed with the
technique presented here is always much smaller than the corresponding
error presented in Sect. 4.1, ${\cal R}_{n,M}^a$. An accuracy of less
than $10\%$ can be achieved with a relatively small value of $M$. In
Table~\ref{tabrhs001} we show also the comparison between percentage
errors with $x_0=0.01$. We observe that, accordingly to what 
we have discussed in Sect. 4.2, the percentage error on the second 
truncated moment is smaller for $x_0=0.01$ than for $x_0=0.1$, 
while it is larger for the first truncated moment. 

Finally, in Fig.~\ref{fig:RHS2}, we show the
$x_0$ dependence of the \rhs of the evolution equation for the first
and the second truncated moments. In both cases, we note that
the approximation of the exact case is very good.

The complete solution of the evolution equation, LO and NLO terms, 
can be written as
\bea  
\label{mynlo}  
&&q_n(x_0,Q^2)= \\ \nonumber
&& R^{-1}\left[e^{\gamma\tau}  
+ae^{(\gamma+b_0)\tau}\int_0^\tau d\sigma\, e^{-(\gamma+b_0)\sigma}  
(\widehat{C}_1+\widehat{D}_1) e^{\gamma\sigma}\right]R\,q_n(x_0,Q_0^2)\,,  
\eea  
with the initial condition  
\bea  
q_n(x_0,Q_0^2)=q_n^{(0)}(x_0,Q_0^2)+a(0)q_n^{(1)}(x_0,Q_0^2)\,,
\label{mytrevol}
\eea  
and
\bea  
R(C_0+D_0)R^{-1}&=&{\rm diag}(\gamma_1,\ldots,\gamma_M)\equiv \gamma\,,
\label{rmatrix}
\\ \nonumber
\widehat{C}_1+\widehat{D}_1&=&R(C_1+D_1)R^{-1}\,. 
\eea
The matrix $R$ that diagonalizes $C_0+D_0$ must be computed numerically.
This is not a problem, since, as we have seen, its dimension $M$ does 
not need to be too large.

The improvement of the numerical efficiency presented here can be
extended straightforwardly to the unpolarized singlet case, and to
polarized partons as well.  The tests we presented are only for the LO
equations. We have checked that the inclusion of NLO terms does not
modify our conclusions. The technique of truncated moments can now be
easily implemented numerically for phenomenological purposes.

\begin{table}[t]  
\begin{center}  
\begin{tabular}{rrrrr} \hline 
\multicolumn{5}{c}{$x_0=0.01$}\\ \hline  
$M$  & ${\cal R}_{1,M}^a$ & ${\cal R}_{1,M,6}^b$
&  ${\cal R}_{2,M}^a$ & ${\cal R}_{2,M,6}^b$ \\
\hline
5  & 0.78 & 0.11 & 0.0042 & 0.0019  \\
10 & 0.70 & 0.06 & 0.0039 & 0.0017  \\
20 & 0.58 & 0.03 & 0.0035 & 0.0014  \\
40 & 0.46 & 0.01 & 0.0030 & 0.0009  \\
\hline
\end{tabular}
\end{center}
\tcaption{}{Comparison between percentage errors for the first and the second 
truncated moment at LO with $N=6$, $x_0=0.1$, $y_0=(1+x_0)/2$
and $q(x,Q^2)=(1-x)^{3.5}$.}
\label{tabrhs001}
\end{table}

\begin{figure}[t]  
\begin{center}  
\epsfig{figure=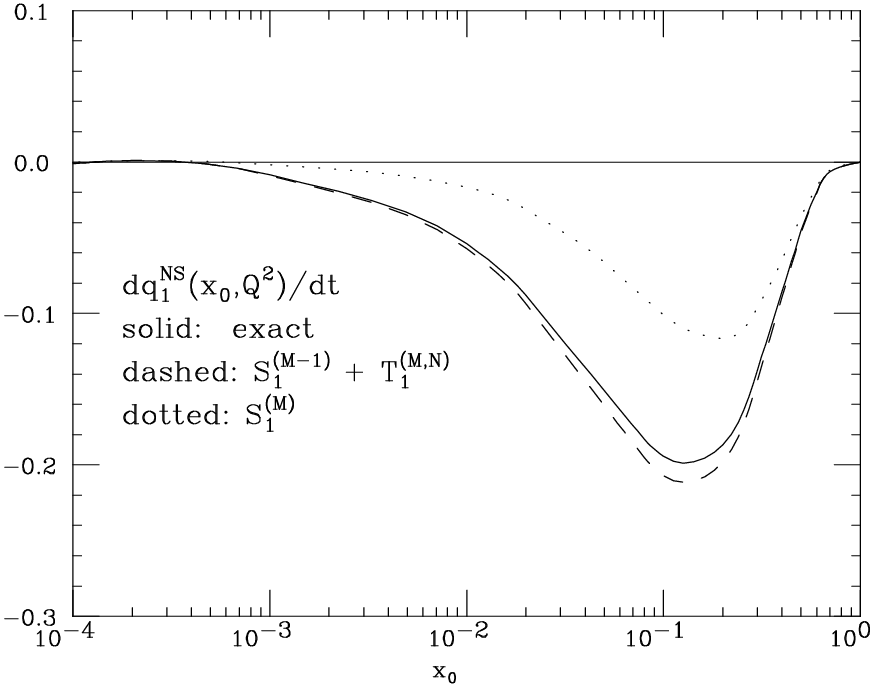,width=0.48\textwidth}  
\epsfig{figure=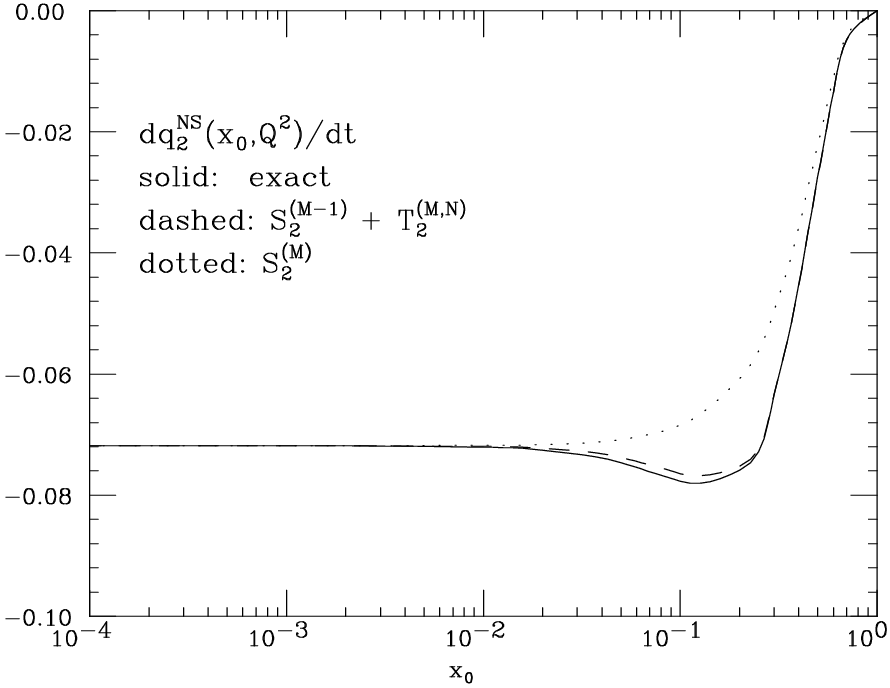,width=0.48\textwidth}  
\end{center}  
\fcaption{}{Right-hand side of the evolution equations for the  
first and the second truncated moments of the non-singlet distribution
with $M=10$ and $N=6$. 
The overall scale is set by $\alpha_s(2\hbox{GeV}^2)$.}  
\label{fig:RHS2}  
\end{figure}  

\chapter{Introduction to Neural Networks}

In the present times {\it artificial neural networks} constitute one
of the most successful and multidisciplinary subjects. People with
very different formation, ranging from physicists to philosophers, and
from biologists to engineers, are working on them trying to
understand better and better how they work. Applications of
artificial neural networks are widely used from images reconstruction
to financial markets predictions. Artificial neural network me\-thods
are well established techniques for high energy physics too, where
they are used for event reconstruction in particle
accelerators. Quark-gluon separation, heavy quark jet tagging, mass
reconstruction, and track finding procedures make use of artificial neural
networks \cite{peterson}.

Here we will give a very brief and general introduction to artificial
neural networks (henceforth simply neural networks). Our main interest
is to describe the very basic algorithms of neural networks, and some
useful practical issues in order to better understand the fits
presented in the following Chapter.

\section{From biology to artificial neural networks}

\noindent
The structure of biological nervous system started to be understood in
1888, when Dr. Santiago Ram\'on y Cajal succeeded in seeing the {\it
synapses} between individual nervous cells, the {\it neurons}. This
discovery was quite impressive as it proved that all the
capabilities of the human brain rest not so much in the complexity of
its constituents as in the enormous number of neurons and connections
between them. To give an idea of these magnitudes, the usual estimate
of the total number of neurons in the human central nervous system is
$10^{11}$, with an average of $10 000$ synapses per neuron. The
combination of both numbers yields a total of $10^{15}$ synaptic
connections in a single human brain.
\begin{figure}[t]
\begin{center}
\epsfig{width=0.7\textwidth,figure=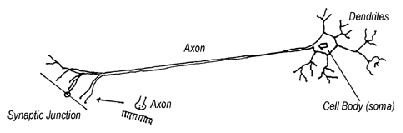}  
\end{center}
\begin{center}
\fcaption{}{Schematic structure of a neuron.}
\label{neuron}
\end{center}
\end{figure}
All the neurons share the common structure schematized in
Fig.~\ref{neuron}. There is a {\it cell body} or {\it soma} where the {\it cell
nucleus} is located, a tree-like set of fibres called {\it dendrites}
and a single long tabular fibre called the {\it axon} which arborize
as its end. Neurons establish connections either to sensory organs
(input signals), to muscle fibres (output signals) or to other
neurons (both input and output). The output junctions are called {\it
synapses}. The inter-neuron synapses are placed between the axon of a
neuron and the soma or the dendrites of the next one.

The way the neuron works is basically the following: a potential
difference of chemical nature appears in the dendrites or soma of the
neuron, and if its value reaches a certain threshold then an
electrical signal is created in the cell body, which immediately
propagates through the axon without decaying in intensity. When it
reaches its end, this signal is able to induce a new potential
difference in the post-synaptic cells, whose answer may or may not be
another {\it firing} of a neuron or a contraction of a muscle fibre,
and so on. Of course a much more detailed overview could be given, but
this suffices for our purposes.

In 1943 W. S. McCulloch and W. Pitts \cite{culloch} suggested a
mathematical model for capturing some of the above characteristics of
the brain. First, an {\it artificial neuron} (or simply a {\it
neuron}) is defined as a processing element whose state $\xi$ at time
$t$ can take two different values only: $\xi(t)=1$, if it is firing,
or $\xi(t)=0$, if it is at rest. The state of, say, the $i^{th}$ unit,
$\xi_i(t)$, depends on the inputs from the rest of the N neurons
through the discrete dynamical equation
\bea
\xi_i(t)=\Theta\l(\sum_{j=1}^N\omega_{ij}\xi_j(t-1)-\theta_i\r)\,,
\eea
where the {\it weights} $\omega_{ij}$ represent the strength of the
synaptic coupling between the $j^{th}$ and the $i^{th}$ neurons,
$\theta_i$ is the {\it threshold} which points out the limit between
firing and rest, and $\Theta$ is the unit step {\it activation
function} defined as
\bea
\Theta (h)\equiv\l\{\matrix{0\,\rm{if} \,{\it h}\leq 0\,, 
\cr 1\,\rm{if} \,{\it h}>1\,.}\r.
\eea
Then, a set of mutually connected McCulloch-Pitts units is what is 
called an {\it artificial neural network}.

In spite of the simplicity of their model, McCulloch and Pitts were
able to prove that artificial neural networks could be used to any
desired computation, once the weights $\omega_{ij}$ and the
thresholds $\theta_i$ were chosen properly. This fact made that the
interest toward artificial neural networks was not limited to the
description of the collective behavior of the brain, but also as a new
paradigm of computing.

\section{Multilayer Neural Networks}

\noindent
Among the different types of neural networks, those in which we have
concentrated our interest are Rosenblatt's {\it perceptrons}, also
known as {\it multilayer feed-forward neural networks}
\cite{rosenblatt}. 
In these networks there is a layer of input units whose only role is to
feed input patterns into the rest of the networks. Next, there are one
or more intermediate or hidden layers of neurons evaluating the same
kind of function of the weighted sum of inputs, which, in turn, send
it forward to units in the following layer. This process goes on until
the output level is reached, thus making it possible to read
off the computation.

In the class of networks one usually deals with, there are no
connections leading from a neuron to units in the previous layers, nor
to neurons further than the next contiguous level, \ie every unit
feeds only the ones contained in the next layer. Once we have updated
all the neurons in the right order, they will not change their states.
For these architectures time plays no role.

\begin{figure}[t]
\begin{center}
\includegraphics[width=0.70\textwidth]{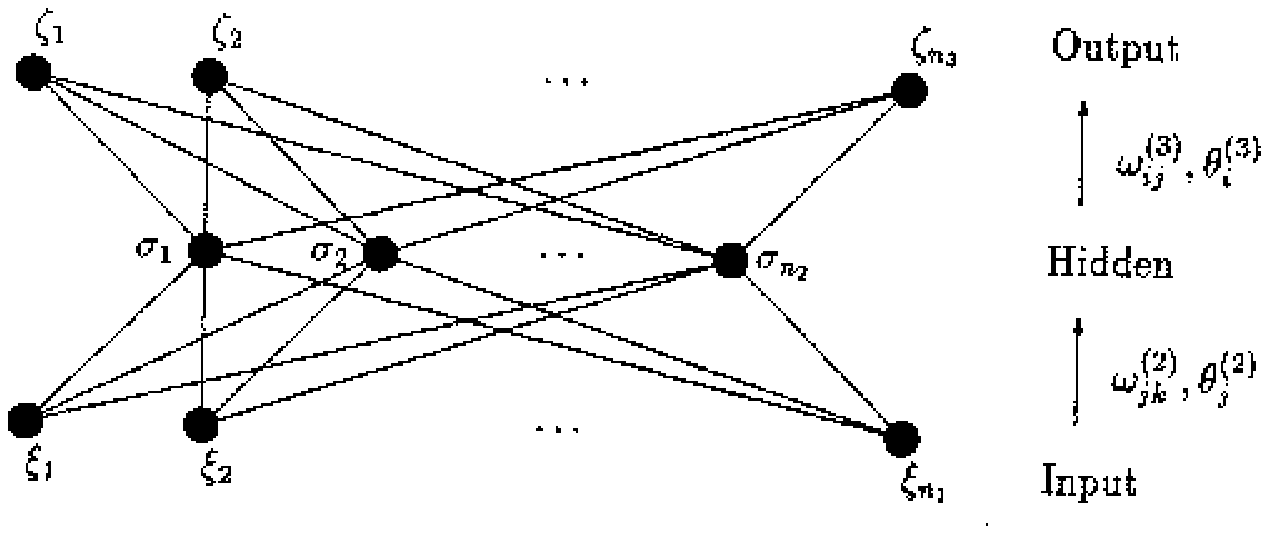}
\end{center}
\fcaption{}{A three-layer perceptron consisting of input, hidden and
output layers.}
\label{hidden}
\end{figure}

In Fig.~\ref{hidden} we have represented a three-layer perceptron with $n_1$
input units, $n_2$ hidden units in a single hidden layer, and $n_3$
outputs. When an input vector $\boldsymbol{\xi}$ is introduced to the 
network, the states of the hidden neurons acquire the values
\bea
\sigma_i=g\l(\sum_{k=1}^{n_1}\omega_{ij}^{(2)}\xi_k 
- \theta_j^{(2)}\r)\,,j=1,\ldots,n_2\,;
\eea
the output of the network is the vector ${\bf \zeta}$ whose 
components are given by
\bea
\zeta_i=g\l(\sum_{j=1}^{n_1}\omega_{ij}^{(3)}\sigma_j 
- \theta_i^{(3)}\r)\,,i=1,\ldots,n_3\,.
\eea
Here we have supposed that the {\it activation function} can be any
arbitrary function $g$, though it is customary to work only with
bounded ones either in the interval $[0,1]$ or $[-1,1]$. If this
transfer function is of the form of the $\Theta$ step function it is
said that the activation is {\it discrete}, since the states of the
neurons are forced to be in one of a finite number of different
possible values. Otherwise, continuous functions commonly used are the
{\it sigmoids} or {\it Fermi functions}
\bea
g(h)\equiv\frac{1}{1+e^{-\beta h}}\,,
\eea
which satisfy
\bea
\lim_{\beta\ra\infty} g(h)=\Theta(h)\,.
\eea
In the terminology of statistical mechanics the parameter $\beta$ is
regarded as the inverse of a temperature. However, for our practical
applications we will set $\beta=1$.

Generally, if we have $L$ layers with $n_1,\ldots,n_L$ units
respectively, the state of the multilayer perceptron is established
by recursive relations
\bea
\xi_i^{(l)}=g\l(\sum_{j=1}^{n_l-1}\omega_{ij}^{(l-1)}\xi_j^{(l-1)} 
- \theta_i^{(l)}\r)\,,i=1,\ldots,n_l\,,l=2,\ldots,L\,,
\eea
where $\boldsymbol{\xi}^{(l)}$ represents the state of the neurons in the
$l^{th}$ layer, $\omega_{ij}^{(l)}$ the weights between units in
the $(l-1)^{th}$ and the $l^{th}$ layers, and $\theta_i^{(l)}$ the
threshold of the $i^{th}$ unit in the $l^{th}$ layer. Then the input is
the vector $\boldsymbol{\xi}^{(1)}$ and the output the 
vector $\boldsymbol{\xi}^{(L)}$.

By {\it simple perceptron} one refers to networks with just two
layers, the input one and the output one, without internal units, in
the sense that there are no intermediate layers. These devices have
been seen to have limitations, such as the XOR problem, which do not
show up in feed-forward networks with hidden layers present. Actually,
it has been proved that a network with just one hidden layer can represent
any Boolean function \cite{denker}. Any continuous function can be
uniformly approximated by a continuous neural network having only one
internal layer, and with an arbitrary continuous sigmoid non-linearity
\cite{cybenko}.

\section{Learning process of Neural Networks}

\noindent
Connecting neurons together may well produce networks that can do
something, but we need to be able to {\it train} them in order 
to make them do anything useful. We also want to find the simplest learning
rule that we can, in order to keep our model understandable. As is often
the case in neural computing, inspiration comes from looking at real
neural systems.

Dogs are given tit-bits to encourage them to come when called. 
Generally, good behavior is reinforced, while bad behavior is
reprimanded. We can transfer this idea to our network. The guiding
principle is to allow neurons to learn from mistakes. If they produce an
incorrect output, we want to reduce the chances of that happening
again; if they come up with correct output, then we need do nothing.
The learning paradigm can be summarized as follows:
\begin{itemize}
\item set the weights and thresholds randomly;
\item present an input;
\item calculate the actual output by taking the thresholded value of the weighted sum of the inputs;
\item alter the weights to reinforce correct decisions and discourage incorrect ones, \ie reduce the error;
\item present the input next time, etc.
\end{itemize}

\subsection{A simple perceptron learning algorithm}

Let us consider the set
\bea
\{({\bf x}^{\mu},{\bf z}^{\mu})
\in\mathbb{R}^{n}\times\mathbb{R}^m,
\,\mu=1,\ldots,p\}
\label{trainset}
\eea
of pairs input-output and a simple perceptron, \ie a neural network
with only two layers. The equations governing the state of the network 
are given by
\bea
\xi_i^{(2)}=g(h_i)\,,i=1,\ldots,m\,,
\eea
where the fields are given by
\bea
h_i=\sum_{j=1}^n \omega_{ij}\,\xi_j^{(1)}-\theta_i\,.
\eea
We shall use the notation for the input and output patterns:
\bea
\l\{\matrix{{\bf x} &=& \boldsymbol{\xi}^{(1)}\,, \cr
            {\bf o}({\bf x}) &=& \boldsymbol{\xi}^{(2)}}\,. \r.
\eea
For any given values of the weights and thresholds it is possible to
calculate the {\it quadratic error} between the actual and the desired
output of the network, measured over the training set:
\bea
E[{\bf o}]\equiv\frac{1}{2}\sum_{\mu=1}^p\sum_{i=1}^m
(o_i({\bf x}^{\mu})-z_i^{\mu})^2\,.
\label{energy}
\eea
Therefore, the last squares estimate minimizes 
$E[{\bf o}]$. Applying the {\it gradient descent minimization} procedure, 
what we have to do is just to look for the direction (in the space of
weights and thresholds) to steepest descent of the error function
(which coincides with minus the gradient), and then modify the
parameters in that direction so as to decrease the actual error:
\bea
\delta\omega_{ij} &=& 
-\eta\frac{\partial E}{\partial\omega_{ij}}=
-\eta (o_i({\bf x}^{\mu})-z_i^{\mu})
g'(\boldsymbol{\omega}_i \cdot {\bf x})x_j\,, 
\nonumber \\
\label{deltarule}
\\ \nonumber 
\delta\theta_{i} &=& 
-\eta\frac{\partial E}{\partial\theta_{i}}=
-\eta (o_i({\bf x}^{\mu})-z_i^{\mu})
g'(\boldsymbol{\omega}_i \cdot {\bf x})
\eea
where $\boldsymbol{\omega}_i \cdot {\bf x}=\sum_{j=0} \omega_{ij}x_j$, 
with the updating rule
\bea
\label{updrule}
\omega_{ij}(t+1)&\longrightarrow &
\omega_{ij}(t)+\delta\omega_{ij}\,, 
\\ \nonumber
\theta_{i}(t+1)&\longrightarrow &
\theta_{i}(t)+\delta\theta_{i}\,.
\eea
The appearance of the derivative $g'$ of the activation function
explains why we have supposed in advance that it has to be continuous
and differentiable. 
The intensity of the change is controlled by the 
{\it learning rate} parameter $\eta$.
When no more changes (within some accuracy) occurs, 
\ie $\delta\omega_{ij}\sim0$, the weights are frozen and 
the network is ready to use for data it has never ``seen''.
The procedure is summarized as follows:
\begin{enumerate}
\item initialize $\omega_{ij}$ with $\pm$ random values;
\item pick pattern $p$ from the training set;
\item feed input {\bf x} to network and calculate the
output {\bf o};
\item update the weights according to 
eq.~(\ref{updrule}) and eq.~(\ref{deltarule});
\item repeat from 2 until 
$\omega_{ij}(t+1)\sim\omega_{ij}(t)$.
\end{enumerate}

\subsection{Learning by error back-propagation}

We now consider a multilayer neural network and generalize the
{\it delta rule} or {\it gradient descent procedure} that we have seen
in the perceptron case, eq.~(\ref{deltarule}) \cite{rumelhart}. 
The equations that describes the multilayer neural network are given by
\bea
\xi_i^{(l)}&=&g(h_i^{(l)})\,,i=1,\ldots,m\,,\\
h_i^{(l)}&=&\sum_{j=1}^{n_l-1} 
\omega_{ij}^{(l-1)}\,\xi_j^{(1)}-\theta_i^{(l)}\,.\\
\delta\omega_{ij}^{(l)} &=& 
-\eta\frac{\partial E}{\partial\omega_{ij}^{(l)}}\,, 
\nonumber \\
\label{deltarule2}
\\ \nonumber
\delta\theta_{i}^{(l)} &=& 
-\eta\frac{\partial E}{\partial\theta_{i}^{(l)}}
\eea
Substituting the first two equations in eq.~(\ref{energy}),
and taking the derivatives, it is easy to get
\bea
\delta\omega_{ij}^{(l)} &=& 
-\eta\sum_{\mu=1}^p
\Delta_i^{(l)\mu}\xi_j^{(l-1)\mu}\,, 
~~i=1,\ldots,n_l,~~j=1,\ldots,n_{l-1}\,,
\nonumber \\  \\ \nonumber
\delta\theta_{i}^{(l)} &=& 
\eta\sum_{\mu=1}^p
\Delta_i^{(l)\mu}\,,~~i=1,\ldots,n_l\,,
\eea
where the error is introduced in the units of the last layer 
through
\bea
\Delta_i^{(L)\mu}=
g'\l(h_i^{(L)\mu}\r)[o_i({\bf x}^{\mu})-z_i^{\mu}]\,
\eea
and then is {\it back-propagated} to the rest of the network by
\bea
\Delta_j^{(l-1)\mu}=g'\l(h_j^{(l-1)\mu}\r)
\sum_{i-1}^{n_l} \Delta_i^{(l)\mu}\omega_{ij}^{(l)}\,.
\label{backprop}
\eea
This result can be easily derived by taking a neural network with only
one hidden layer. Summarizing the {\it batched back-propagation}
algorithm for the learning of the training set (\ref{trainset})
consists in the following steps
\begin{enumerate}
\item initialize all the weights and thresholds randomly,
and choose a small value for the learning rate $\eta$;
\item run a pattern ${\bf x}^{\mu}$ of the training set 
and store the activations of all the units 
(\ie $\l\{\xi_{i}^{(l)\mu},\forall l\forall i\r\}$);
\item calculate the $\Delta_i^{(L)\mu}$ and then 
back-propagate the error using eq.(\ref{backprop});
\item compute contributions to $\delta\omega_{ij}^{(l)}$
and $\delta\theta_{i}^{(l)}$, induced by this input-output pair
$({\bf x}^{\mu},{\bf z}^{\mu})$;
\item update weights and thresholds;
\item go to the second step unless enough training 
epochs have been carried out.
\end{enumerate}
The adjective ``batched'' refers to the fact that the update of the
weights and thresholds is done after all patterns have been presented to
the network. Nevertheless, simulations show that, in order to speed up the
learning, it is usually preferable to perform this update each time a
new pattern is processed, choosing them in random order: this is known
as {\it non-batched} or {\it on-line back-propagation}.
Generally, we choose the on-line mode if the number of patterns is small
(\ie $\sim 10^3$), and the batched one if it is large.

It is clear that back-propagation seeks minima of the er\-ror func\-tion
given in eq.~(\ref{energy}), but it cannot ensure that it ends in a global one,
since the procedure may get stucked in a {\it local minimum}. Several
modifications have been proposed to improve the algorithm so as to
avoid local minima and to accelerate its
convergence. One of the most successful, simple and commonly used
variants is the introduction of a {\it momentum term} to the updating
rule, either in the batched or the on-line schemes.
It consists in the substitution of eq.~(\ref{deltarule2}) by
\bea
\delta\omega_{ij}^{(l)} &=& 
-\eta\frac{\partial E}{\partial\omega_{ij}^{(l)}}
+\alpha \delta\omega_{ij}^{(l)}(\rm{last})\,, 
\\ \nonumber
\delta\theta_{i}^{(l)} &=& 
-\eta\frac{\partial E}{\partial\theta_{i}^{(l)}}
+\alpha \delta\theta_{i}^{(l)}(\rm{last})
\eea
where the ``last'' means the values of the $\delta\omega_{ij}^{(l)}$ and
$\delta\theta_{i}^{(l)}$ used in the previous updating of the weights
and thresholds. The parameter $\alpha$ is called momentum of the
learning, and it has to be a positive number smaller than 1.

\section{Practical issues}

\subsection{Rules of thumb}

\subsubsection*{Number of layers} 

The number of layers issue depends on the specific task one wants the
networks to perform, but a general statement is that no more than two
hidden layers are needed, even though very many units might be needed
in these layers.

Any function, no matter how complex, can be represented by a multilayer
neural network of no more than three layers; the inputs are fed through
an input layer, a middle hidden layer, and an output layer. This is an
important result in that it proves that whatever is done in four or
more layers could also be done in three.  However, we note that
with only one hidden layer one may need a very large number of units
in the hidden layer. Thus, it could be more useful to have two hidden
layers and a smaller number of neurons on each. Notice also that, as the
sigmoid is very close to a step function, if we have a neural network
with two hidden layers and a small number of units it could be useful
to adopt a linear activation function between the last two layers to
have a smoother output.

\subsubsection*{Number of hidden units}

The number of hidden units needed to approximate a given function 
${\cal F}$ is related to how many terms are needed in an expansion
of ${\cal F}$ in the function $g()$.
There are several techniques to determine the optimal number
of units. We can reduce a large number units by the {\it weight decay}
approach, where weights which are rarely updated are allowed to decay
according to
\bea
\delta_{ij}=-\eta\frac{\partial E}{\partial\omega_{ij}} 
-\epsilon\omega_{ij}\,,
\eea
where $\epsilon$ is the decay parameter, typically a very small
number, ${\cal O}(10^{-4})$. This corresponds to adding an extra
complexity term to the energy function \cite{pruning}
\bea
E\ra E+\frac{\epsilon}{2\eta}\sum_{i,j}\omega_{ij}
\eea
imposing a ``cost'' for large weights. A more advanced complexity
term \cite{pruning} is
\bea
E\ra E+\lambda\sum_{i,j}
\frac{\omega_{ij}^2/\omega_0^2}{1+\omega_{ij}^2/\omega_0^2}\,,
\eea
where the sum extends over all weights. For large
$|\omega_{ij}|$ the cost is $\lambda$, whereas for
small weights it is zero. The scale of weights is set by $\omega_0$.
In this way the cost reflects the number of weights, instead of the
size, hence the network gets {\it pruned} to only contain weights
that are really needed to represent the problem.
One can also begin with a small number of units, and then increases it
one by one by {\it trial-and-error}
till a stability of the error function or of a given indicator
is reached. 

\subsubsection*{Learning parameters}

If the activation $\lan \xi_i \ran$ of output node $i$ is large, the
optimal $\eta$ for weights connecting to that unit will be small. The
natural thing is to rescale the input data such that 
$\lan \xi_i \ran\sim{\cal O}(1)$ for all $i$, in which case they 
will have approximately the same optimal $\eta$ (even the thresholds). 
This also simplifies for the case of the hidden units where 
$\lan h_i \ran\sim{\cal O}(1)$ for sigmoidal units, and the same 
learning rate can be used for all weight layers.

The momentum term $\alpha$ controls the ``averaging'' of the
updatings and it is closely connected to the learning rate. 
An increase of $\alpha$ means an increase of the ``effective'' 
learning rate. The optimum $\alpha$ depends on the updating
procedure used. For the batched method $\alpha$ is very useful 
and should be a number close to unity ($0.5<\alpha<1$).
For the on-line updating, on the other hand, $\alpha$ is often
(but not always) useless.

\subsubsection*{Time dependence}

One can choose a learning parameter that varies with time. In this way
after the region of the minimum is reached with a large $\eta$,
the learning rate is reduced as to avoid to get out of there
and to refine the research of the minimum. Notice however that if
a local minimum is reached in this way one can not get out of there.

\subsubsection*{Choosing patterns}

If patterns are shown sequentially, there may result a bias, \eg due
to a regularity of some sequential input patterns in giving a positive 
variations of weights. This is the reason why it is a good rule to
choose patterns randomly. It is also a good rule to show a pattern 
as an input at least $10^{4}$ times, in order to allow compensations
of different variations.

\subsection{Generalization}

One of the major features of neural networks is their ability to
generalize, \ie to classify successfully patterns that have not been
previously presented. Multilayer neural networks generalize by
detecting features of the input pattern that have been learned to be
significant, and so coded into the internal units. Thus an unknown
pattern is classified with others that share the same distinguishing
features. This means that learning by example is a feasible
proposition, since only a representative set of patterns have to be
taught the network, and the generalization properties will allow
similar inputs to be classified as well. It also means that noisy
inputs will be classified, by virtue of their similarity with the pure
input. It is generalization ability that allows multilayer neural
networks to perform more successfully on real-world problems that other
pattern recognition or expert systems.

Generally, neural networks are good at interpolation, but not so good
at extrapolation. They are able to detect the patterns that exist in
the inputs they are given, and allow for intermediate states that have
not been seen. However, inputs that are extensions of the range patterns
are less well classified, since there is little which compare them
with. Put another way, given an unseen pattern that is an intermediate
mixture of two previously taught patterns, the net will classify it as
an example of the predominant pattern. If the pattern does not
correspond to anything similar to what the neural network has seen before, then
the classification will be much poorer. Notice also that the
generalization performance is usually lower than its performance on
the training set, although for very large data sets the performance
can be approximately equal. For a feed-forward network with one hidden
layer, the generalization error is of order ${\cal O}(N_{\omega}/N_p)$
where $N_{\omega}$ is the number of weights and $N_p$ the number of
training patterns \cite{baum}.
\begin{figure}[t]
\begin{center}
\epsfig{width=0.32\textwidth,figure=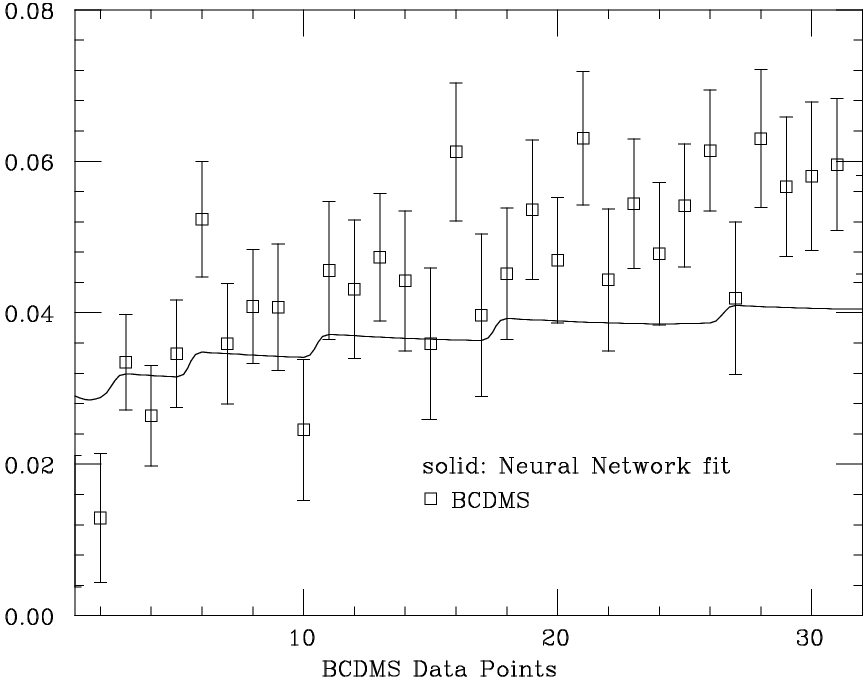}  
\epsfig{width=0.32\textwidth,figure=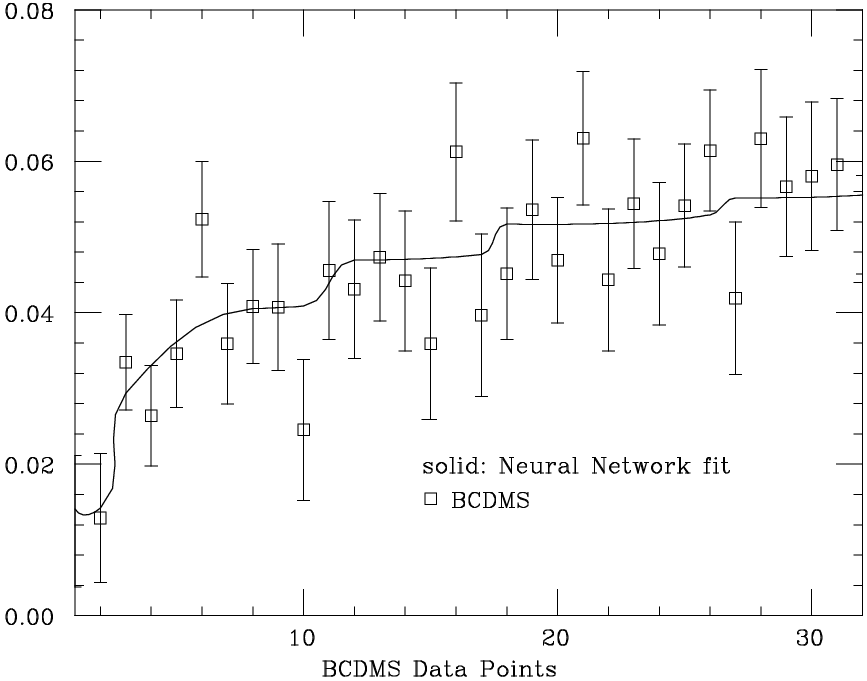}  
\epsfig{width=0.32\textwidth,figure=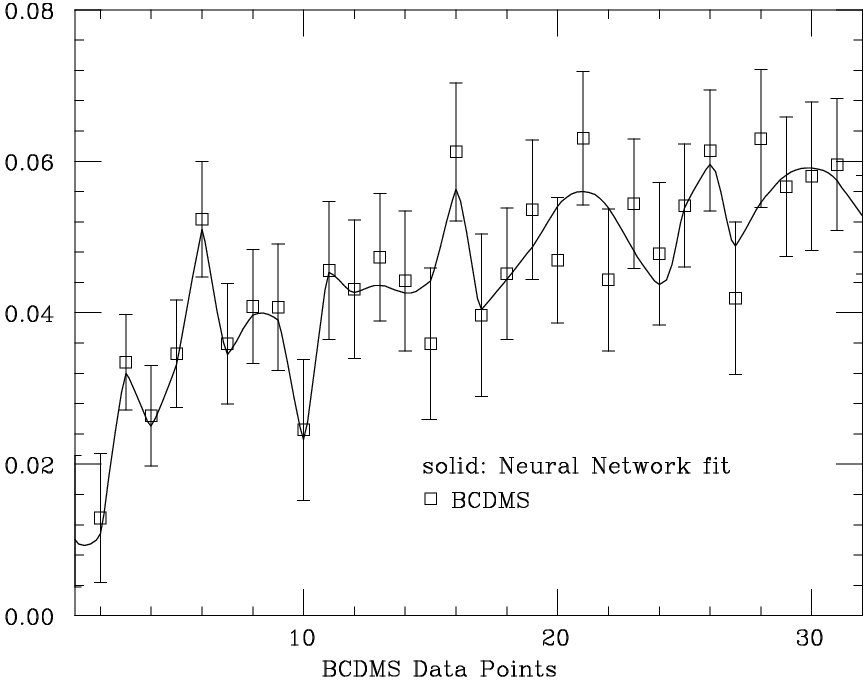}  
\end{center}
\fcaption{}{Different trainings for
a fit of the non-singlet structure function $F_2^p-F_2^d$ 
for a subset of BCDMS data points.}
\label{netcomp}
\end{figure}

In Fig.~\ref{netcomp} we show, as an example, a plot with three
different trainings of the structure function $F_2^p-F_2^d$ (details will
be given in the next chapter). We have chosen the architecture (4,5,3,1)
that will be used in the following, and we have trained the neural network
on a very small arbitrary chosen subset of data, \ie 30 experimental 
points from the BCDMS data (see later).

In the first plot we can see a very smooth almost constant function
which corresponds to a very short training. Here we see that the
neural network strongly correlates data points, but since there has been
no enough training it does not reproduce the behavior of data. The
second plot corresponds to a
longer training. Here we see that the neural network has found an
underlying law from data; also in this case it strongly
correlates data points, but now it reproduce their behavior. The last plot
shows the result of a very long training.
Here the neural network goes on top of many data point loosing
its generalization ability. This is what is called {\it over-learning}.
If the number of weights of the neural network is equal to the number
of data points we can have over-learning for a large enough number of
training epochs. If the number of weights is less than the number of
points and the number of training epochs is large enough we can have
only partial over-learning.

\chapter{Neural Network fit of $F_2$}
\def\ngen{rep}

The DIS cross section is expressed in terms of the nucleon
structure function $F_2$, that carries informations about the
inner structure of the nucleon.
If we were able to solve QCD in the non-perturbative domain,
we could calculate $F_2$ with quark masses and $\Lambda_{QCD}$
as the only inputs. Unfortunately this is not the case at present
times. However, we need a more and more detailed knowledge
of the structure of nucleons as they are essential ingredients
for present and future hadron colliders \cite{qcd}. Note that
the unpolarized structure function is necessary also to determine
the polarized structure function from the spin asymmetry in 
polarized deep inelastic scattering. Although the accuracy
of such experiments is not yet very high, it could be
anyway useful for future tasks to minimize the sources of errors.
We have thus to extract as much precise information as possible
from experiments. For this purpose we present here an alternative approach
to extract $F_2$ from data.
Results will also be published in Ref.~\cite{fglp}. 

The Chapter is organized as follows. First we will describe the
experimental data used in our fit with details on correlated
systematic uncertainties. Then, we will discuss techniques to
calculate errors and correlations from the fitted $F_2$, and we will
show how we can minimize theoretical assumptions on the shape of the
structure functions with the help of neural networks. A detailed
description of the behavior of neural networks will be given. Finally,
we will show our results for the non-singlet structure function
$F_2^p-F_2^d$, and preliminary results on the proton and the deuteron
structure functions.

\section{Experimental data}

\noindent
We have used experimental data given by the New Muon Collaboration
(NMC) \cite{NMC} and the BCDMS (Bologna-CERN-Dubna-Munich-Saclay)
Collaboration \cite{BCDMS} as they measure both the proton and
the deuteron structure functions. We have not considered data given by
the E665 Collaboration \cite{e665}. E665 data cover a kinematic range
with small $x$ and $Q^2$, that is almost entirely excluded in the
analysis performed in Chapter 7. The inclusion of these data would
have implied a further effort, and would not have been very
significant to our purpose. A future analysis may consider these data
as well. Note, however, that since some of these data cover a kinematic
range similar to that of NMC, we could use them to test the prediction
ability of our neural networks
\footnote{Remember that neural networks are good in interpolation,
but their are not in extrapolation. Thus, it is reasonable to test them 
on a kinematic range similar to the one they have been trained on.}.
Details on errors of both the experiments are given below. In
Fig. \ref{fig:kinrange} we show the kinematic region explored by 
the experiments.

\subsection{NMC}

We have used data and data tables given in \cite{NMC}. 
NMC data consist of four data sets for the proton
and the deuteron structure functions corresponding to beam energies of
$90,120,200$ and $280$ GeV, covering the kinematic range
$0.002\le x\le 0.60$ and $0.5\,\rm{GeV}^2\le Q^2 \le 75\,\rm{GeV}^2$. 
The systematic errors are:
\begin{itemize}
\item the incoming (E) and outgoing beam (E') energies, fully correlated 
between the proton and the deuteron data, but independent 
for the data taken at different beam energies;
\item the radiative correction (RC), fully correlated between all
energies, but independent for the proton and the deuteron;
\item the acceptance (AC) and the reconstruction efficiency (RE) fully
correlated for all data sets;
\item the normalization uncertainty ($2\%$), correlated between 
the proton and the deuteron data, but independent for data taken at 
different beam energies.
\end{itemize}

The uncertainties due to acceptance range from 0.1 to 2.5\% and reach
at most 5\% at large $x$ and $Q^2$. The uncertainty due to radiative
corrections is highest at small $x$ and large $Q^2$ and is at most 2\%.
The uncertainty due to reconstruction efficiency is estimated to be 4\%
at most. The uncertainties due to the incoming and the scattered muon 
energies contribute to the systematic error by at most 2.5\%.

Finally note that each bin of data has larger statistical errors in
the extremes and lower in the middle. In particular the worse points
are those of larger $Q^2$. The experimental reason for this
effect is related to the details of the geometry of the detector and
to the explored kinematic region. As one pushes the detection to some
areas at the limit of the acceptance region many points are lost and
the statistics goes down significantly.

\begin{figure}[t]
\begin{center}
\epsfig{width=0.7\textwidth,figure=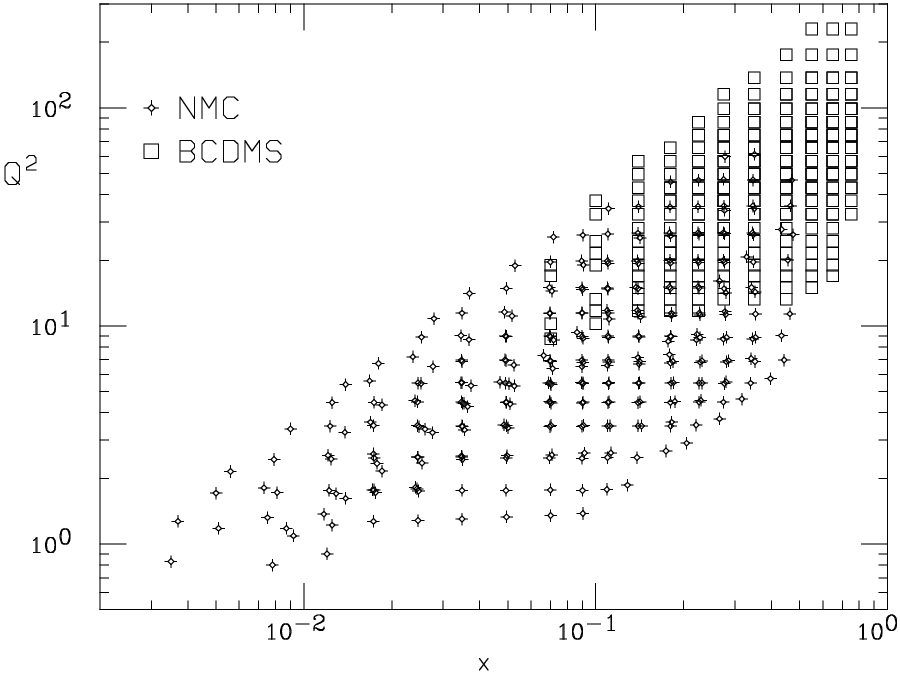}  
\end{center}
\begin{center}
\fcaption{}{NMC and BCDMS kinematic range.}
\label{fig:kinrange}
\end{center}
\end{figure}

\subsection{BCDMS}

We have used data given in \cite{BCDMS} and data tables given in 
\cite{BCDMS-pre}. BCDMS data consist of four data sets for the proton
structure function corresponding to beam energies of
$100,120,200$ and $280$ GeV and three data sets for the deuteron
structure function corresponding to beam energies of 
$120,200$ and $280$ GeV covering the kinematic range
$0.06\le x\le 0.80$ and $7 \,\rm{GeV}^2\le Q^2 \le 280\,\rm{GeV}^2$.
The systematic errors are:
\begin{itemize}
\item $f_{N}=3\%$, the absolute normalization error totally
correlated for all energies and the two targets;
\item $f_b<0.15\%$, the calibration of the incident beam energy;
\item $f_s<0.15\%$, the calibration of the spectrometer 
magnetic field;
\item $f_r<1\%$, the calibration of the outgoing muon energy;
\item $f_d$, inefficiencies of the detector (negligible).
\end{itemize}
The first two systematic errors sources have large effects at large
$x$ and small $Q^2$. Even if the experiments for the two targets
where done at different times, the systematic errors are fully
correlated for all targets and for all beam energies \cite{milsztajn}:
\begin{itemize}
\item the calibration of the incoming beam energy E was 
realized in a magnet on the beam line, and it was dominated by 
a systematic uncertainty due to scintillators that measured the
muon beam. This position was more stable in time than the precision 
of measurements;
\item the calibration of the outgoing muon energy was reproducible
with a relative value of about 0.02\%. This remained the same
in the four years of the experiment, and thus remained independent
of the used target or the beam energy. On the other hand the
absolute calibration had an uncertainty at a level of 0.1-0.2\%;
\item the resolution of the spectrometer depended on the 
constituent material, which had not changed in the four years, thus
same detector, same magnet and same positions. It could have depended
on the hadronic production absorbed in the magnet, as these particles
perturbed the recognition of the scattered muon. However, the hadronic
productions from proton and deuteron are so similar that this
effect is negligible.
\end{itemize} 

BCDMS data have a further source of uncertainty due to the relative
normalization of data between different beam energies and different
targets. Specifically, there is a $2\%$ cross section normalization of
data taken with different targets, a $1\%$ relative cross section
normalization of data taken at different beam energies for the proton,
a $1\%$ relative cross section normalization between data taken 
at 120 GeV and 200 GeV and $1.5\%$ between data taken 
at 200 GeV and 280 GeV for the deuteron. 

\section{Fitting procedure}

\noindent
We consider the case where we have $M$ measurements of 
a nucleon structure function $F_2$. 
The central problem is to determine
$F_2$ based on observations $F_2^{(1)},\ldots,F_2^{(M)}$. 
Specifically, we can introduce a 
hypothesis for the structure function $F_2$ which depends on
unknown parameters
$\boldsymbol{\theta}=(\theta_1,\ldots,\theta_m)$. The goal is then to
estimate parameters by comparing the hypothesis with 
experimental data.

We first discuss as an example a functional parametrization 
of $F_2$ (see \cite{Tulay} and references therein).
A QCD inspired parametrization of $F_2$ can be constructed by
observing that the evolution equations (\ref{AP}) introduce 
a logarithmic dependence on $Q^2$, while a polynomial function could
fit the dependence on $x$. Moreover, from the parton model we have
a further constraint on the $x$ behavior for $x=1$.
We can then take 
\bea
F_2(x,Q^2)=x^{a_1}\,f(x,Q^2)\,,
\label{f2tulay}
\eea
where
\bea
f(x,Q^2)=A(x)\l[\frac{\log Q^2/\Lambda^2}{\log Q_0^2/\Lambda^2}\r]^{B(x)}
\l[1+\frac{C(x)}{Q^2}\r]\,,
\label{f2tulay2}
\eea
and
\bea
A(x)&=&(1-x)^{a_2}[a_3+a_4\,(1-x)+a_5\,(1-x)^2 \nonumber \\
&+& a_6\,(1-x)^3+a_7\,(1-x)^4]\,, 
\nonumber \\
B(x)&=& b_1+b_2\,x+\frac{b_3}{x+b_4}\,,
\\ \nonumber 
C(x)&=& c_1\,x+c_2\,x^2+c_3\,x^3+c_4\,x^4\,.
\eea
The small $x$ behavior
of the function is described by the $a_1$ parameter, while the $a_2$
term carries most of the information on the large $x$
behavior. The reference scale is fixed arbitrarily at
$Q_0^2=20\,\mathrm{GeV}^2$, while the value
$\Lambda=0.250\,\mathrm{GeV}$ is extracted from $\as$ measurements at
different $Q^2$ \cite{BCDMSlambda}.  The last term of
eq.~(\ref{f2tulay2}) takes account of {\it higher twist} effects (see
the next Chapter for details).

The total number of parameters in eq.~(\ref{f2tulay})
is 15, and they can be
estimated by minimizing
the $\chi^2$ on a given set of experimental data. Once the parameters
are known we can evaluate any quantity that depends on $F_2$,
\eg the asymmetry $A_1$ or a Mellin moment. Since parameters are
themselves stochastic variables, they are determined with an error
and with a correlation matrix. In this way a value of the structure
function is given with an error that is the combination of the
parameter errors. Note that because of the non-linearity of
$F_2$ some linearization is necessary to determine the
structure function error, and this may be a source of uncertainty.
With this approach it may be non-trivial to obtain the errors on
two given data points and especially their correlations given by
\bea
&&\mathrm{cov}[F_2(x_1,Q_1^2),F_2(x_2,Q_2^2)]= \\ \nonumber
&&\sum_{i,j=1}^m \frac{\partial F_2(x,Q^2)}{\partial\theta_i}
\frac{\partial F_2(x,Q^2)}{\partial\theta_j}
\Bigg|_{(x_1,Q_1^2),(x_2,Q_2^2)}
\mathrm{cov}[\theta_i,\theta_j]\,.
\eea 
Even more complications arise if we would like to calculate functions
of $F_2$, as for example a Mellin moment, while it would be a very hard
task to evaluate the correlation between two Mellin moments. A
possible way to overcome this problem was given in
\cite{Tulay}. Indeed, since $F_2$ is provided with an ``estimated
error band'' one might hope to get a qualitative idea of the error by
taking the integrals of the upper and lower curves of the band as
estimates of the error on a Mellin moment. This procedure is however
meaningless. In fact, if we use the fitted $F_2$ to extract, 
as an example, $\as$, we note that the error on $\as$
could be made arbitrarily small by increasing the number of
values of $Q^2$ at which the moment is evaluated (even within a fixed
range in $Q^2$) \cite{FM}. This apparently paradoxical result is of
course due to the fact that the procedure neglects correlations
between the values of the moment extracted from the fit at two
different scales, which tend to one as the scales get closer.

The Monte Carlo method is an alternative approach for
calculating errors and correlations by using
sequences of random numbers. A sequence of Monte Carlo generated values
of $F_2$ may be used to evaluate estimators for errors and correlations of
the values of the parameters of $F_2$. Techniques for
constructing estimators are discussed in Appendix B.3.
An important feature of properly constructed estimators is that their
statistical accuracy improves as the number of values $N$ in the data
sample increases. A question which we will address is how large
$N$ must be to achieve an accuracy at the percent level on
estimators.

A functional parametrization as the one given in eq.~(\ref{f2tulay})
introduces an uncertainty due to the imposed dependence on $x$ and
$Q^2$. However, nobody knows which is the real behavior of $F_2$, and
any assumption on its functional form may be a source of
uncertainty, whose exact size is very hard to assess. 
For this purpose we will consider a fit with a neural
network. Indeed, a neural network with a given architecture can
describe a structure function as well as, say, a demographic
distribution having very different behaviors: the difference only
depends on the input and the output data which it is trained with. In
this case however we will have a larger number of parameters than with
the functional parametrization. The number of parameters included in a
fit corresponds more or less at the number of terms included in a
Taylor expansion of $F_2$. As nobody knows the exact expression of
$F_2$, nobody also knows how many terms we have to include in its
Taylor expansion or the exact number of parameter that we need to fit
it. The space of parameters is an infinite dimensional space, and the
arbitrary choice of a fixed number of parameters corresponds to an
arbitrary reduction of this space without assessing the
uncertainty with which this reduction is done. As we have seen in
Section 5.4.1 the number of neurons, and then of the parameters, in a
neural network is chosen only by looking at the stability of the error
function without making any theoretical assumption on their number.
We also observe that once the stability of a neural network is reached,
information is maintained even in the case a neuron dies. In this way a
neural networks guarantees a more robust and less arbitrary
parametrization of a structure function.  The only request is to
determine the most stable and economic architecture for the problem at
hand. Thus, a neural network fit of $F_2$ will avoid both theoretical
assumptions on the functional behavior of the structure function and
an arbitrary choice of the number of parameters used in the fit.

Specifically, we will proceed as follows. First, we will
generate $N$ replicas of artificial data with the given experimental
statistical and systematic correlated errors starting from the
original ones. Then, we will perform a neural network fit of each
replica. Finally, we will take the average over the number of neural
networks, reproducing central values, errors and correlations of the
original data with an uncertainty that depends on the number of
generated replicas $N$. In this way errors and correlations between
two quantities depending on the structure functions (\eg both two
values of the $A_1$ asymmetry or two Mellin moments) can be 
calculated keeping under control all the theoretical biases.

\section{Generation of artificial data}
\noindent
Artificial data for the NMC experiment are generated with
\bea
F_i^{(art)}&=& (1+r_7\,\sigma_{N}) \l[F_i^{(exp)} \r. \\ \nonumber 
&+&\l. \frac{r_1\,E+r_2\,E'+r_3\,AC+r_4\,RC+r_5\,RE}{100}F_i^{(exp)}+
r_6\,\sigma_{stat}\r]\,,
\eea
where the systematic errors have been discussed in the previous
section and $r_i$ are random Gaussian numbers. In order to reproduce
the correct correlations we choose $r_3$ and $r_5$ equal for different
energies and target, $r_4$ equal for data of the same target, $r_1$,
$r_2$ and $r_7$ equal for different targets.
Gaussian random numbers are generated with the {\tt gasdev} 
routine given in \cite{NumRec} and reported in Appendix C.

Artificial data for the BCDMS experiment are generated with
\bea
F_i^{(art)}&=& (1+r_5\,\sigma_{N}) \sqrt{1+r_6\,\sigma_{N_t}}\sqrt{1+r_7\,\sigma_{N_b}}
\l[F_i^{(exp)} \r. 
\label{genbcdms}
\\ \nonumber 
&+&\l. \frac{r_1\,f_b+r_2\,f_s+r_3\,f_r}{100}F_i^{(exp)}+
r_4\,\sigma_{stat}\r]\,,
\eea
where the systematic errors have been discussed in the previous
section, $\sigma_{N}$ is the absolute normalization, $\sigma_{N_t}$ is
the relative normalization between different targets, $\sigma_{N_b}$
is the relative normalization between different beam energies, $r_i$ are
random Gaussian numbers as above. In particular, $r_1$,$r_2$,$r_3$ and
$r_5$ are equal for different energies and different targets, $r_6$ is
the same for data of a given target, $r_7$ is the same for data of a given
energy. 

A relative normalization uncertainty $\sigma_{N_m}$
between two measurements $m_{1}$ and $m_{2}$ is taken into account by
multiplying one measure by $\sqrt{1+\sigma_{N_m}\,r}$ and the other by
$1/\sqrt{1+\sigma_{N_m}\,r}$, where $r$ is a random Gaussian number. 
In this way the product $m_1 m_2$ is equal to one, while their ratio gives 
$1+\sigma_{N_m}$. Thus, when the normalization of $m_1$ increases, that of
$m_2$ decreases.
The same result is obtained by multiplying the two measurements by
$\sqrt{1+\sigma_{N_m}\,r}$ and $\sqrt{1-\sigma_{N_m}\,r}$
respectively. If we generalize to many measurements having 
a relative normalization uncertainty with each other, 
we obtain as an averaged effect
an additional uncertainty on the measurements, see
eq.~(\ref{genbcdms}). The error introduced by the relative
normalization uncertainties is $\sqrt{1+\sigma_{N_m}}-1\sim
\frac{1}{2}\sigma_{N_m}$.

We now address the problem of understanding how the central values, 
errors, and point-to-point correlations computed from the 
artificially generated data compare to the corresponding 
input experimental values, as a
function of the number of generated replicas.
\begin{figure}[t]
\begin{center}
\includegraphics[width=0.50\textwidth]{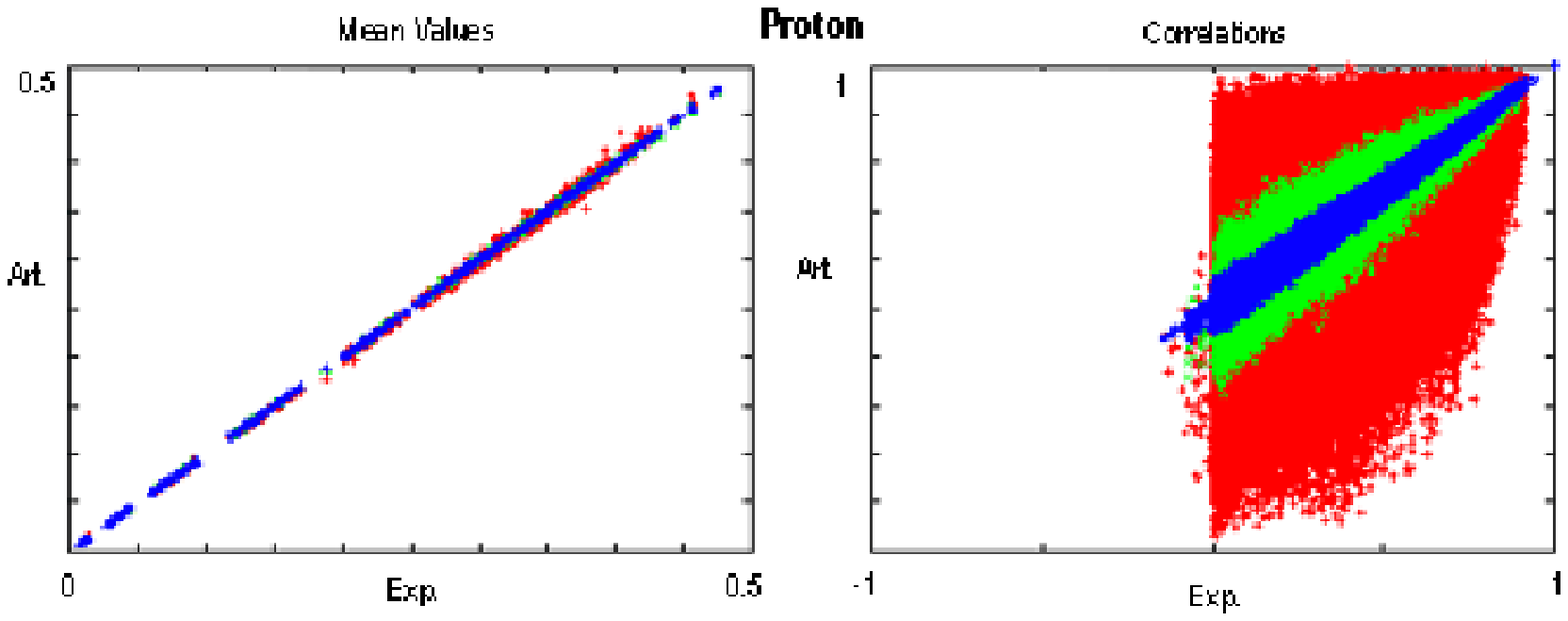}
\includegraphics[width=0.50\textwidth]{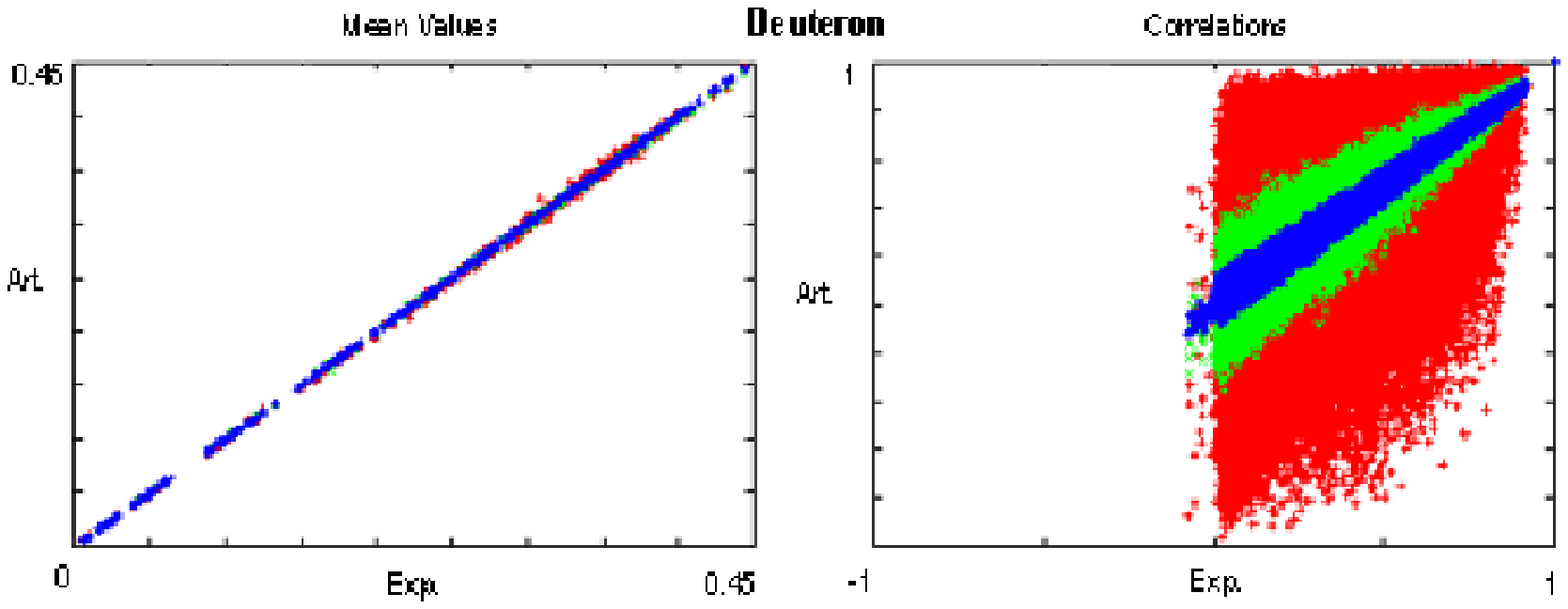}
\includegraphics[width=0.50\textwidth]{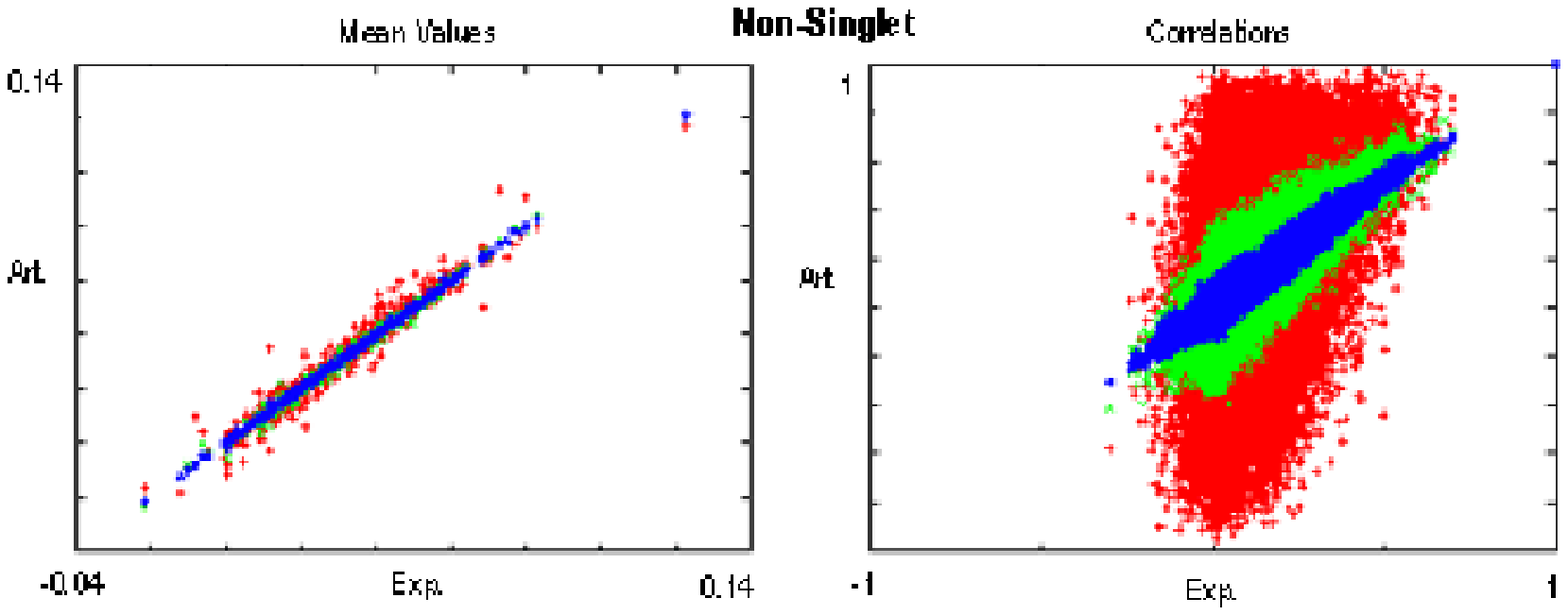}
\end{center}
\fcaption{}{$\Big\lan F_i\Big\ran_{\ngen}$ vs $F_i$ and
$r^{(art)}$ $vs$. $r^{(exp)}$ for the different 
structure functions with $N_{\ngen}=$10, 100 and 1000.}
\label{fig:genexp}
\end{figure}
A qualitative answer can be obtained from Fig.~\ref{fig:genexp},
where we show the scatter plots for mean values and correlations
for the proton, the deuteron and the non-singlet structure functions.
The reason why we take into account also the non-singlet structure
function will be explained in the next section.

\begin{table}[t]  
\begin{center}  
\begin{tabular}{cccc} 
\multicolumn{4}{c}{$F_2^p$}\\ 
\hline  
$N_{\ngen}$ & 10 & 100 & 1000 \\
\hline  
$\Big\lan V\l[\Big\lan F \Big\ran_{\ngen}\r]\Big\ran_{points}$ 
  & $1.2\times 10^{-5}$ & $1.4\times 10^{-6}$ & $2.5\times 10^{-7}$\\
$\Big\lan PE\l[\Big\lan F \Big\ran_{\ngen}\r]\Big\ran_{points}$ 
  & 0.9\% & 0.3\% & 0.1\% \\
$\widetilde{r}[F]$ 
  &  0.99966 & 0.99996 & 0.99999 \\
\hline
$\Big\lan V[\sigma ]\Big\ran_{points}$ 
  & $8.0\times 10^{-5}$ & $2.4\times 10^{-5}$ & $8.3\times 10^{-6}$ \\
$\Big\lan PE[\sigma ]\Big\ran_{points}$ 
  & 41\% & 12\% & 4\% \\
$\Big\lan \sigma^{(art)} \Big\ran_{points}$ 
  & 0.0114 & 0.0122 &  0.0122\\
$\widetilde{r}[\sigma ]$ 
  & 0.859 & 0.988 &  0.999\\
\hline
$\Big\lan V[r]\Big\ran_{points}$ 
  & 0.0904 & 0.0072 &  0.0007\\
$\Big\lan r^{(art)} \Big\ran_{points}$ 
  & 0.364 & 0.321 &  0.319\\
$\widetilde{r}[r]$ 
  & 0.723 & 0.952 & 0.995 \\
\hline
$\Big\lan V[\mathrm{cov}]\Big\ran_{points}$ 
  & $5.4\times 10^{-9}$ & $6.7\times 10^{-10}$ & $ 5.6\times 10^{-11}$ \\
$\Big\lan \mathrm{cov}^{(art)} \Big\ran_{points}$ 
  & $4.0\times 10^{-5}$ & $4.0\times 10^{-5}$ & $ 3.8\times 10^{-5}$\\
$\widetilde{r}[\mathrm{cov}]$ 
  & 0.529 & 0.891 & 0.987 \\
\hline
\end{tabular}
\end{center}
\tcaption{}{Comparison between experimental and generated artificial data for
the proton structure function. 
Experimental data yield: $\Big\lan \sigma^{exp}
\Big\ran_{points}=0.0123$, $\Big\lan r^{exp} \Big\ran_{points}=0.323$
and $\Big\lan \mathrm{cov}^{exp} \Big\ran_{points}=3.9\times 10^{-5}$.}
\label{Tgenexpp}
\end{table}
\begin{table}[t]  
\begin{center}  
\begin{tabular}{cccc} 
\multicolumn{4}{c}{$F_2^d$}\\   
\hline
$N_{\ngen}$ & 10 & 100 & 1000 \\
\hline  
$\Big\lan V\l[\Big\lan F \Big\ran_{\ngen}\r]\Big\ran_{points}$ 
  & $9.2\times 10^{-6}$ & $9.1\times 10^{-7}$ & $9.7\times 10^{-8}$\\
$\Big\lan PE\l[\Big\lan F \Big\ran_{\ngen}\r]\Big\ran_{points}$ 
  & 1.1\% & 0.3\% & 0.1\% \\
$\widetilde{r}[F]$ 
  & 0.99976  & 0.99998 & 0.99999 \\
\hline
$\Big\lan V[\sigma ]\Big\ran_{points}$ 
  & $5.3\times 10^{-5}$ & $1.5\times 10^{-5}$ & $4.0\times 10^{-6}$ \\
$\Big\lan PE[\sigma ]\Big\ran_{points}$ 
  & 39\% & 11\% & 3\% \\
$\Big\lan \sigma^{(art)} \Big\ran_{points}$ 
  & 0.0095 & 0.0102 & 0.0102 \\
$\widetilde{r}[\sigma ]$ 
  & 0.857 & 0.990 & 0.999 \\
\hline
$\Big\lan V[r]\Big\ran_{points}$ 
  & 0.0923 & 0.0075 & 0.0007 \\
$\Big\lan r^{(art)} \Big\ran_{points}$ 
  & 0.374 & 0.310 & 0.310 \\
$\widetilde{r}[r]$ 
  & 0.641 & 0.934 & 0.993 \\
\hline
$\Big\lan V[\mathrm{cov}]\Big\ran_{points}$ 
  & $2.6\times 10^{-9}$ & $2.5\times 10^{-10}$ & $ 2.3\times 10^{-11}$ \\
$\Big\lan \mathrm{cov}^{(art)} \Big\ran_{points}$ 
  & $3.3\times 10^{-5}$ & $3.3\times 10^{-5}$ & $ 3.2\times 10^{-5}$\\
$\widetilde{r}[\mathrm{cov}]$ 
  & 0.568 & 0.932 & 0.992 \\
\hline
\end{tabular}
\end{center}
\tcaption{}{Comparison between experimental and generated artificial data for
the deuteron structure function. 
Experimental data yield: $\Big\lan \sigma^{exp}
\Big\ran_{points}=0.0102$, $\Big\lan r^{exp} \Big\ran_{points}=0.313$
and $\Big\lan \mathrm{cov}^{exp} \Big\ran_{points}=3.3\times 10^{-5}$.}
\label{Tgenexpd}
\end{table}
\begin{table}[t]  
\begin{center}  
\begin{tabular}{cccc} 
\multicolumn{4}{c}{$F_2^p-F_2^d$}\\   
\hline  
$N_{\ngen}$ & 10 & 100 & 1000 \\
\hline  
$\Big\lan V\l[\Big\lan F \Big\ran_{\ngen}\r]\Big\ran_{points}$ 
  & $1.4\times 10^{-5}$ & $1.8\times 10^{-6}$ & $2.7\times 10^{-7}$\\
$\Big\lan PE\l[\Big\lan F \Big\ran_{\ngen}\r]\Big\ran_{points}$ 
  & 35\% & 11\% & 4\% \\
$\widetilde{r}[F]$ 
  &  0.980 & 0.998 & 0.999 \\
\hline
$\Big\lan V[\sigma ]\Big\ran_{points}$ 
  & $6.6\times 10^{-5}$ & $2.19\times 10^{-5}$ & $7.8\times 10^{-6}$ \\
$\Big\lan PE[\sigma ]\Big\ran_{points}$ 
  & 35\% & 12\% & 4\% \\
$\Big\lan \sigma^{(art)} \Big\ran_{points}$ 
  & 0.0101 & 0.0114 & 0.0114 \\
$\widetilde{r}[\sigma ]$ 
  & 0.927 & 0.991 & 0.999 \\
\hline
$\Big\lan V[r]\Big\ran_{points}$ 
  & 0.1133 & 0.0094 & 0.0010 \\
$\Big\lan r^{(art)} \Big\ran_{points}$ 
  & 0.1112 & 0.0990 & 0.0946 \\
$\widetilde{r}[r]$ 
  & 0.405 & 0.816 & 0.971 \\
\hline
$\Big\lan V[\mathrm{cov}]\Big\ran_{points}$ 
  & $5.1\times 10^{-9}$ & $5.8\times 10^{-10}$ & $ 6.0\times 10^{-11}$ \\
$\Big\lan \mathrm{cov}^{(art)} \Big\ran_{points}$ 
  & $7.3\times 10^{-6}$ & $9.0\times 10^{-6}$ & $ 8.7\times 10^{-6}$\\
$\widetilde{r}[\mathrm{cov}]$ 
  & 0.346 & 0.791 & 0.972 \\
\hline
\end{tabular}
\end{center}
\tcaption{}{Comparison between experimental and generated artificial data for
the non-singlet structure function. 
Experimental data yield: $\Big\lan \sigma^{exp}
\Big\ran_{points}=0.0114$, $\Big\lan r^{exp} \Big\ran_{points}=0.090$
and $\Big\lan \mathrm{cov}^{exp} \Big\ran_{points}=8.4\times 10^{-5}$.}
\label{TgenexpNS}
\end{table}
A more detailed description can be obtained by defining 
the following quantities:
\begin{itemize}
\item average over the number of replicas for each experimental point $i$
(\ie~a pair of values of $x$ and $Q^2$)
\bea
\Big\lan F_i\Big\ran_{\ngen}=\frac{1}{N_{\ngen}}\sum_{k=1}^{N_{\ngen}}
F_i^{(k)}\,.
\eea
The error on $\lan F_i\ran_{\ngen}$ is given by
\bea
\sigma_i=\sqrt{\Big\lan F_i^2\Big\ran_{\ngen}-\Big\lan F_i\Big\ran_{\ngen}^2}\,;
\eea
the mean correlation between two points is given by
\bea
r_{ij}=\frac{\Big\lan F_i\,F_j\Big\ran_{\ngen} - \Big\lan F_i\Big\ran_{\ngen}
\Big\lan F_j\Big\ran_{\ngen}}{\sigma_i\,\sigma_j}\,;
\eea
\item mean variance and mean percentage error on central values over 
the number of points $N_{points}$ 
\bea
\Big\lan V\l[\Big\lan F \Big\ran_{\ngen}\r]\Big\ran_{points} &=& 
\frac{1}{N_{points}} \sum_{i=1}^{N_{points}}\l(\Big\lan F_i\Big\ran_{\ngen}-F_i\r)^2 \\
\Big\lan PE\l[\Big\lan F \Big\ran_{\ngen}\r]\Big\ran_{points}
&=& 
\frac{1}{N_{points}} \sum_{i=1}^{N_{points}}\l|\frac{\Big\lan F_i\Big\ran_{\ngen}-F_i}{F_i}\r|\,. 
\eea
$\Big\lan V[\sigma]\Big\ran_{points}$, 
$\Big\lan PE[\sigma]\Big\ran_{points}$, $\Big\lan V[r]\Big\ran_{points}$
and $\Big\lan V[\mathrm{cov}]\Big\ran_{points}$ are defined in a similar way. 
These estimators 
indicate how close the averages over generated data are to the
experimental values. Specifically, they will indicate if and how
the slopes of the scatter plots in Fig.~\ref{fig:genexp} differ from one; 
\item {\it averaged correlation}: 
\bea
\widetilde{r}[F]=\frac{\Big\lan F\, \Big\lan F\Big\ran_{\ngen}  
\Big\ran_{points} - \Big\lan F\Big\ran_{points}
\Big\lan \Big\lan F\Big\ran_{\ngen}  
\Big\ran_{points}}{\sigma^{(exp)}\,\sigma^{(art)}}\,.
\eea
Similarly we define $\widetilde{r}[\sigma]$, $\widetilde{r}[r]$ and
$\widetilde{r}[\mathrm{cov}]$. This estimator indicates which is the
spread of data around the {\it art. vs. exp.} line 
in the scatter plots of Fig.~\ref{fig:genexp}.
\end{itemize}
We expect \eg the variance on central values to scale as
$1/N_{\ngen}$, while the variance on the errors should scale as
$1/\sqrt{N_{\ngen}}$ (see Appendix B.3). From
Tables~\ref{Tgenexpp},~\ref{Tgenexpd} and \ref{TgenexpNS} we see that
this is approximately the case. Note that the exact scaling
behavior is observed for $N_{\ngen}\sim 10^5$. We see also that one
needs about 100 artificial data to get an accuracy at the percent
level on central values, and about 1000 to get the same accuracy on
errors and correlations.

\section{Building and Training Neural Networks}
\noindent
We decided to fit separately $F_2^p$, $F_2^d$ and $F_2^p-F_2^d$.
Fitting the difference $F_2^p-F_2^d$ is safer than taking the
difference of $F_2^p$ and $F_2^d$ after they have been fitted separately.
Indeed, if we want a precision on $F_2^p-F_2^d$ of $10^{-2}$,
we must have a precision of at least $10^{-3}$ on $F_2^p$ and $F_2^d$
separately. However, since the precision on $F_2^p$ and $F_2^d$ is
$10^{-2}$, in this way we would have a very poor
precision, $10^{-1}$, on $F_2^p-F_2^d$.
The fit of $F_2^p-F_2^d$ is also delicate, as
the non-singlet combination is very close to zero on a wide range of
$x$, and errors are summed in quadrature; thus, lots of care must be
taken to avoid fitting noise. The problem does not arise for the
singlet combination $F_2^p+F_2^d$. In the following we outline the 
common features of the different fits, and then we will focus on 
details for each case.

We have constructed a neural network with the architecture (4,5,3,1), and we 
have used as
inputs $x$,$Q^2$,$\log x$ and $\log Q^2$. The choice of taking 
$\log x$ and $\log Q^2$ as inputs does not introduce 
a theoretical bias, as the neural networks could ``decide'' to
ignore these inputs if they are useless. One could even take as an
input, say, a temperature, and if this variable is  
useless for fitting the given function the neural networks would have zero
weights on the paths corresponding to this variable.

We have used sigmoid activation functions between the first three layers, 
while for the last layer we have chosen a linear activation function. As
explained in the previous Chapter, this guarantees a smoother
behavior of the neural network.

The number of units per layer has been chosen by trial-and-error, 
\ie by adding a unit to a layer and looking for the
stability of the output. In particular, we looked for the stability of
correlation between central values by asking that it was more than
$80\%$.

We have adopted the on-line training since the number of data is reasonably
small. As we pointed out in the previous Chapter, it is worth  
pre-processing data in order to avoid fitting biases. In particular, we 
should not take data in the sequential given order, but it is better
to pick them randomly. For this purpose we have used the
{\tt idirty} random number generator of Numerical Recipes
\cite{NumRec} reported in Appendix C.
Such a generator has a periodicity that may cause oscillations in the
sampling of data. However, we have checked that 
oscillations due to the periodicity of the
random number generator do not sensibly affect the fit.

The only theoretical assumption on the shape of $F_2$ has been the
kinematic request $F_2(x=1, Q^2)=0$. This has been done
by artificially adding 10 points at $x=1$ with equally spaced values
of $Q^2$. The choice of the error on these points is very delicate,
because if it is too small the neural networks would spend a lot of
their training time in learning these points.  One would obtain a very
precise fit of the kinematical constraint $F_2(x=1, Q^2)=0$, and a
worse fit of the experimental data. We have thus taken an error of
the same order of the smallest experimental error.  In particular, we
have taken $10^{-3}$ for $F_2^p$ and $F_2^d$ where data are very
different from zero within errors. For $F_2^p-F_2^d$ since the
structure function is very close to zero, we have asked a higher
precision on these points by setting the error equal to
$\sqrt{2}\,10^{-4}$.

As explained in Appendix B.4, the quantity we want to minimize 
in a fit is the {\it Covariance Matrix Estimator} (CME) given by
\bea
E[{\bf o}]\equiv\frac{1}{2}\sum_{\mu=1}^p\sum_{i,j=1}^m
(o_i({\bf x}^{\mu})-z_i^{\mu})V_{ij}^{-1}
(o_j({\bf x}^{\mu})-z_j^{\mu})\,,
\label{energy3}
\eea
as in this way not only statistical errors, but also correlated
systematic uncertainties are taken into account. The updating rules
given in eq.~(\ref{deltarule2}) that govern the neural network
learning, show that the variations on a single data point very
weakly affect a given weight. Thus, to have a stronger effect we can
switch to the batched training mode in which all the weights are
updated after all patterns have been presented to enhance the effect
of variations. However, the energy definition given in
eq.~(\ref{energy3}) makes a sum over all patterns appears on the
\rhs of the updating rule eq.~(\ref{deltarule2}) and its evaluation
deeply affects the rapidity of the training.
We can then minimize the simpler case where the energy is
defined as the {\it Simplest $\chi^2$ Estimator} (SCE),
\bea
E[{\bf o}]=\frac{1}{2}\sum_{\mu=1}^p\sum_{i=1}^m
\l(\frac{o_i({\bf x}^{\mu})-z_i^{\mu}}{\sigma_i}\r)^2\,.
\label{energy2}
\eea
If systematic errors are not predominant with respect to the 
statistical ones, eq.~(\ref{energy2}) can be a good estimator
of eq.~(\ref{energy3}) (see Appendix B.4). We will not take
care of correlations when we fit each neural network to
a replica of artificial data, since they are produced
anyway by the Monte Carlo. 
In order to increase the efficiency of the training we can perform it
in two cycles. First we minimize the error function
eq.~(\ref{energy}), and then we minimize eq.~(\ref{energy2}). This
procedure corresponds to starting with a coarse search for the
minimum, and then refining it once its neighborhood 
has been located. In the first
example, for the first cycle we have used $\eta=0.004$ and a number of
epochs of $2\times 10^6$. For the second cycle we have taken $\eta=4\times
10^{-8}$ and a number of epochs of $4\times 10^6$. For every cycle the
value of the momentum term was $\alpha=0.9$. As the number of epochs is
larger than $6\times 10^6$, every pattern may be seen at least $10^4$
times. Henceforth we will label a training by a shorthand for the
number of epochs of the second cycle; in the present case it is 4M.

As we have seen in the previous chapter, we expect neural networks to
produce lower errors because they interpolate smoothly. Actually we 
expect them to reduce more the statistical errors than the
systematic ones. An example is the problem of interpolating data randomly
distributed around a horizontal line. If we fit these data with, 
say, a parabola, the result of the fit will be a
parabola and not a horizontal line. If we instead take a set of 
neural networks, that do not have a definite functional behavior, we expect
that all distributions will have fits
very similar and close to the horizontal line. Average over fits will
give smaller errors to each point. If on top of the statistical error
a global shift is added, then the fits will follow this shift.  It is
clear that infinite training will make the fit to go on top of each
point and then all errors would be reproduced.  But then the neural
network is no longer assuming continuity or capacity of
generalization. For that purpose a cubic spline fit would do the job.

The question which now we address is how the central values, errors,
and point-to-point correlations computed from the neural networks
compare to the corresponding data (experimental and artificial), both
as a function of the number of replicas and of the length of the
training, and both when comparing neural networks one by one, or 
their average.  We then define the following quantities
($N_{\ngen}=N_{net}$):
\begin{itemize}
\item we define an averaged $\chi^2$ over the number of
points for practical reasons. As the number of points is of order 500 and the
number of parameters is of order 50, this quantity differs
for a 10\% from the $\chi_{\mathrm d.o.f.}^2$. In the last section
we will show results also for $\chi_{\mathrm d.o.f.}^2$. We thus define:
\bea
\chi_i^{2\,(net-art)} &=& \frac{1}{N_{net}}\sum_{k=1}^{N_{net}} \l(
\frac{ F_i^{(k,net)} -  F_i^{(k,art)}}{\sigma_i}\r)^2\,, \\
\chi_i^{2\,(net-exp)} &=& \frac{1}{N_{net}}\sum_{k=1}^{N_{net}} \l(
\frac{ F_i^{(k,net)} -  F_i^{(exp)}}{\sigma_i}\r)^2\,,
\eea
and 
\bea
\chi_{SCE}^{2\,(net-art)} &=& \frac{1}{N_{points}}\sum_{i=1}^{N_{points}} \l(
\frac{\Big\lan F_i^{(net)}\Big\ran_{net} -  
\Big\lan F_i^{(art)}\Big\ran_{net}}{\sigma_i}\r)^2 \\
\chi_{SCE}^{2\,(net-exp)} &=& \frac{1}{N_{points}}\sum_{i=1}^{N_{points}} \l(
\frac{\Big\lan F_i^{(net)}\Big\ran_{net} -  
F_i^{(exp)}}{\sigma_i}\r)^2\,,
\eea
where $\sigma_i$ is the experimental statistical error;
$\chi_{CME}^2$ is defined in an analogous way with the inverse of 
the covariance matrix. In the following we will use always
$\chi_{SCE}^2$, unless otherwise explicitly stated.
The reason why we have given two different definition of $\chi^2$ will
become clear in the following. Note that the average of 
$\chi_i^{2\,(net-art,exp)}$ over the number of points gives
a result different from $\chi^{2\,(net-art,exp)}$, \ie
the averages are not commutative. Specifically, 
$\chi_i^{2\,(net-art,exp)}$ 
indicate how each neural network
fits the corresponding replica, while $\chi^{2\,(net-art,exp)}$
reflect the quality of the fit of experimental data
or the average over the replicas as obtained from the average over
the neural networks;
\item {\it averaged correlation}:
\bea
\widetilde{r}[F^{(net-art)}]&=& \frac{1}{\sigma^{(net)}\,\sigma^{(art)}}
\l[\Big\lan \Big\lan F^{(net)}\Big\ran_{net}
\Big\lan F^{(art)}\Big\ran_{net}  \Big\ran_{points} \r. \nonumber \\
&-& \l.\Big\lan \Big\lan F^{(net)}\Big\ran_{net} \Big\ran_{points}
\Big\lan \Big\lan F^{(art)}\Big\ran_{net} \Big\ran_{points}\r]\,.
\eea
Similarly we define $\widetilde{r}[\sigma^{(net-art)}]$,
$\widetilde{r}[r^{(net-art)}]$ and
$\widetilde{r}[\mathrm{cov}^{(net-art)}]$.
\item variance and percentage error are defined as in the previous 
section with obvious replacements
\end{itemize}

As an example we first consider a training of 1000 neural networks on
non-singlet data with the learning parameters discussed above. Results
are collected in Table~\ref{Tgennetst}. First, we observe that one can
reach a $\chi^{2\,(net-exp)}\sim\chi^{2\,(net-art)}\sim 1$ with an
80\% correlation between neural networks and data (experimental or
artificially generated), but the mean percentage error on the errors
remains about 90\%. The errors and correlations are only
correlated to about 40\%. Furthermore neural networks errors are
systematically lower and correlations systematically higher by about a
factor 4-5 than the experimental ones. Neural networks are producing a
statistically good fit, but they are smoothing.
\begin{table}[t!]  
\begin{center}  
\begin{tabular}{cccc} 
\multicolumn{4}{c}{$F_2^p-F_2^d$}\\  
\hline  
$N_{net}$ & NMC+BCDMS & NMC & BCDMS\\
\hline  
$\chi_{SCE}^{2\,(net-exp)}$ 
  & 0.97 &  0.57 &  1.41 \\ \\
$\Big\lan \chi^{2\,(net-art)} \Big\ran_{points}$ 
  & 2.05 & 1.64 & 2.50 \\
$\Big\lan \chi^{2\,(net-exp)} \Big\ran_{points}$ 
  & 1.22 & 0.73 & 1.74 \\
${\cal R}$ 
  & 0.60 & 0.45 & 0.70\\
$\widetilde{r}\l[F^{(net-exp)}\r]$ 
  & 0.85 & 0.74 &  0.96 \\
\hline
$\Big\lan PE[\sigma^{(net-exp)}]\Big\ran_{points}$ 
  & 94\%  & 95\% & 93\%\\
$\Big\lan \sigma^{(exp)} \Big\ran_{points}$ 
  & 0.011 & 0.016 &  0.006\\
$\Big\lan \sigma^{(net)} \Big\ran_{points}$ 
  & 0.003 & 0.004 & 0.002\\
$\widetilde{r}[\sigma^{(net-exp)}]$ 
  & 0.33 & 0.16 &  0.49\\
\hline
$\Big\lan r^{(exp)} \Big\ran_{points}$ 
  & 0.09 & 0.04 &  0.16\\
$\Big\lan r^{(net)} \Big\ran_{points}$ 
  & 0.59 & 0.45 &  0.77\\
$\widetilde{r}[r^{(net-exp)}]$ 
  & 0.42 & 0.28 &  0.48\\
\hline
$\Big\lan \mathrm{cov}^{(exp)} \Big\ran_{points}$ 
  & $8.4\times 10^{-6}$ & $9.7\times 10^{-6}$  & $6.8\times 10^{-6}$ \\
$\Big\lan \mathrm{cov}^{(net)} \Big\ran_{points}$ 
  & $6.5\times 10^{-6}$ & $7.6\times 10^{-6}$  & $5.1\times 10^{-6}$ \\
$\widetilde{r}[\mathrm{cov}^{(net-exp)}]$ 
  & 0.30 & 0.24 & 0.62\\
\hline
\end{tabular}
\end{center}
\tcaption{}{Comparison between neural networks and experimental data
for the non-singlet structure function with a 4M training.}
\label{Tgennetst}
\end{table}

In order to understand whether the smoothing
reproduces an underlying law or not, we will describe a toy model.
Let us first consider a measured value $m_i$ of $F_2$ where $i$
represents a pair of values $(x, Q^2)$; we have
\bea
m_i=t_i+\sigma_i\,s_i\,
\eea
where $s_i$ is a univariate Gaussian number, 
$t_i$ is the true value of $F_2$, and $\sigma_i$ its error.
The $k^{th}$ replica of generated data gives
\bea
g_i^{(k)}=m_i+r_k\,\sigma_i=t_i+(s_i+r_k)\sigma_i\,,
\eea
where $r_k$ is a univariate Gaussian random number. 
If the neural networks succeed in finding the true value $t_i$
with an error $\hat\sigma_i$ smaller than the experimental one,
for the $k^{th}$ neural network we have
\bea
n_i^{(k)}=t_i+r_k'\,\hat\sigma_i\,.
\eea
The variance between the neural networks values 
and the experimental data or the replicas is given by
\bea
&&\frac{1}{N_{\ngen}}\sum_{k=1}^{N_{\ngen}} \l(m_i-n_i^{(k)}\r)^2
=s_i^2\,\sigma_i^2+\hat\sigma_i^2\,.
\\
&&\frac{1}{N_{\ngen}}\sum_{k=1}^{N_{\ngen}} \l(g_i^{(k)}-n_i^{(k)}\r)^2
=(1+s_i^2)\,\sigma_i^2+\hat\sigma_i^2\,,
\eea
If we now take the $\chi^2$ divided by the number of points, we get
\bea
\Big\lan \chi^{2\,(net-exp)} \Big\ran_{points}&=&
\Bigg\lan \l(\frac{m-n^{(k)}}{\sigma}\r)^2  \Bigg\ran_{points}
=\lan\sigma^2\ran+\lan\hat\sigma^2\ran\,,
\nonumber \\ \\ \nonumber 
\Big\lan \chi^{2\,(net-art)} \Big\ran_{points}&=&
\Bigg\lan \l(\frac{g^{(k)}-n^{(k)}}{\sigma}\r)^2  \Bigg\ran_{points}
=2\lan\sigma^2\ran+\lan\hat\sigma^2\ran\,. 
\eea
The ratio yields
\bea
{\cal R} = 
\frac{1+\frac{\lan\hat\sigma^2\ran}{\lan\sigma^2\ran}}
{2+\frac{\lan\hat\sigma^2\ran}{\lan\sigma^2\ran}}\,,
\label{netratio}
\eea
thus, ${\cal R}\approx\frac{1}{2}$ if 
$\lan\hat\sigma\ran\ll\lan\sigma\ran$.
From Table~\ref{Tgennetst} we have that 
$\lan\hat\sigma\ran/\lan\sigma\ran\approx 0.4$, 
and ${\cal R}\approx 0.54$. As a consequence
\bea
\Big\lan \chi^{2\,(net-exp)} \Big\ran_{points}\sim
2\,\Big\lan \chi^{2\,(net-art)} \Big\ran_{points}\,.
\eea
If we first take the average over the replicas and then the 
variance between the neural networks values and the replicas ones
or the experimental data, we find
\bea
(m_i -\lan n_i\ran_{\ngen})^2=\sigma_i^2\,s_i^2\,,~~~~~
(\lan g_i\ran_{\ngen} -\lan n_i\ran_{\ngen})^2=\sigma_i^2\,s_i^2\,,
\eea
and
\bea
\chi^{2\,(net-exp)}=\chi^{2\,(net-exp)}=1\,.
\eea
Table~\ref{Tgennetst} beautifully shows this behavior.

To understand why the error of neural networks can be smaller than the
experimental one, and how we can recover this loss of information, as
a toy model let us assume that the experimental data on $F_2$ 
satisfy an underlying linear law
as a function of, say, the Bjorken variable $x$,
\bea
m_i=Cx_i+D\,.
\eea
Given $M+2$ data points, two points will be sufficient to determine
the linear parameters $C$ and $D$, whose error is $\lan\sigma\ran$.
All the other points will only play the role of
reducing the error on $C$ and $D$ by a factor
$\frac{1}{\sqrt{M}}$ since they are all measurements of the same
quantities $C$ and $D$. If we take
$\lan\hat\sigma\ran=\frac{1}{\sqrt{M}}\lan\sigma\ran$, 
from eq.~(\ref{netratio}) we have
\bea
{\cal R} = \frac{M+1}{M}\frac{M}{2M+1}=\frac{M+1}{2M+1}\ge\frac{1}{2}\,.
\eea
If $M=0$, the neural networks obviously correlate a point only with itself,
giving ${\cal R}=1$ and an undefined uncertainty; if
$M\ra\infty$, the neural networks correlate all points giving a null
error and ${\cal R}=1/2$. 

Obviously $F_2$ is neither a function of $x$ only nor is it linear.
Anyway the neural networks exhibit a finite correlation length
$\sim\frac{1}{\sqrt{M}}$ and reduce the error on each point.
Indeed, as $\lan\hat\sigma\ran/\lan\sigma\ran\approx 0.4$, 
we have that the neural networks correlate at about 10 points 
lowering their error at about by a factor $1/3$.
From Fig.~\ref{fig:corlen} we see the way the
neural networks behave. We have that BCDMS points
coming from data sets with different beam energies, but with the
same values of $x$ and $Q^2$ are seen as the same point. On the 
other hand we see also that points of NMC and BCMDS that should have
zero correlation between each other, are strongly correlated with all 
the points with similar values of $x$ and $Q^2$ independently of the
experiment. 
\begin{figure}[t]
\begin{center}
\epsfig{width=0.46\textwidth,figure=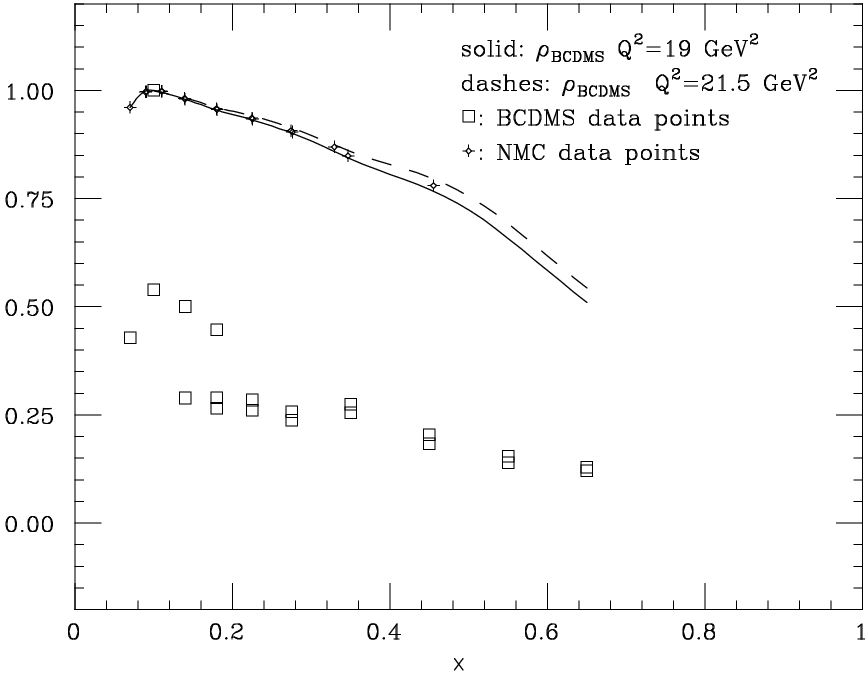}  
\epsfig{width=0.46\textwidth,figure=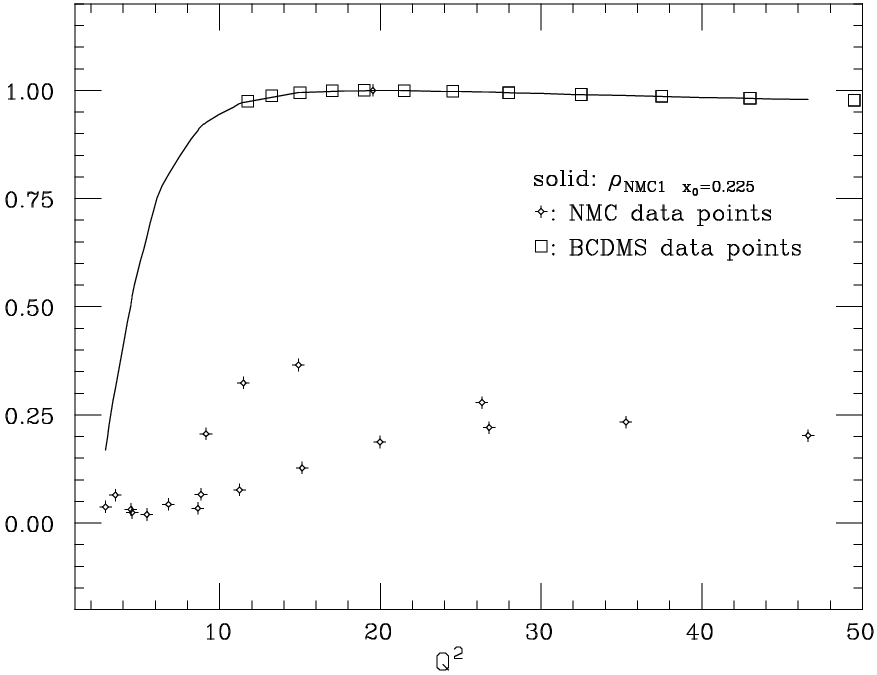}  
\end{center}
\begin{center}
\fcaption{}{Neural network correlation length for
$x$ and $Q^2$.}
\label{fig:corlen}
\end{center}
\end{figure}

\subsection{Non-singlet}

We now turn to specific details on the fit of the non-singlet structure
function. From the short training example given in
Table~\ref{Tgennetst} we see that $\chi^{2\,(net-exp)}$ for NMC is
excellent ($\sim 0.6$), but correlations and errors are very poorly
reproduced, while $\chi^{2\,(net-exp)}$ for BCDMS is good ($\sim
1.4$), but not as good as for NMC, and correlations and errors are
well reproduced. In order to better explain the neural networks
behavior we will study the relative impact of the two experiments on
the training of the neural networks. Here we will consider
the training of only one neural network on the experimental data.

First we find that the length of the training on NMC does not
improve errors and correlations for NMC. Then increasing the
training for BCDMS one would expect to improve $\chi^{2\,(net-exp)}$
for BCDMS while deteriorating $\chi^{2\,(net-exp)}$ for NMC. However,
while BCDMS does improve, NMC does not deteriorate.
If we now fit the neural networks to BCDMS only plus the points at
$x=1$, we find that $\chi^{2\,(net-exp)}$ for NMC is just as good as
before. We conclude that the NMC points have essentially no impact on
these fits: the corresponding central values are ``predicted'' by the
neural networks fitted to BCDMS.  Thus neural networks errors and
correlations for NMC have nothing to do with their experimental
values. The neural networks trained on both experiments only exploit
the BCDMS data, while the NMC data have little or no impact on them.

We then address the problem of reducing $\chi^{2\,(net-exp)}$ for
BCDMS where the mediocre quality of the BCDMS fit is
mostly due to the large $x$ region that is not very well
reproduced. Notice that if we increase the error on the
points at $x=1$, this does not help in making the fit of the remaining
points easier (\ie the $\chi^{2\,(net-exp)}$ for the remaining points
remains more or less the same). If we now increase the training, say at
40M, and keep fitting to BCDMS only, the ability of the
neural networks to predict NMC deteriorates very fast as
$\chi^{2\,(net-exp)}$ for BCMDS improves. Roughly, when BCDMS is down
to $\chi^{2\,(net-exp)}=1$, NMC is already at $\chi^{2\,(net-exp)}=2.6$.

However, it turns out that we could get an equally good fit for the
two experiments by not excluding the NMC data completely, but also not
showing them the same number of times as the BCDMS points
(Fig.~\ref{fig:chi2bcd_nmc10}). We find that a rather long training
(of order 150M) is required in order to get a good
$\chi^{2\,(net-exp)}$ for BCDMS. We obtain a good
equilibrium between the two experiments by showing the NMC data 10\%
of times. Specifically, with 180M training cycles, we get
\bea
\chi^{2\,(net-exp)}_{BCDMS}=\chi^{2\,(net-exp)}_{NMC}
=\chi^{2\,(net-exp)}_{BCDMS+NMC}=0.88\,.
\eea

In order to further understand the relative impact of the two
experiments we can show some more fits. 
Specifically, to the fit training the neural network only on
BCDMS data (Fig.~\ref{fig:chi2bcd}), we add a fit training the neural
network to NMC data only (plus the 10 points at $x=1$), see
Fig.~\ref{fig:chi2nmc}. The interesting result is that in both cases
the neural network trained on one data set successfully predicts the
other data set. The $\chi^{2\,(net-exp)}$ of the predicted data set is
in both cases about 1.5 after a long enough training. In particular
the very bad behavior of the predicted NMC as the
$\chi^{2\,(net-exp)}$ for BCDMS improves is a feature of short-medium
training, 40M, as the training goes on also the predicted NMC improves
again. Thus, the neural network can use either of the two experiments
to predict the other one. Of course, if the experiment with smaller
errors is used to train the neural network, the other one is
predicted also with smaller error.
\begin{figure}[t!]
\begin{center}
\begin{tabular}{lr}
\begin{minipage}[t]{0.48\textwidth}
    \includegraphics[width=\textwidth,clip]{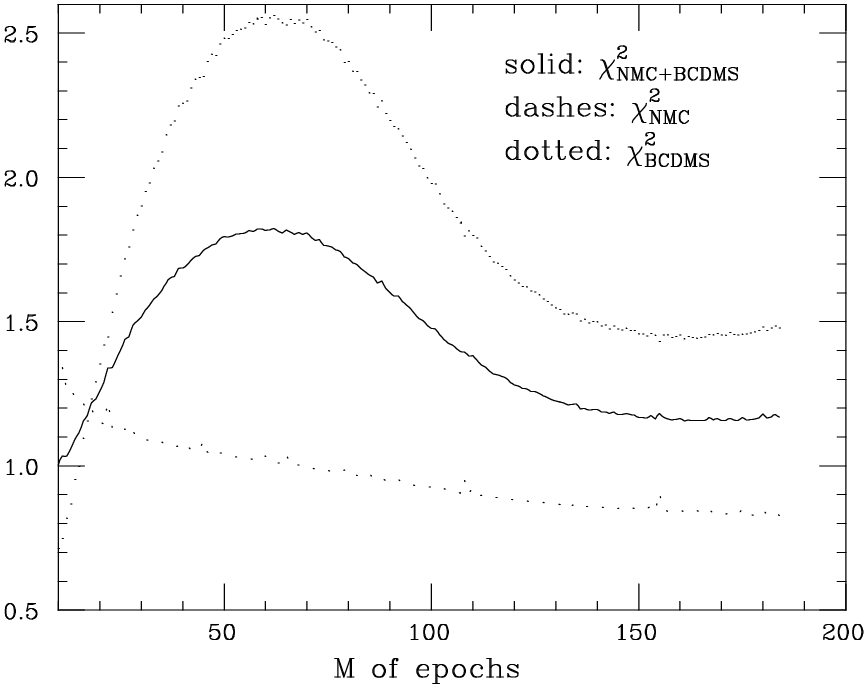}
\fcaption{}{Training with BCDMS data.
    \label{fig:chi2bcd}}
\vskip-0.5cm
\end{minipage}
&
\begin{minipage}[t]{0.48\textwidth}
    \includegraphics[width=\textwidth,clip]{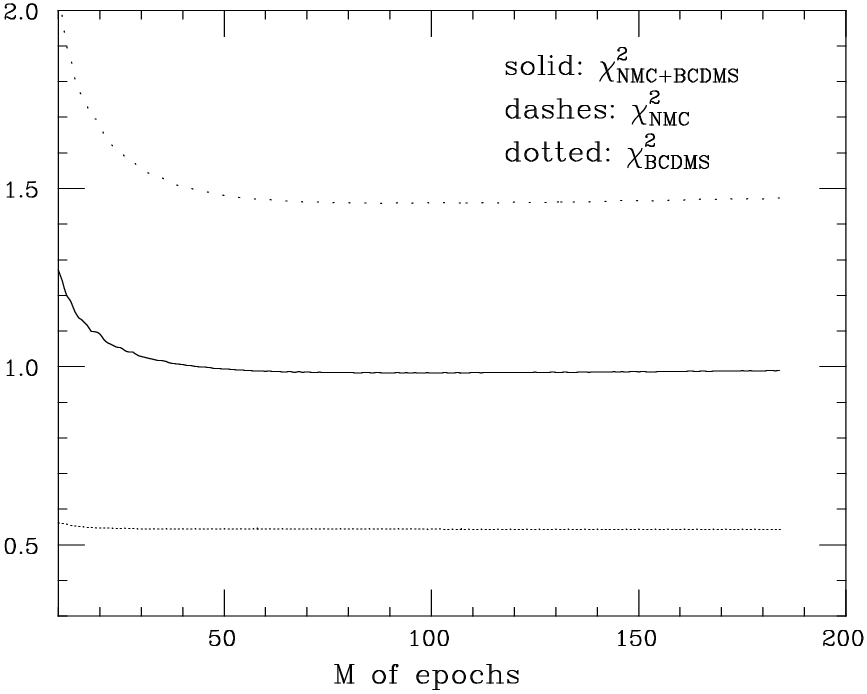}
    \fcaption{}{Training with NMC data.
    \label{fig:chi2nmc}}
\vskip-0.5cm
\end{minipage}
\end{tabular}
  \end{center}
\vspace{-0.5cm}
\end{figure}
\begin{figure}[t!]
\begin{center}
\begin{tabular}{lr}
\begin{minipage}[t]{0.47\textwidth}
    \includegraphics[width=\textwidth,clip]{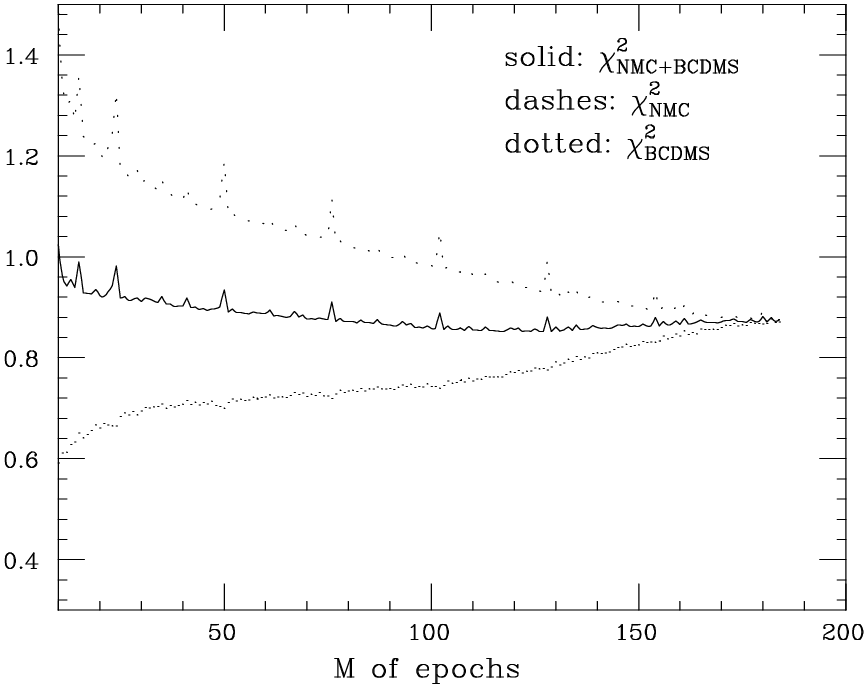}
    \fcaption{}{Training with BCDMS data and 10\% of times with NMC data
    \label{fig:chi2bcd_nmc10}}
\end{minipage}
&
\begin{minipage}[t]{0.47\textwidth}
    \includegraphics[width=\textwidth,clip]{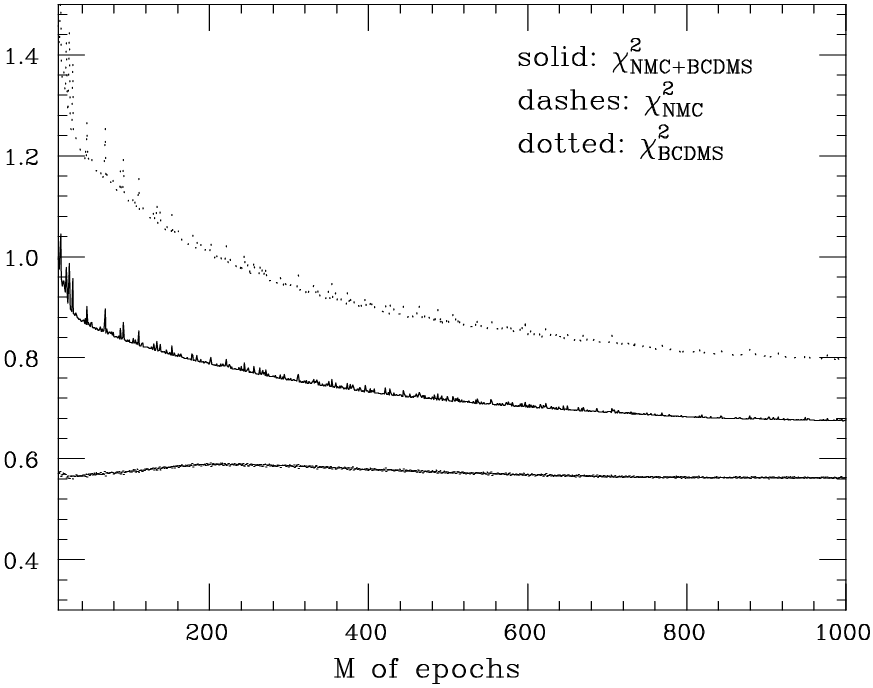}
    \fcaption{}{Long training with NMC and BCDMS data with equal weight  
    \label{fig:chi2all}}
\end{minipage}
\end{tabular}
  \end{center}
\vspace{-0.5cm}
\end{figure}
In order to show the relevance of the fitting procedure, where
the NMC data are only shown 10\% of times, we can now consider the
training where data are given equal weight, but letting it run for a
very long training, 1000M training epochs, see Fig.~\ref{fig:chi2all}.
We see that the BCDMS $\chi^{2\,(net-exp)}$ does improve constantly,
so that also in this case the $\chi^{2\,(net-exp)}$ for the two
experiments would eventually intersect for long enough
training. However, this way of training is both inefficient (it takes
very long to find the intersection) and also subject to the criticism
that at the intersection point the neural network is almost certainly
over-learning (presumably at the intersection $\chi^2\sim 0.6$). So
training to NMC only 10\% of times is a trick which helps both in
making the training faster, and in obtaining the same
$\chi^{2\,(net-exp)}$ for the two experiments at a value where there
is no over-learning yet.

Finally, we look again at the good training, \ie 180M and 10\% of times
on NMC. As we have noted already the $\chi^{2\,(net-exp)}$ for the average over
the neural networks is good and essentially equal for both
experiments. The errors and correlations are well reproduced for
BCDMS, but very poorly for NMC (no correlation between neural networks
average and data and so on). The reason why the
errors and correlations are poorly reproduced for NMC is that that the
local information provided by each NMC data point is very weak, and
individual NMC points carry little information on the shape of $F_2$.
In other words, a few NMC points are sufficient to
train the neural network and the remaining ones do not provide
significant extra information. As a consequence, the values of error and
correlations for each individual point (or pair of points) have little
or no impact on the neural network. This fact can be seen in
the fit where we use all BCDMS data, but only 20 NMC
points (7\% of NMC points arbitrarily chosen among those where the
systematics is less than the statistical error), Fig.~\ref{fig:chi2bcd_nmc20}. 
This fit is as good as the fit where all NMC data are kept. 
\begin{figure}[t]
\begin{center}
\epsfig{width=0.48\textwidth,figure=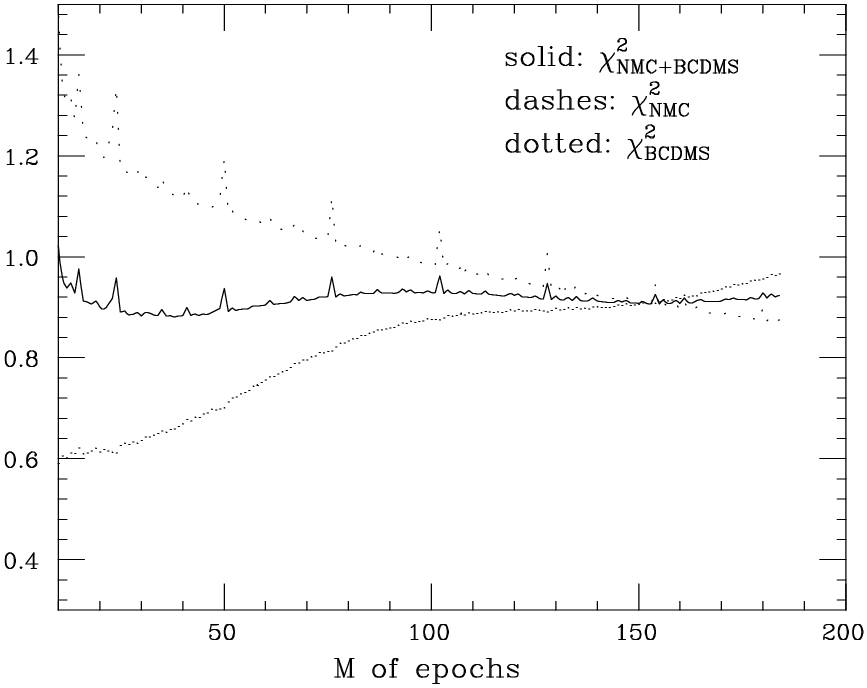}  
\end{center}
\begin{center}
\fcaption{}{Training with BCDMS data and 20 points of NMC
    \label{fig:chi2bcd_nmc20}}
\label{fig:chi2.2}
\end{center}
\end{figure}

\subsection{Proton and Deuteron}

Fits of proton and deuteron structure functions are more complicate.
Here we will outline only some interesting features.
Tables~\ref{Tgennetstp} and~\ref{Tgennetstd} show the results for a
10M training of 100 neural networks for the two structure functions.
We have used the same learning parameters for the non-singlet
structure function fit, while the number of epochs on the first cycle
has been taken equal to $4.6\times 10^6$. The data sets of the two
experiments have been shown an equal number of times. 

First we observe that the behavior of neural networks is approximately
the same for $F_2^p$ and $F_2^d$.  We note also that there are
differences from the fit of $F_2^p-F_2^d$.  Indeed, the ratio 
${\cal R}$ for the two experiments is of order 1.  Specifically, we have
that ${\cal R}_{NMC}=0.77$ and ${\cal R}_{BCDMS}=1.51$ for the
proton and ${\cal R}_{NMC}=0.84$ and ${\cal R}_{BCDMS}=1.05$ for the
deuteron. This means
that while for NMC the neural networks perform a smoothing, for BCDMS
they do not. In particular, we find that the neural network errors
for NMC are smaller than the experimental ones, while correlations are
larger. However, their ratio is smaller than in the non-singlet case.
Thus, neural networks are smoothing, but less than in the non-singlet case.
For BCDMS neural network errors and correlations are very close to
the experimental ones showing that no additional correlation has been
added. BCDMS data are all necessary to constraint the neural network
behavior. Some other features can be summarized as follows:
\begin{itemize}
\item when we increase the training for both the proton and the
  deuteron, $\chi_{BCDMS}^2$ always decreases (it can become less than
  1), while $\chi_{NMC}^2$, although initially decreases faster,
  saturates at a value of about 1.3 for the proton and 1.2 for the
  deuteron;
\item for both the proton and the deuteron $\chi_{NMC}^2$ is already
  minimized by the first cycle; it very little decreases with the
  second cycle, and sometimes if the learning rate is too large (order
  $10^{-8}$) it increases. The situation is the same if we fit only
  NMC excluding BCDMS data points;
\item systematic correlated errors do not play a significant role, as
  $\chi_{CME}^2$ is not much less than $\chi_{SCE}^2$;
\item correlations for central values, errors and correlations, are
  better then in the non-singlet case, although here we have only 100
  neural networks; the scatter plots in Figs.~\ref{fig:fitprot} and 
  \ref{fig:fitdeut} show
  that for both the proton and the deuteron we have a good agreement
  on central values. However, the statistical experimental errors are
  very small compared with the differences between the experimental
  values and the fitted ones, and this is why the $\chi^2$ is still
  bad although all the other estimators are good. 
\end{itemize}
\begin{table}[t!]
\begin{center}
\begin{tabular}{c}
\begin{minipage}[t]{0.9\textwidth}
\begin{center}  
\begin{tabular}{cccc} 
\multicolumn{4}{c}{$F_2^p$}\\  
\hline
\end{tabular}
\end{center}
\vskip-0.5cm
\end{minipage}
\\
\begin{minipage}[t]{0.9\textwidth}
    \includegraphics[width=0.47\textwidth,clip]{fitpnmc.ps}
    \includegraphics[width=0.47\textwidth,clip]{fitpbcd.ps}
\fcaption{}{Comparison between neural networks and experimental data
for the proton structure function with a 10M training.\label{fig:fitprot}}
\vskip 1.5cm
\end{minipage}
\\ \\
\begin{minipage}[t]{0.9\textwidth}
\begin{center}  
\begin{tabular}{cccc} 
\hline
$N_{net}$ & NMC+BCDMS & NMC & BCDMS\\
\hline  
$\chi_{SCE}^{2\,(net-exp)}$ 
  & 1.48 & 1.53 & 1.43 \\ \\
$\chi_{CME}^{2\,(net-exp)}$ 
  & 1.38 & 1.44 & 1.31 \\ \\
$\Big\lan \chi^{2\,(net-art)} \Big\ran_{points}$ 
  & 3.01 & 3.22 & 2.78 \\
$\Big\lan \chi^{2\,(net-exp)} \Big\ran_{points}$ 
  &  3.30 & 2.49 & 4.19 \\
${\cal R}$ 
  &  1.10 & 0.77 & 1.51\\
$\widetilde{r}\l[F^{(net-exp)}\r]$ 
  & 0.995 & 0.959 & 0.999 \\
\hline
$\Big\lan PE[\sigma^{(net-exp)}]\Big\ran_{points}$ 
  &  59\%  & 73\% & 44\%\\
$\Big\lan \sigma^{(exp)} \Big\ran_{points}$ 
  &  0.012 & 0.017 &  0.007 \\
$\Big\lan \sigma^{(net)} \Big\ran_{points}$ 
  &  0.006 &0.008  & 0.005 \\
$\widetilde{r}[\sigma^{(net-exp)}]$ 
  &  0.51 & 0.09 & 0.92 \\
\hline
$\Big\lan r^{(exp)} \Big\ran_{points}$ 
  & 0.32 & 0.17 &  0.52\\
$\Big\lan r^{(net)} \Big\ran_{points}$ 
  &  0.64 & 0.54 & 0.76 \\
$\widetilde{r}[r^{(net-exp)}]$ 
  & 0.59 & 0.32 &  0.77\\
\hline
$\Big\lan \mathrm{cov}^{(exp)} \Big\ran_{points}$ 
  & $3.9\times 10^{-5}$ & $4.5\times 10^{-5}$  & $3.3\times 10^{-5}$ \\
$\Big\lan \mathrm{cov}^{(net)} \Big\ran_{points}$ 
  & $ 2.7\times 10^{-5}$ & $3.0\times 10^{-5}$  & $2.3\times 10^{-5}$ \\
$\widetilde{r}[\mathrm{cov}^{(net-exp)}]$ 
  & 0.67 & 0.57 & 0.92\\
\hline
\end{tabular}
\end{center}
\tcaption{}{Comparison between neural networks and experimental data
for the proton structure function with a 10M training.}
\label{Tgennetstp}
\vskip-0.5cm
\end{minipage}
\end{tabular}
  \end{center}
\vspace{-0.5cm}
\end{table}
\begin{table}[t!]
\begin{center}
\begin{tabular}{c}
\begin{minipage}[t]{0.9\textwidth}
\begin{center}  
\begin{tabular}{cccc} 
\multicolumn{4}{c}{$F_2^d$}\\  
\hline
\end{tabular}
\end{center}
\vskip-0.5cm
\end{minipage}
\\
\begin{minipage}[t]{0.9\textwidth}
    \includegraphics[width=0.46\textwidth,clip]{fitdnmc.ps}
    \includegraphics[width=0.46\textwidth,clip]{fitdbcd.ps}
\fcaption{}{Comparison between neural networks and experimental data
for the deuteron structure function with a 10M training.\label{fig:fitdeut}}
\vskip 1.5cm
\end{minipage}
\\ \\
\begin{minipage}[t]{0.9\textwidth}

\begin{center}  
\begin{tabular}{cccc} 
\hline
$N_{net}$ & NMC+BCDMS & NMC & BCDMS\\
\hline  
$\chi_{SCE}^{2\,(net-exp)}$ 
  & 1.30 & 1.42 & 1.16 \\ \\
$\chi_{CME}^{2\,(net-exp)}$ 
  & 1.25 & 1.33 & 1.16 \\ \\
$\Big\lan \chi^{2\,(net-art)} \Big\ran_{points}$ 
  & 3.01 & 3.40 & 2.58 \\
$\Big\lan \chi^{2\,(net-exp)} \Big\ran_{points}$ 
  & 2.78 & 2.84 & 2.72 \\
${\cal R}$ 
  & 0.93 & 0.84 & 1.05\\
$\widetilde{r}\l[F^{(net-exp)}\r]$ 
  & 0.997 & 0.985 &  0.999\\
\hline
$\Big\lan PE[\sigma^{(net-exp)}]\Big\ran_{points}$ 
  & 61\%  & 69\% & 52\%\\
$\Big\lan \sigma^{(exp)} \Big\ran_{points}$ 
  & 0.010 & 0.014 & 0.006 \\
$\Big\lan \sigma^{(net)} \Big\ran_{points}$ 
  & 0.006  & 0.007 & 0.004 \\
$\widetilde{r}[\sigma^{(net-exp)}]$ 
  & 0.67  & 0.31 &  0.93\\
\hline
$\Big\lan r^{(exp)} \Big\ran_{points}$ 
  & 0.31 & 0.22 & 0.43 \\
$\Big\lan r^{(net)} \Big\ran_{points}$ 
  & 0.61 & 0.55 &  0.69\\
$\widetilde{r}[r^{(net-exp)}]$ 
  & 0.48 & 0.27 &  0.66\\
\hline
$\Big\lan \mathrm{cov}^{(exp)} \Big\ran_{points}$ 
  & $3.3\times 10^{-5}$ & $4.0\times 10^{-5}$  & $2.2\times 10^{-5}$ \\
$\Big\lan \mathrm{cov}^{(net)} \Big\ran_{points}$ 
  & $2.0\times 10^{-5}$ & $2.6\times 10^{-5}$  & $1.3\times 10^{-5}$ \\
$\widetilde{r}[\mathrm{cov}^{(net-exp)}]$ 
  & 0.69 & 0.62 & 0.88\\
\hline
\end{tabular}
\end{center}
\tcaption{}{Comparison between neural networks and experimental data
for the deuteron structure function with a 10M training.}
\label{Tgennetstd}
\vskip-0.5cm
\end{minipage}
\end{tabular}
\end{center}
\vspace{-0.5cm}
\end{table}

\section{Results}

\noindent
We consider the case of neural networks trained over 1000 replicas
of the experimental data. The learning parameters are the same
used in the previous examples, \ie $\eta=0.004$ for the
first cycle, $\eta=4\times 10^{-8}$ for the second, the momentum
term $\alpha=0.9$. We have trained the neural networks
$4.6\times 10^6$ times for the first cycle, $180\times 10^6$ for
the second one. We have shown the NMC data 10\% of times. As
we have discussed in the previous Section, this is only a trick to
improve the training time.
Results for the non-singlet structure function are collected in 
Table~\ref{Tgennet}. We observe that:
\begin{itemize}
\item $\chi_{CME}^2\le 1$ for each experiment, \ie our fit differs
  from data for less than one standard deviation, even properly
  keeping into account correlations between systematic uncertainties.
  The same result occurs if we consider the $\chi_{\mathrm d.o.f.}^2$
  whose values are collected in Table~\ref{chi2dof}. As we have used
  552 experimental points and 47 parameters (38 weights and 9
  thresholds), the number of {\it degrees of freedom} is 505;
\item central values are correlated each other at the level of 81\%;
  the fit over BCDMS data is more correlated (95\%) than that of NMC
  data (69\%), according to what we have shown in the previous section;
\item the loss of information due to the smaller error given by neural
  networks is regained in the increasing of correlations. As we have
  shown this effect is due to the ability of neural networks of finding
  an underlying law, ${\cal R}=0.58$;
\item the mean value of the covariance between data points is 
approximately the same for the fit over NMC and NMC+BCDMS data, 
while it is exactly the same for BCDMS data.
\end{itemize}

Results for proton and deuteron structure functions are
forthcoming. Once they are ready, they will be available through a
package to be embedded in a code for practical purposes. A web site will
also be realized for an on-line distribution of the fitted data.
\begin{table}[t]  
\begin{center}  
\begin{tabular}{cccc} 
\multicolumn{4}{c}{$F_2^p-F_2^d$}\\  
\hline
$N_{net}$ & NMC+BCDMS & NMC & BCDMS\\
\hline  
$\chi_{SCE}^{2\,(net-exp)} $ 
  & 0.83 & 0.80 &  0.86\\ \\
$\chi_{CME}^{2\,(net-exp)} $ 
  & 0.79 & 0.78 &  0.80\\ \\
$\Big\lan \chi^{2\,(net-art)} \Big\ran_{points}$ 
  & 1.95 &  2.00 &  1.89\\
$\Big\lan \chi^{2\,(net-exp)} \Big\ran_{points}$ 
  & 1.14 &  1.05 &  1.25\\
${\cal R}$ 
  & 0.58 & 0.53 & 0.66\\
$\widetilde{r}\l[F^{(net-exp)}\r]$ 
  & 0.81 & 0.69 &  0.95\\
\hline
$\Big\lan V[\sigma^{(net-exp)}]\Big\ran_{points}$ 
  & $1.7\times 10^{-4}$ & $2.9\times 10^{-4}$ & $3.5\times 10^{-5}$  \\
$\Big\lan PE[\sigma^{(net-exp)}]\Big\ran_{points}$ 
  & 80\%  & 82\% & 77\%\\
$\Big\lan \sigma^{(exp)} \Big\ran_{points}$ 
  & 0.011  &  0.016 & 0.006 \\
$\Big\lan \sigma^{(net)} \Big\ran_{points}$ 
  & 0.004  & 0.005 &  0.003\\
$\widetilde{r}[\sigma^{(net-exp)}]$ 
  & 0.50 & -0.02 &  0.92\\
\hline
$\Big\lan V[r^{(net-exp)}]\Big\ran_{points}$ 
  & 0.34 & 0.33 &  0.35\\
$\Big\lan r^{(exp)} \Big\ran_{points}$ 
  & 0.09 & 0.04 &   0.16\\
$\Big\lan r^{(net)} \Big\ran_{points}$ 
  & 0.57 & 0.45 &  0.73\\
$\widetilde{r}[r^{(net-exp)}]$ 
  & 0.37 & 0.12 &  0.56\\
\hline
$\Big\lan V[\mathrm{cov}^{(net-exp)}]\Big\ran_{points}$ 
  & $9.7\times 10^{-10}$ & $ 1.7\times 10^{-9}$ & $ 2.1\times 10^{-11}$ \\
$\Big\lan \mathrm{cov}^{(exp)} \Big\ran_{points}$ 
  & $8.4\times 10^{-6}$ & $ 9.7\times 10^{-6}$  & $ 6.8\times 10^{-6}$ \\
$\Big\lan \mathrm{cov}^{(net)} \Big\ran_{points}$ 
  & $9.0\times 10^{-6}$ & $ 1.1\times 10^{-5}$  & $ 6.8\times 10^{-6}$ \\
$\widetilde{r}[\mathrm{cov}^{(net-exp)}]$ 
  & 0.26 & 0.21 &  0.86\\
\hline
\end{tabular}
\end{center}
\tcaption{}{Comparison between neural networks and experimental data
for the non-singlet structure function with a 180M training.}
\label{Tgennet}
\end{table}

\begin{table}[t]  
\begin{center}  
\begin{tabular}{cccc} 
 & NMC+BCDMS & NMC & BCDMS \\
\hline
SCE & 0.91 & 0.96 & 1.04 \\
CME & 0.86 & 0.93 & 0.97 \\
\hline
\end{tabular}
\end{center}
\tcaption{}{$\chi_{\mathrm d.o.f.}^2$ for SCE and CME with 505 
degrees of freedom for all the experimental
points, and for each experiment.}
\label{chi2dof}
\end{table}

\chapter{Determination of $\as$ with truncated moments}
\label{chapt:as}

The strong coupling constant $\as$ is the only free parameter in the
QCD Lagrangian, and more and more accuracy is needed in its
determination. As shown in Fig.~\ref{fig:pdata} there are
several experimental processes from which we can extract $\as$.
Deep inelastic scattering is one of the prime ways of
determining the strong coupling constant $\as$.

Here we will show how the technique of truncated moments can be used
for a determination of the strong coupling constant by keeping
under control theoretical biases. We will take the fit of the
non-singlet structure function given in the previous Chapter to obtain
the truncated moments of the non-singlet parton distribution at the
initial and the final scale. Then the value of $\as$ will be extracted 
from the evolution of truncated moments \cite{fglmp}.

\begin{figure}[t]
\begin{center}
\epsfig{width=0.6\textwidth,figure=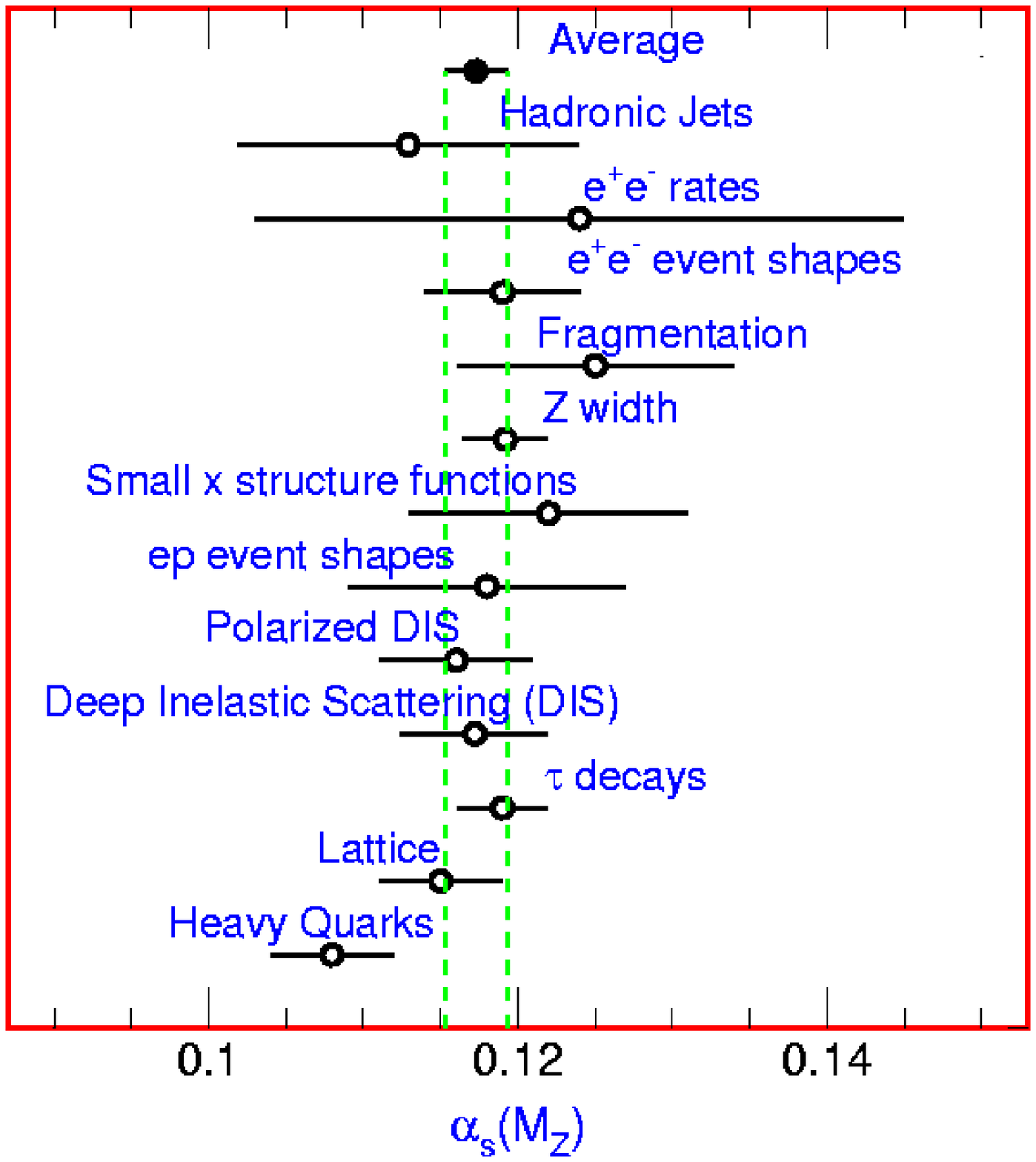}  
\end{center}
\tcaption{}{Present status of the determination of $\as$ 
as given in Ref.~\rm{\cite{pdata}}}.
\label{fig:pdata}
\end{figure}

\section{The DIS phenomenology}

\noindent
In this Section we will complete the phenomenology of Deep Inelastic
Scattering processes introduced in Chapter~\ref{chapt:parton}.
Specifically, we will introduce and describe possible effects of
theoretical uncertainty, and the techniques we adopt to keep them under
control.

\subsection{Target Mass Corrections}

Within the operator product expansion \cite{Wilson} 
the structure functions are given by the sum of contributions 
coming from operators of different twists. For unpolarized 
lepton scattering only even twists larger or equal to two contribute.
Thus keeping the twist-4 contributions into account, we have
\bea
F_{2}(x,Q^2)=F_{2}^{\rm LT,TMC}(x,Q^2)
+\frac{H(x)}{Q^2},
\label{f2ht}
\eea
where $F_{2,\rm L}^{\rm LT,TMC}$ gives the leading twist (LT) 
including target mass corrections, as calculated in 
Ref.~\cite{GP}:
\bea
F_{2,\rm L}^{\rm LT,TMC} (x,Q^2)
&=& \frac{x^2}{r^3}F_2(\xi,Q^2)+6\frac{m^2}{Q^2}
                  \frac{x^3}{r^4}\int_{\xi}^{1}d\xi 'F_2(\xi ',Q^2) \nonumber \\
     &+& 12\frac{m^4}{Q^4} \frac{x^4}{r^5}\int_{\xi}^{1}d\xi '
            \int_{\xi '}^{1}d\xi ''F_2(\xi '',Q^2), 
\label{f2tmc}
\eea
where 
\bea
\xi=\frac{2x}{1+\sqrt{r}},~~~~r=\sqrt{1+\frac{4x^2 m^2}{Q^2}}\,  
\eea
and $F_2$ is the structure function of twist 2 eq.~(\ref{f2LT}).
This approach allows us to separate pure kinematical corrections, 
so that the functions $H(x)$ correspond to 
``genuine'' or ``dynamical'' contribution of the twist 4 operators.

There is a well-known difficulty in eqs.~(\ref{f2tmc}) at $x=1$
(see \eg discussion and references in \cite{pr}): in
fact, when $x=1$ the structure functions should vanish for kinematical
reasons, while the \rhs of eq.~(\ref{f2tmc}) is clearly nonzero, since
$\xi(x=1)<1$. Indeed, in the large $x$ region dynamical
higher twist corrections become important and cannot be neglected any
more.  This is because the twist $2+2k$ contribution to the $n^{th}$
moment of a generic structure function has the form
\bea
B_{kn}(Q^2)\left(\frac{n\Lambda^2}{Q^2}\right)^k,
\label{ht}
\eea
where $\Lambda$ is a mass scale of the order of a few hundreds MeV, and
the coefficients $B_{kn}(Q^2)$ have no power dependence on $n,k$ or $Q^2$.
The crucial feature of eq.~(\ref{ht}) is the presence of a factor $n^k$,
which arises because there are at least $n$ twist $2+2k$ operators
of a given dimension for each leading twist operator of the same dimension.
One can prove that the behavior of structure functions in the
$x \sim 1$ region is governed by moments $n={\cal O}(Q^2/m^2)$;
in fact, when $x=1$ and $m^2/Q^2\ll 1$ we have
\bea
\xi\simeq 1-\frac{m^2}{Q^2};
\label{xi1}
\eea
on the other hand, if we assume a $(1-\xi)^{a_2}$ behavior for the structure
function, with 
$a_2$ of order 1, its $n^{th}$ moment receives the dominant contribution from
the region
\bea
\xi\simeq \frac{n}{a_2+n} \simeq 1-\frac{a_2}{n}.
\label{xi2}
\eea
Comparing eqs.~(\ref{xi1}) and (\ref{xi2}), we obtain that the relevant 
moments for the $x=1$ region are of order
\bea
n=\frac{a_2\,Q^2}{m^2}
\eea
as announced. Inserting this in eq.~(\ref{ht}), one immediately realizes that
the contribution of dynamical higher twists is no longer suppressed 
by inverse powers of $Q^2$ when $x$ is close to 1, and we cannot expect 
eqs.~(\ref{f2ht}) and (\ref{f2tmc}), to hold in this region.
A solution to this problem is that of ex\-pan\-ding the result
in powers of $m^2/Q^2$ up to any finite order. In this way, the dangerous
contribution of terms with large powers of $m^2/Q^2$ is not included.
The expansion remains reliable even when $Q^2$ is as low as $m^2$,
provided $x$ is not too large; in fact, powers of $m^2/Q^2$ always appear multiplied
by an equal power of $x^2$. The expanded result of course cannot be
reliable at $x\simeq 1$, but this would not be the case even without expanding
in $m^2/Q^2$, since we are not including the contributions of eq.~(\ref{ht}),
which are important in this region. 

There are several attempts to give a theoretical estimation of
dynamical HT (see \eg \cite{dht}), as well as to extract
HT contribution from data taken at low $Q^2$ (see \eg
\cite{alekhinfit}). Meanwhile, due to the fast fall of the HT
contribution with $Q^2$, it is significant for $Q^2\le 10~{\rm GeV}^2$
only. Thus, if we choose a kinematic range given by $x_0=0.01$ and
$Q^2\ge 20\,\mathrm{GeV}^2$, target mass corrections and dynamical HT
will be numerically negligible, even in the large-$x$ limit.

\subsection{Renormalization and factorization scales}

As we have seen in Sect.~\ref{sect:fact} $F_2$ is given by the
convolution of coefficient functions and parton distributions which
both depend on a factorization scale $\mu_F^2$. A dependence on
the renormalization scale $\mu_R^2$ is introduced by NLO QCD
approximation of $\as$ and by the NLO splitting functions, when we
consider the scale evolution of parton distributions with the
Altarelli-Parisi equations.  Since we use the truncated perturbative
series, the results depend on the factorization scale $\mu_F^2$
and the renormalization scale $\mu_R^2$.

Only a limited set of Mellin moments of the NNLO Altarelli-Parisi
splitting functions \cite{Larin}, as well as some asymptotes, are known
\cite{vanNeerven}.  Nevertheless, there are attempts to analyze the
DIS data in the NNLO QCD approximation considering the
available moments only \cite{Kataev}, or modeling splitting functions
\cite{Vogt}. Our analysis is performed in the NLO QCD approxi\-mation
with the use of the splitting and coefficient functions in $x$-space
as they are given in Refs.~\cite{buras,Furmanski}, and the DIS scheme
change given in \cite{fmpr}. We also set $\mu_F^2=\mu_R^2=Q^2$.

Since we work in the DIS factorization scheme $F_2$ is given by
eq.~(\ref{pdfsc}). The parton distributions are defined at the
physical scale of $F_2$, and the factorization scale is identified
with the renormalization scale.  As we will consider only the
non-singlet structure function, we will not need a prescription for
the gluon (see Sects. 3.4.1 and 4.3).  We are then left to estimate
the theoretical uncertainty due to the choice of $\mu_R^2$ in the
evolution equations.  This will be done using the approach described
in Ref.~\cite{MRS91}. In accordance with this approach the
renormalization scale $\mu_R^2$ is chosen equal to $k_R\,Q^2$. The
effect of a variation of the scale at LO in $\as$ is given by
\bea
\as (k_R\,Q^2) &=& \frac{\as (M_Z^2)}{1 + 
\frac{b_0}{2\pi}\,\as (M_Z^2)\,\log \frac{k_R\,Q^2}{M_Z^2}} \\
\nonumber
&\approx& \as (Q^2) - \frac{b_0}{2\pi}\as^2 \log k_R\,.
\eea
Thus, the NLO evolution equations are modified in the following way
\bea
&&\frac{d}{dt}q^{(NS)}(x,k_R\,Q^2)= \\ \nonumber 
&&\frac{\as(t)}{2\pi}\l[P_{qq}^{(0)} 
+\frac{\as(t)}{2\pi}\l(P_{qq}^{(1)}-b_0\,P_{qq}^{(0)}\log k_R\r)\r]\otimes q^{(NS)}(x,Q^2).
\label{APxren}
\eea
where $P^{(0)}$ and $P^{(1)}$ are respectively the LO and NLO 
splitting functions.
This contribution can be compensated at NLO by a redefinition
of the NLO splitting function
\bea
P_{qq}^{(1)}\ra P_{qq}^{(1)}+b_0\,P_{qq}^{(0)}\log k_R\,.
\eea
In the following analysis we will take a variation of $k_R$ from
1/4 to 4 which gives an estimate of the error due to renormalization
scale uncertainty. By definition, this uncertainty is connected with
the impact of higher order terms of the perturbative series.
 
\subsection{Elastic contribution}

Elastic contributions are taken into account with the nucleon structure function, 
\eg of the proton, given by
\cite{bodek}
\bea
F_2^{(el)}(x,Q^2)=\frac{G_E^2+G_M^2\tau}{1+\tau}\frac{1}{(1+Q^2 r_0^2)^4}\,
\delta(x-1)\,,
\eea
where $G_E$ and $G_M$ are the elastic form factors, 
$\tau=Q^2/4 M^2$ with $M$ the proton mass, $r_0=1/0.71\,\mathrm{GeV}^{-2}$
is the scale of the proton radius.

This contribution may be significant in our analysis
thus spoiling its accuracy. Indeed, if we take the lower integration bound of
truncated moments to be large, and we fit a higher moment, the elastic contribution
to the fitted truncated moment may reach 20\%. To make this effect
negligible we will consider only the fit of the first 10 moments. 

\subsection{Evolution uncertainties}

In Sect. 4.4, we have given an estimation of the theoretical uncertainty
introduced by truncated moments. There we used a toy model PDF to have
an upper bound on these uncertainties. Here we reproduce those checks
for the non-singlet PDF for all moments that will be considered in our
analysis. As the input PDF we take the one obtained from the neural
networks fit of $F_2$ discussed in the previous Chapter. In the DIS factorization
scheme we have
\bea
q^{NS}(x,Q^2)=2\frac{F_2^p(x,Q^2)-F_2^d(x,Q^2)}{x}\,.
\eea
In Table~\ref{rhsnn} we show the percentage errors on the \rhs of the
LO evolution equations for $x_0=0.1$ that exhibits the largest errors
on the evolution. As expected the values of the percentage errors are less than
those given with the toy PDF, even if of the same order and higher moments have
negligible errors on their evolution.
We checked that the results hold also at NLO. The
value of $x_0$ that we will use for our final analysis is $x_0=0.01$.
In this case even for the fitted non-singlet PDF, we have 
${\cal R}_{n,M}^a \sim {\cal R}_{n,M,N}^b \sim 0.1\%$ as reported in 
Table~\ref{tabrhs}.
\begin{table}[t]  
\begin{center}  
\begin{tabular}{ccc} \hline 
\multicolumn{3}{c}{$x_0=0.1$}
\\ \hline  
$n$  & ${\cal R}_{n,11}^a$ & ${\cal R}_{n,11,6}^b$ \\
\hline
1  & $4.5 \times 10^{-1}$ & $5.4 \times 10^{-2}$  \\
2  & $9.9 \times 10^{-2}$ & $1.3 \times 10^{-2}$ \\  
3  & $1.9 \times 10^{-2}$ & $5.7 \times 10^{-3}$ \\
4  & $3.2 \times 10^{-3}$ & $1.3 \times 10^{-3}$ \\
5  & $5.1 \times 10^{-4}$ & $2.4 \times 10^{-4}$ \\
6  & $7.5 \times 10^{-5}$ & $3.9 \times 10^{-5}$ \\
7  & $1.0 \times 10^{-5}$ & $5.7 \times 10^{-6}$ \\
8  & $1.4 \times 10^{-6}$ & $7.9 \times 10^{-7}$ \\  
9  & $1.8 \times 10^{-7}$ & $1.1 \times 10^{-7}$ \\ 
10 & $2.3 \times 10^{-8}$ & $1.4 \times 10^{-8}$ \\
11 & $2.8 \times 10^{-9}$ & $1.7 \times 10^{-9}$ \\
\hline
\end{tabular}
\end{center}
\tcaption{}{Comparison between percentage errors on the LO evolution
equation for $M=11$ and $N=6$.}
\label{rhsnn}
\end{table}
\section{Fitting procedure and results}

\noindent
The technique of truncated moments requires a high numerical precision in
the computation of the evolution matrices (see Sect. 4.4). 
Thus, the following results are obtained with 11 truncated
moments, \ie the maximum number of truncated moments we can use
to avoid numerical uncertainties on the evolution. The evolution is
performed between an initial scale of 20 $\mathrm{GeV}^2$ and a final
scale of 70 $\mathrm{GeV}^2$. A logarithmically equally spaced
intermediate scale is used.
Details on these choices will be discussed at the end of the section.
We will show both results with $x_0=0.01$ and $x_0=0.1$. We checked
that with $x_0=0.1$ higher twists effects as well as elastic
contributions are negligible. We do not take into account the fit of
the first truncated moment. This is because the exact first Mellin
moment does not evolve and does not affect a fit of $\as$. The same
holds for the first truncated moment with a small value of $x_0$. We
also observe that the percentage error on the evolution of the first
truncated moment is significant for large values of $x_0$, thus spoiling
the precision of our analysis.

We generate truncated moments of the LO and NLO splitting functions,
as well as the matrices $R$ and $R^{-1}$ that diagonalize the
evolution equation given in eq.~(\ref{rmatrix}), with a {\it 
Mathematica} code \cite{wolfram}. Results are passed to a FORTRAN code that
reads the parameters of the neural network fit of $F_2$ and that performs
the minimization of the $\chi^2$ with the {\tt MINUIT} routine \cite{minuit}. 
The errors on $\as$
reported in this section are the statistical errors given by {\tt MINUIT}.

We will start our analysis by fitting $\as$ and a single truncated moment.
\begin{table}[t]  
\begin{center}  
\begin{tabular}{lccc} \hline 
\multicolumn{3}{c}{$\as$} \\ 
\hline  
$n$ & $x_0=0.1$ & $x_0=0.01$ \\
\hline  
2  & 
     0.0914 $\pm$ 0.0469 & 0.0886 $\pm$ 0.0798 \\
3  & 
     0.1002 $\pm$ 0.0240 & 0.1059 $\pm$ 0.0311 \\
4  & 
     0.1131 $\pm$ 0.0187 & 0.1153 $\pm$ 0.0193 \\
5  & 
     0.1222 $\pm$ 0.0151 & 0.1225 $\pm$ 0.0153 \\
6  & 
     0.1266 $\pm$ 0.0142 & 0.1266 $\pm$ 0.0142 \\
7  & 
     0.1286 $\pm$ 0.0146 & 0.1286 $\pm$ 0.0146 \\
8  & 
     0.1294 $\pm$ 0.0160 & 0.1294 $\pm$ 0.0160 \\
9  & 
     0.1292 $\pm$ 0.0184 & 0.1292 $\pm$ 0.0184 \\
10 & 
     0.1282 $\pm$ 0.0217 & 0.1282 $\pm$ 0.0217 \\
11 & 
     0.1264 $\pm$ 0.0260 & 0.1264 $\pm$ 0.0260 \\
\hline
\end{tabular}
\end{center}
\tcaption{}{Values of the fitted $\as(M_Z^2)$ for the fit with a single moment for different values
of $x_0$.}
\label{assinglemom}
\end{table}
From Table~\ref{assinglemom} we first observe that all central values,
whenever different from the world average, are compatible with it at
least within $1\sigma$.  We see also that lower moments have a central
value lower than 0.119 and large errors. The error is minimum for
$n=6$. As we expect, higher truncated moments are independent from the
lower integration bound $x_0$. In the chosen kinematic range
from eq.~(\ref{xi2}) we have
\bea
x\simeq \frac{n}{a_2+n} \simeq 1-\frac{a_2}{n}.
\eea
Thus, only the lower truncated moments are sensitive to the
small-$x$ region. Notice that if we fit a single truncated 
moment we can not take into account all the available information
we have from data.

Let us consider a fit of pairs of moments. As an example, we
have collected results with $x_0=0.1$ in Table~\ref{aspairmom}.
\begin{table}[t]  
\begin{center}  
\begin{tabular}{lc} \hline 
\multicolumn{2}{c}{$x_0=0.1$}\\ \hline  
Fitted moments  & $\as$  \\
\hline
3+4  &  0.1369  $\pm$ 0.0106 \\
3+5  &  0.1360  $\pm$ 0.0115 \\
3+6  &  0.1337  $\pm$ 0.0125 \\
3+7  &  0.1306  $\pm$ 0.0135 \\
3+8  &  0.1273  $\pm$ 0.0143 \\
\hline
\end{tabular}
\end{center}
\begin{center}  
\tcaption{}{Values of the fitted $\as(M_Z^2)$ for the fit with a pair of moments.}
\label{aspairmom}
\end{center}  
\end{table}
We observe that $\as$ obtained by fitting simultaneously two 
truncated moments slightly differs from the weighted average 
of the $\as$ given by the case in which we fit the corresponding
truncated moments one by one. The central values are different
and the errors are smaller. Thus correlations play an
important role. As we expect, the effect of correlations is higher
for neighboring moments, while it is smaller for distant ones. 
In particular, we have that if moments are weakly correlated the 
fit is closer to the fit of the single more precise moment.
We have checked that this behavior is independent of 
the chosen value of $x_0$.

Results for different combination of fitted moments are given in 
Tables~\ref{as02},\ref{as01} and \ref{as001}. In this case we also show 
for completeness the results for $x_0=0.2$ that is close to the $x$
cut of the fit given in \cite{Virchaux}. 
\begin{table}[t]  
\begin{center}  
\begin{tabular}{lc} \hline 
\multicolumn{2}{c}{$x_0=0.2$}\\ \hline  
Fitted moments  & $\as$  \\
\hline
2+4+5    &   0.1099 $\pm$ 0.0079  \\
3+5+7    &   0.1169 $\pm$ 0.0079  \\
2+5+7    &   0.1130 $\pm$ 0.0079  \\
2+3+4    &   0.0949 $\pm$ 0.0081  \\
3+6+9    &   0.1152 $\pm$ 0.0092  \\
\hline
2+4+5+6  &   0.1046 $\pm$ 0.0077  \\
2+3+5+7  &   0.0874 $\pm$ 0.0100  \\
\hline
\end{tabular}
\end{center}
\tcaption{}{Values of the fitted $\as(M_Z^2)$ for the fit with different combinations of moments.}
\label{as02}
\end{table}
\begin{table}[t]  
\begin{center}  
\begin{tabular}{lc} \hline 
\multicolumn{2}{c}{$x_0=0.1$}\\ \hline  
Fitted moments  & $\as$  \\
\hline
2+3+4    &  0.1279 $\pm$  0.0082 \\
2+3+5    &  0.1188 $\pm$  0.0097 \\
2+4+5    &  0.1214 $\pm$  0.0104 \\
2+4+6    &  0.1200 $\pm$  0.0104 \\
3+4+5    &  0.1202 $\pm$  0.0117 \\
3+5+7    &  0.1207 $\pm$  0.0153 \\
\hline
2+3+4+5  &  0.1160 $\pm$  0.0067 \\
2+4+5+6  &  0.0953 $\pm$  0.0096 \\
\hline
\end{tabular}
\end{center}
\tcaption{}{Values of the fitted $\as(M_Z^2)$ for the fit with different combinations of moments.}
\label{as01}
\end{table}
\begin{table}[t]  
\begin{center}  
\begin{tabular}{lc} \hline 
\multicolumn{2}{c}{$x_0=0.01$}\\ \hline  
Fitted moments  & $\as$  \\
\hline
3+6+9          &  0.1395  $\pm$     0.0096\\
2+5+7          &  0.1352  $\pm$     0.0097\\
2+3+4          &  0.1301  $\pm$     0.0105\\
\hline         
3+5+7+9        &  0.1248  $\pm$     0.0074\\
3+4+6+8        &  0.1200  $\pm$     0.0082\\
2+4+6+8        &  0.1207  $\pm$     0.0097\\
2+3+5+7        &  0.1350  $\pm$     0.0121\\
\hline         
3+5+6+7+8      &  0.1271  $\pm$     0.0050\\
2+3+5+6+7      &  0.1284  $\pm$     0.0051\\
3+5+6+7+9      &  0.1272  $\pm$     0.0051\\
2+3+5+7+9      &  0.1279  $\pm$     0.0059\\
3+4+5+7+9      &  0.1261  $\pm$     0.0060\\
2+3+4+6+8      &  0.1264  $\pm$     0.0107\\
\hline         
2+3+4+5+6+7    &  0.1279  $\pm$     0.0047\\
2+4+5+6+7+9    &  0.1281  $\pm$     0.0049\\
2+3+4+5+6+8    &  0.1281  $\pm$     0.0049\\
1+3+4+5+6+9    &  0.1279  $\pm$     0.0051\\
1+2+4+5+6+7    &  0.1243  $\pm$     0.0055\\
\hline         
2+3+4+5+7+8+9  &  0.1279  $\pm$     0.0049\\
2+3+4+5+6+8+9  &  0.1282  $\pm$     0.0050\\
2+3+4+5+6+7+9  &  0.1286  $\pm$     0.0050\\
\hline
\end{tabular}
\end{center}
\tcaption{}{Values of the fitted $\as(M_Z^2)$ for the fit with different combinations of moments.}
\label{as001}
\end{table}
At first sight we note that we have the smallest error on $\as$ when
the number of fitted moments is the largest, \ie $x_0=0.01$ and 6 or 7
truncated moments.  However, from Table~\ref{as001} we observe also
that there is a trade-off due to the fact that we have a limited amount
of information: 6 truncated moments and 3 scales is the number of
parameters sufficient to extract the whole information from data. 
Increasing the number of parameters does not improve the fit.

The number of fitted moments is sometimes different for different
values of $x_0$. When $x_0$ is large the $x$-range is very narrow and
moments are strongly correlated. As a consequence the correlation
matrix is almost singular since all its elements are close to one.
Thus, the numerical precision on the inversion of the correlation
matrix may fail.  In Appendix B.4 we show that under these conditions
also pathological effects on the minimization of the $\chi^2$ may
arise, \ie values of $\as$ with extremely small errors and pathological
best fits of moments. When $x_0=0.01$ the $x$-range is wider and even
if correlations still play an important role, they are generally not
so strong as to give rise to pathological effects. The best value of $\as$ is
given by
\bea
\as(M_Z^2)=0.128\pm0.005\,.
\eea
As an example, for the case of the fit with 7 moments, in
Table~\ref{as001asym} we give also asymmetric errors. In order to be sure that
no local minima are present in the parameter space, we explore the
region of the minimum for $\as$.  From Fig.~\ref{fig:scanning} we can
see that we have a well located parabolic minimum as confirmed by
asymmetric errors.
\begin{figure}[t]
\begin{center}
\epsfig{width=0.7\textwidth,figure=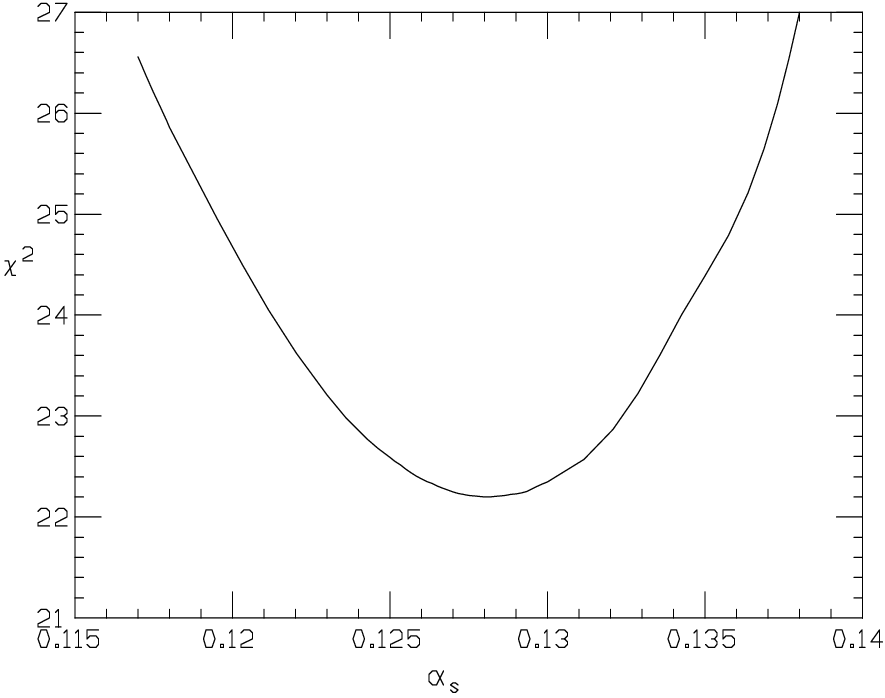}  
\end{center}
\fcaption{}{$\chi^2$ vs. $\as(M_Z^2)$ with $x_0=0.01$, fit 2+3+4+5+6+8+9 and 15 d.o.f..}
\label{fig:scanning}
\end{figure}
\begin{table}[t]  
\begin{center}  
\begin{tabular}{lc} \hline 
\multicolumn{2}{c}{$x_0=0.01$}\\ \hline  
Fitted moments  & $\as$  \\
\hline\\
\vspace{0.1cm}
2+3+4+5+6+8+9  &  $0.1282_{-0.0051}^{+0.0047}$\\\vspace{0.1cm}
2+3+4+5+7+8+9  &  $0.1279_{-0.0050}^{+0.0046}$\\\vspace{0.1cm}
2+3+4+5+6+7+9  &  $0.1286_{-0.0052}^{+0.0048}$\\
\hline
\end{tabular}
\end{center}
\tcaption{}{Values of the fitted $\as(M_Z^2)$ with asymmetric errors
for the fit with 7 moments.}
\label{as001asym}
\end{table}

We then address the choice of the number of scales and of
the $Q^2$ range. We need at least two scales to fit one moment and $\as$.
Additional intermediate scales increase the number of points of the fit, and thus
reduce the error. However, we can not increase arbitrarily the number of
scales, especially when we fit several moments simultaneously. 
We have seen already that we have a limited amount of information
from data, and, as a result, we must use a limited number of truncated moments
as well as of intermediate scales.

As it is shown in Table~\ref{asscales} when we take the simultaneous
fit of several moments, the error is smaller for 3 scales, than for 2
scales. Moreover, increasing the number of scales to 4 has a worse
effect than increasing the number of fitted moments. Since a moment at
different scales is more correlated than two neighboring moments at
the same scale, a large number of scales does not add significant
information and only causes pathological effects in the minimization
of the $\chi^2$ as discussed above. Thus, for the problem at hand the
best choice is to use 3 scales.
\begin{table}[t]  
\begin{center}  
\begin{tabular}{lcc} 
\hline 
\multicolumn{3}{c}{$x_0=0.01$}\\ 
\hline  
Fitted moments & N. of Scales & $\as$  \\
\hline
2+3+5+6+7     & 2 & 0.1193  $\pm$ 0.0077 \\
2+3+5+6+7    & 3 & 0.1284  $\pm$ 0.0050 \\
2+3+5~~~~+7   & 4 & 0.1051  $\pm$ 0.0071 \\
\hline
\end{tabular}
\end{center}
\tcaption{}{Values of the fitted $\as(M_Z^2)$ for the fit with a different number of scales 
and moments with $x_0=0.01$}
\label{asscales}
\end{table}

The plot of the kinematic region Fig.~\ref{fig:kinrange} indicates how
we have to choose $x$ and $Q^2$, if we want to avoid extrapolating
outside the experimental kinematic region. Such a request is needed as
we know (see Chapter 5) that neural networks are good in
interpolation, but they are not very trustworthy in extrapolation.  When
$x_0=0.01$ the maximum allowed value of $Q^2$ is $70\,\mathrm{GeV}^2$.
Thus, in order to understand the effects of varying the $Q^2$ range we
will take $x_0=0.1$ that allows us more flexibility in the choice of
the final scale (see Fig.~\ref{fig:kinrange}). Table~\ref{as01q2} shows
that the effect of varying the $Q^2$ range is small and within the
statistical errors. 
\begin{table}[t]  
\begin{center}  
\begin{tabular}{lc} 
\hline  
\multicolumn{2}{c}{$x_0=0.1$}\\ 
\hline  
$Q^2$ range & $\as$  \\
\hline
20-70    & 0.1131 $\pm$ 0.0187 \\
20-100   & 0.1091 $\pm$ 0.0176 \\
30-100   & 0.1071 $\pm$ 0.0183 \\
30-120   & 0.1064 $\pm$ 0.0181 \\
\hline
\end{tabular}
\end{center}
\tcaption{}{Values of the fitted $\as(M_Z^2)$ for the fit with 
different ranges of $Q^2$ and $x_0=0.01$}
\label{as01q2}
\end{table}

\section{The theoretical uncertainties}
\noindent
The theoretical uncertainties of a phenomenological analysis can not be
ultimately defined. The progress of studies may increase or decrease
such biases. Throughout this thesis we have shown how we can avoid
some of them, specifically:
\begin{itemize}
\item with the technique of truncated moments we can avoid
theoretical assumptions on the shape of parton distributions in the
small-$x$ region (see Chapter 4). As we have shown the
evolution uncertainty is negligible;
\item with a neural network fit we can avoid theoretical assumptions on the 
shape of the structure function $F_2$ when we fit experimental data (see Chapter 6);
\item with a careful choice of the kinematical region over which we
perform our analysis we can avoid kinematical and dynamical higher twists, as well
as elastic contributions. 
\end{itemize}
We are then left with only two sources of uncertainties, \ie the matching of
$\as$ at quark mass thresholds and the effects NNLO unaccounted contributions.

It is worth noting that since combinations with several moments may be
sensitive to the numerical precision (see previous Section) of
calculations and since we are looking for small effects, we prefer to
adopt in the following analysis the truncated moments combination
2+3+5+6+7 with $x_0=0.01$ that yields the same central value and the
same error of combinations with more moments, but with a more stable
numerical precision.

We have seen in Sect.~\ref{sect:massth} how we can define $\as$ in a scheme
where it varies with the number of active quarks. We have shown also
how we can match the different couplings to let $\as$ be continuous when it
crosses quark mass thresholds. As result we may have a dependence
on the matching conditions we have used to connect the couplings.
As explained we will assess this ambiguity by requiring that
$\as$ is continuous at $Q_i^2=k_{th}\,m_i^2$, with $m_i$ the heavy 
quark mass. Then, we will let $k_{th}$ vary from $1/2$ to 2.
In the kinematic range we have chosen, we have only the $b$-quark
threshold, and we have used $m_b=5\,\mathrm{GeV}$. 
From Tab.~\ref{as001thresh} we observe that matching conditions effects are negligible.
For completeness in Fig.~\ref{fig:renscale} we show the effects of varying
$k_{th}$ over a wider range. Almost all points lay in the systematic
error band of the value with $k_{th}=1$ and their spread is very narrow.

The uncertainty due to the renormalization scale has been discussed in
previous section. Tab.~\ref{as001ren} shows the results of varying
$k_R$ from $1/4$ to 4. Again in Fig.~\ref{fig:renscale} we show the
effects of varying $k_R$ over a wider range, analogous to that used in
\cite{Virchaux}. We observe that the uncertainties are relevant.  Note
also that they are asymmetric and that we have a larger effect when
the central value is closer to the world average. Thus, we can conclude
that NNLO contributions play an important role in the evolution
of truncated moments, and an analysis taking into account these
contributions may be explored. Our final result is
\bea
\as(M_Z^2)=0.128\pm0.005\,(stat.)_{-0.006}^{+0.004}\,(theor.).
\eea
\begin{figure}[t]
\begin{center}
\epsfig{width=0.48\textwidth,figure=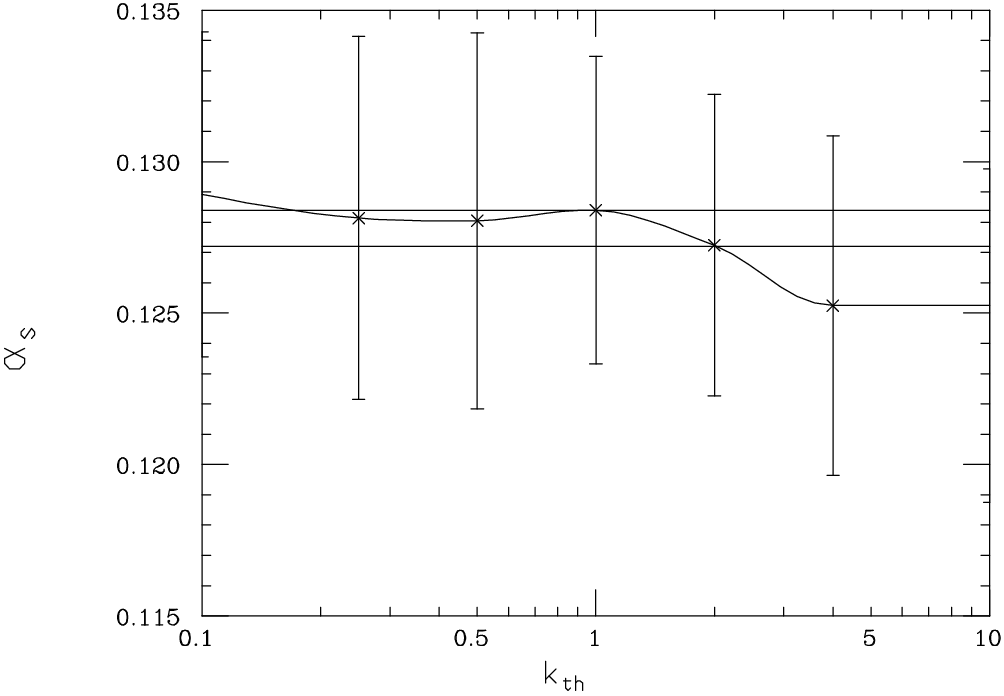}  
\epsfig{width=0.48\textwidth,figure=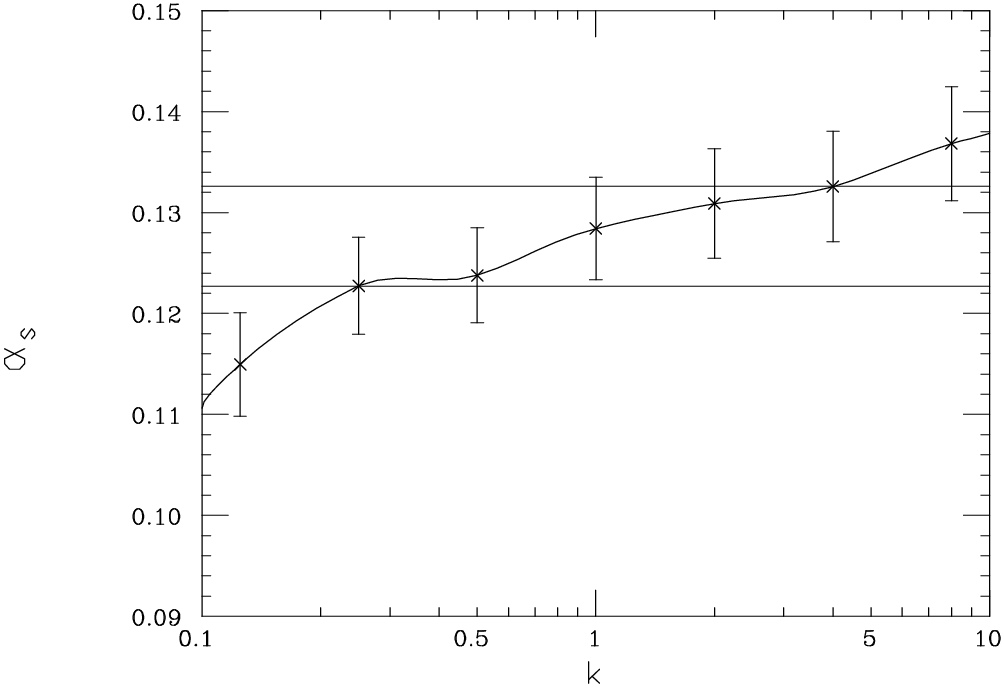}  
\end{center}
\fcaption{}{Variation of $\as(M_Z^2)$ as a function of $k_{th}$ and $k_R$ 
with $x_0=0.01$, fit 2+3+5+6+7 with theoretical systematic error bands obtained from
\rm{Tables~\ref{as001thresh}} {\it and} \rm{\ref{as001ren}} {\it and statistical error 
bars on each point}.}
\label{fig:renscale}
\end{figure}
\begin{table}[t]  
\begin{center}  
\begin{tabular}{lc} 
\hline 
\multicolumn{2}{c}{$x_0=0.01$}\\ 
\hline  
$k$ & $\as$  \\
\hline
2   & 0.1272 $\pm$ 0.0050 \\
1   & 0.1284 $\pm$ 0.0051 \\
1/2 & 0.1281 $\pm$ 0.0062 \\
\hline
\end{tabular}
\end{center}
\tcaption{}{Values of the fitted $\as(M_Z^2)$ for the fit with different quark mass thresholds and 
moments 2+3+5+6+7.}
\label{as001thresh}
\end{table}
\begin{table}[t]  
\begin{center}  
\begin{tabular}{lc} 
\hline 
\multicolumn{2}{c}{$x_0=0.01$}\\ 
\hline  
$k$ & $\as$  \\
\hline
4   & 0.1326  $\pm$ 0.0055 \\  
1   & 0.1284  $\pm$ 0.0051 \\  
1/4 & 0.1227  $\pm$ 0.0048 \\ 
\hline
\end{tabular}
\end{center}
\tcaption{}{Values of the fitted $\as(M_Z^2)$ for the fit with different 
renormalization scales and moments 2+3+5+6+7.}
\label{as001ren}
\end{table}

\chapter{Conclusions}

We have performed a phenomenological analysis of DIS data in the 
context of perturbative QCD with the purpose of reducing theoretical
uncertainties, and taking into proper account all experimental
errors and correlations. Specifically:
\begin{itemize}
\item we have extended the method of truncated moments suggested in \cite{FM}
  for the case of non-singlet parton distributions to the singlet 
  case \cite{fmpr}.  
  Truncated moments of parton distributions are defined by
  restricting the integration range over the Bjorken variable to an
  experimentally accessible subset $x_0<x<1$ of the allowed
  kinematic range $0<x<1$.  This method provides a way
  to avoid theoretical biases on the shape of parton distributions.
  We have shown how to
  increase the numerical efficiency of the method, and how to use it as 
  a way to solve the Altarelli-Parisi
  evolution equations \cite{mine};
\item we have performed a fit of the non-singlet structure function
  data. Preliminary work on individual (proton and deuteron)
  structure function fits has also been performed.  The neural network
  approach we have suggested allows to avoid sources of theoretical
  bias, such as the choice of a given shape of the fitted function,
  which are difficult to keep under control. We have adopted a Monte
  Carlo technique to estimate errors and correlations for any quantity
  which can be extracted from our fit, such as \eg the error on $F_2$
  for given values of $x$ and $Q^2$, or the correlation between two
  Mellin moments. Our results will be published in \cite{fglp};
\item using the techniques outlined above, we have performed a 
  determination of the strong coupling constant $\as$ \cite{fglmp}. 
  We have taken into account all correlated systematic experimental 
  errors. As a result of the adopted methods the theoretical bias is
  lower than in the case of previous analyses.
  Our result for $\as$ is compatible within errors with the world average;
  the analysis may be refined in order to reduce the statistical errors. 
  We have also pointed
  out the need of a more precise knowledge of NNLO contributions when
  all other theoretical biases are correctly taken into account.
\end{itemize}

\subsubsection*{Outlook}

The work presented in this thesis can be extended in different directions.

The neural network fit of $F_2$ should be extended to include HERA,
and E665 data. This would cover the whole experimental range presently
explored. In particular, this would allow a detailed analysis of the
small-$x$ region. An analogous fit of the structure functions with deep
inelastic neutrino scattering data may be performed as well.

The approach we have suggested to fit the unpolarized structure
functions may also be adopted to fit the polarized structure function
$g_1$. A neural network fit of $g_1$ would allow an analysis of the
spin structure of the nucleon, as well as an independent extraction of
$\as$.

Finally, a determination of $\as$ may be performed with truncated
moments of singlet parton distributions. In this case, the
theoretical uncertainties are generally larger than in the non-singlet case,
but on the other hand the amount of available data is also larger.

\subsubsection*{Acknowledgments}

I am greatly indebted to Giovanni Ridolfi and Stefano Forte for
reducing the theoretical biases on my knowledge. 
\\
I thank Lorenzo Magnea for opening my mind to truncated moments.
\\
I thank Jos\'e I. Latorre and L. Garrido for 
training my neural networks on neural networks. 
\\
I thank Carlo Becchi for wise advice at the right time.
\\
I thank Beppe Carlino, Franz Montalenti and Stefano Giusto for
teaching me how a good PhD. fellow should behave.  
\\
I thank the
Departament d'Estructura i Constituents de la Mat\'eria de 
la Universitat de Barcelona for kind hospitality during part 
of this work. 
\\
I thank the INFN Sezione di Roma III and PhD. 
students of the Universit\`a di Roma III for
kind hospitality during the last part of this work.

\appendix

\chapter{Diagonalization of triangular matrices}  
\setcounter{section}{1}
\label{triangular}  
In this Appendix, we show how to construct the matrix $R$ which  
diagonalizes a generic $n\times n$ triangular matrix $T$ by means of  
the recursion relations eqs.~(\ref{trir},\ref{invtrir}). The matrix  
$R$ is defined by the requirement that  
\beq   
R T R^{-1}=\mbox{diag}(\gamma_1,\dots,\gamma_n)~,  
\label{tridiag}  
\eeq  
where the matrix $T$ is upper triangular, {\it i.e.} $T_{ij}=0$ if $i>j$.  
It is easy to see, by solving the secular equation, that the eigenvalues  
$\gamma_i$ of $T$ coincide with its diagonal elements,   
\beq  
\gamma_i=T_{ii}~.  
\label{evals}  
\eeq  
Now, define eigenvectors $v^j$ associated to the $j^{th}$ eigenvalue  
$T_{jj}$, with components $v_i{}^j$:   
\beq  
\sum_{k=1}^n T_{i k} v_k{}^j=\gamma_j v_i{}^j~.  
\label{eveccon1}  
\eeq  
Clearly, the matrix $R^{-1}$ coincides with the matrix of right  
eigenvectors, $(R^{-1})_{ij}= v_i{}^j$, while the matrix $R$ coincides  
with the matrix of left eigenvectors $\sum_{k=1}^n \hat v^j{}_k T_{k  
i} =\gamma_j \hat v^j{}_i$, $R_{ij}={\hat v}^i{}_j$.  The eigenvector  
condition eq.~(\ref{eveccon1}) immediately implies that the $j^{th}$  
component of the $j^{th}$ eigenvector is equal to one:  
$v_j{}^j=1$. Furthermore, it is clear that eq.~(\ref{eveccon1}) can  
only be satisfied if all components $v_k{}^j$ of the $j^{th}$  
eigenvector with $k>j$ vanish,  
\beq  
v_j{}^j=1~; \qquad v_k{}^j=0 \quad \mbox{if} \quad k>j~.  
\label{evecs}  
\eeq  
Using eq.~(\ref{evecs}) and the fact that the matrix $T$ is  
triangular, eq.~(\ref{eveccon1}) can be written as  
\beq  
\sum_{k=i}^j T_{i k} v_k{}^j=\gamma_j v_i{}^j~.  
\label{eveccon}  
\eeq  
Substituting the explicit form of the eigenvalues, eq.~(\ref{evals}),  
and identifying $v_i{}^j =(R^{-1})_{ij}$, this is immediately seen to  
coincide with eq.~(\ref{invtrir}). Furthermore, using the condition  
$v_j{}^j=1$, this equation can be viewed as a recursion relation which  
allows the determination of the $(k-1)^{th}$ element of $v^j$ once the  
$k^{th}$ element is known, which is what we set out to prove. The same  
argument, applied to the left eigenvectors, leads to the expression in  
eq.~(\ref{trir}) for $R$.  
%
%
\chapter{Tools of statistics}

The aim of this appendix is to report basic and useful definitions of 
statistic and introduce the notations adopted in the thesis. 
Here we will follow \cite{dagostini,cowan}.
%
%
\section{Distribution of several random variables}

\noindent
We only consider the case of two continuous variables (X and Y).
The extension to more variables is straightforward. The infinitesimal
element of probability is $dF(x,y)=f(x,y)~dx~dy$, and the probability 
density function
\bea
f(x,y) = \frac{\partial^2 F(x,y)}{\partial x \partial y}\,.
\eea
The probability of finding the variable inside a certain area A is
\bea
\int_{A} f(x,y)~dx~dy\,.
\eea

\subsubsection*{Expected Value}

\bea
\mu_X = E[X] = \int_{-\infty}^{\infty} x f(x,y)~dx~dy\,,
\eea
and analogously for Y. Generally,
\bea
E[g(X,Y)] = \int_{-\infty}^{\infty} g(x,y) f(x,y)~dx~dy\,.
\eea

\subsubsection*{Variance}

\bea
V[X] = \sigma_X^2=\int_{-\infty}^{\infty} (x-\mu)^2 f(x,y)~dx~dy
=E[X^2] - E[X]^2\,,
\eea
and analogously for Y. 

\subsubsection*{Covariance}

\bea
V_{XY} &=& \rm{cov}[X,Y]=E\l[(X-E[X])(Y-E[Y])\r]=E[XY]-\mu_X\mu_Y 
\nonumber \\
&=&\int_{-\infty}^{\infty} xy f(x,y)~dx~dy-\mu_X\mu_Y\,,
\eea
where $\mu_X=E[X]$ and $\mu_Y=E[Y]$. The covariance matrix is also
called the error matrix, as $V_{XX}=V[X]=\sigma_X^2$. If X and Y are
independent, then $E[XY]=E[X]E[Y]$ and hence $V_{XY}=0$ (the opposite
is true only if X and Y have the same normalization).

\subsubsection*{Correlation coefficient}

\bea
\rho_{XY}=\frac{\rm{cov}[X,Y]}{\sqrt{V[X]V[Y]}}=
\frac{V_{XY}}{\sigma_X\sigma_Y}\,,
\eea
The correlation coefficient gives a dimensionless measure of the level
of correlation between two random variables X and Y. One can show that
the correlation coefficient lies in the range $-1\leq \rho_{XY} \leq 1$.

\subsubsection{Linear combination of random variables}

If $Y=\sum_i c_i X_i$ with $c_i$ real, then
\bea
\mu_Y &=& E[Y] = \sum_i c_i E[X_i] = \sum_i c_i \mu_i \\
V[Y] &=& \sigma_Y^2 = \sum_i c_i V[X_i] + 2\sum_{i<j} c_ic_j 
\,\rm{cov}[X_i,X_j] \nonumber \\
&=& \sum_i c_i c_j\,\sigma_{ij}\,,
\eea
where $\sigma_{ij}=\rm{cov}[X_i,X_j]=V_{ij}$, $\sigma_{ii}=V[X_i]$
and $\rho_{ii}=1$.

\section{Gaussian distribution}
%
%
\noindent
The N-dimensional Gaussian distribution is defined by
\bea
f({\bf X};\boldsymbol{\mu},V)= \frac{1}{(2\pi)^{N/2}|V|^{1/2}}
\exp\l[\frac{1}{2}({\bf X}-\boldsymbol{\mu})^T V^{-1}
({\bf X}-\boldsymbol{\mu}) \r]\,,
\label{gaussdist}
\eea
where ${\bf X}$ and $\boldsymbol{\mu}$ are column vectors containing
$X_1,\ldots,X_N$ and $\mu_1,\ldots,\mu_N$, ${\bf X}^T$ and 
$\boldsymbol{\mu}^T$ are the corresponding row vectors, and $|V|$ is the
determinant of a symmetric $N\times N$ matrix V, thus containing 
$N(N+1)/2$ parameters.

The importance of the Gaussian distribution stems from the central
limit theorem. The theorem states that the sum of $N$ independent
continuous random variables $X_i$ with means $\mu_i$ and variance
$\sigma_i^2$ becomes a Gaussian random variable with mean
$\mu=\sum_{i=1}^N \mu_i$ and variance $\sigma^2=\sum_{i=1}^N \sigma_i^2$ in the
limit that $N$ approaches infinity. This holds regardless of the form
of the individual probability densities functions of the $X_i$. This
is the formal justification for treating measurements errors as Gaussian
random variables, and holds to the extent that the total error is the
sum of a large number of small contributions.
The expectation values, variances and covariances can be computed to be
\bea
E[X_i]&=&\mu_i 
\nonumber \\
V[X_i]&=&V_{ii} \\
\rm{cov}[X_i,X_j]&=&V_{ij}\,. 
\nonumber 
\eea

\section{Estimators for mean, variance, covariance}
%
%
\noindent
Consider the case where one has made $N$ measurements of a random
variable $X$ whose probability density function $f(x)$ is not
known. Our task is to infer properties of
$f(x)$ based on observations $x_1,\ldots,x_N$. Specifically, we would
like to construct functions of the $x_i$ to estimate the various
properties of the probability density function $f(x)$. Usually we have a
hypothesis for the probability density function $f(x)$ which depends on
an unknown parameter (or parameters
$\boldsymbol{\theta}=(\theta_1,\ldots,\theta_m)$). The goal is then to
construct a function of the observed $x_i$ to estimate parameters.

A function of the observed measurements $x_1,\ldots,x_N$ which
contains no unknown parameters is called {\it statistic}. In
particular, a statistic used to estimate some property of a
probability density function (\eg its mean, variance or other
parameters) is called an {\it estimator}. The estimator of a quantity
$\theta$ is usually written with a hat, $\hat\theta$, to distinguish it
from the true vale $\theta$ whose exact value is unknown.

If $\hat\theta$ converges to $\theta$ in the limit of large $N$, the
estimator is said to be {\it consistent}. Here convergence is meant in
the sense of probability: for any $\epsilon >0$, one has
\bea
\lim_{N\ra\infty} P(|\hat\theta-\theta|>\epsilon)=0\,.
\eea
Consistency is usually a minimum requirement for a useful estimator. 
Other features of estimators are {\it bias} (see later) and 
{\it robustness}, \ie the property of being insensitive to 
departures from assumptions in the probability density function owing to
factors such as noise.
The procedure of estimating the value of a parameter, given the data
$x_1,\ldots,x_N$, is called {\it parameter fitting}.

The expectation value of an estimator $\hat\theta$ with the sampling 
probability density function $g(\hat\theta,\theta)$ is
\bea
E[{\bf \hat\theta}] = \int \hat\theta g(\hat\theta,\theta)
\,d\hat\theta\,.
\eea
Recall that this is the expected mean value of $\hat\theta$ from an
infinite number of similar experiments, each with a sample of size
$N$. We define the {\it bias} of an estimator $\hat\theta$ as
\bea
b=E[\hat\theta] - \theta\,.
\eea
Note that the bias does not depend on the measured values of the
sample but rather on the sample size, the functional form of the
estimator and on the true properties of the probability density
function $f(x)$, including the true value of $\theta$. A parameter for
which the bias is zero independent of the sample size $N$ is said to
be unbiased; if the bias vanishes in the limit $N\ra\infty$ then it is
said to be asymptotically unbiased.

We now consider the case where one has a sample of size $N$ of a
random variable $X, (x_1,\ldots,x_N)$. It is assumed that $X$ is
distributed according to some probability density function $f(x)$
which is not known, not even as a parametrization. We would like to
construct a function of the $x_i$ to be an estimator for the
expectation value of $X$, $\mu$. One possibility is the arithmetic
mean of the $x_i$, defined by
\bea
\lan x \ran_N=\frac{1}{N}\sum_{i=1}^N x_i\,.
\eea
The arithmetics mean of the elements of a sample is called the {\it
sample mean}, and is denoted by $\lan x\ran$ (where we 
generally omit the index $N$) or by a bar, \eg $\overline{x}$.  This
should not be confused with the expectation value (population mean or
central value of $f(x)$) of $x$, denoted by $\mu$ or $E[x]$, for which
$\lan x\ran$ is an estimator. The fist important property of the
sample mean is given by the weak law of large numbers. This states
that if the variance of $x$ exists, then $\lan x\ran$ is a
consistent estimator for the population mean $\mu$. That is, for
$N\ra\infty$, $\lan x\ran$ converges to $\mu$ in the sense of
probability.

The expectation values of the sample mean $E[\lan x\ran]$ is given by
\bea
E[\lan x\ran]=E\l[\frac{1}{N}\sum_{i=1}^N x_i\r] = 
\frac{1}{N}\sum_{i=1}^N E[x_i]=\mu\,,
\eea
for all $i$. Thus one can see that the sample mean $\lan x\ran$ is an
unbiased estimator for the population mean $\mu$.

If the mean $\mu$ is known, then the quantity $\widehat{\sigma^2}$ defined by
\bea
\widehat{\sigma^2}=\frac{1}{N-1}\sum_{i=1}^N (x_i-\mu)^2=\lan x^2\ran-\mu^2\,,
\eea
is an unbiased estimator of the variance $\sigma^2$. In a similar way one 
can show that the quantity
\bea
\widehat{V}_{XY}
=\frac{1}{N-1}\sum_{i=1}^N (x_i-\lan x\ran)(y_i-\lan y\ran)
\eea
is an unbiased estimator for the covariance $V_{XY}$ of two random variables
$X$ and $Y$ of unknown mean. This can be normalized by the square root
of the estimators for the sample variance to form an estimator $r$ for the
correlation coefficient $\rho$:
\bea
r=\frac{\lan xy\ran-\lan x\ran\lan y\ran}{\sqrt{\lan x^2\ran-
\lan x\ran^2}
\sqrt{\lan y^2\ran-\lan y\ran^2}}\,.
\label{rhoest}
\eea

Given an estimator $\hat\theta$ one can compute its variance
$V[\hat\theta]=E[\hat\theta^2]-(E[\hat\theta])^2$. Recall that
$V[\hat\theta]$ (or equivalently its square root
$\sigma_{\hat\theta})$ is a measure of the variation of $\hat\theta$
about its mean in a large number of similar experiments each with
sample size $N$, and as such is often quoted as the statistical error
of $\hat\theta$. For example, the variance of the sample mean
$\lan x\ran$ is
\bea
V[\lan x\ran] = \frac{\sigma^2}{N}\,,
\eea
where $\sigma^2$ is the variance of $f(x)$. In a similar way, the
variance of the estimator $\widehat{\sigma^2}$ for a Gaussian distribution, can be
computed to be
\bea
V\l[\widehat{\sigma^2}\r]=\frac{2}{N-1}\sigma^4\,.
\eea
The expectation value and variance of the estimator of the correlation
coefficient for a two dimensional Gaussian are found to be
\bea
E[r] &=& \rho-\frac{\rho(1-\rho^2)}{2N} + O(N^2)\,,
\\
V[r] &=& \frac{(1-\rho^2)^2}{N}+ O(N^2)\,.
\label{rhovar}
\eea
The estimator $r$ given by eq.~(\ref{rhoest}) is only
asymptotically unbiased.
Thus, we should be very careful when applying eq.~(\ref{rhovar}) 
to evaluate the significance of an observed correlation.

\section{Least  squares minimization}
%
%
\noindent
Consider now a set of N independent Gaussian random variables $m_i$,
$i=1,\dots,N$, each related to another variable $x_i$, which is
assumed to be known without error.  Assume that each value $m_i$ has
a different unknown mean, $t_i$, and a different but known variance,
$\sigma_i$. The N measurements of $m_i$ can be equivalently regarded
as a single measurement of an N-dimensional random vector, for which
the join probability distribution function is the product of N
Gaussians,
\bea
f(m_1,\ldots,m_N;t_1,\ldots,t_N,\sigma_1^2,\ldots,\sigma_N^2) 
&=&\\ \nonumber 
&&\prod_{i=1}^N\frac{1}{\sqrt{2\pi\sigma_i^2}}
\exp\l(-\frac{(m_i-t_i)^2}{2\sigma_i^2}\r)\,.
\eea
We suppose also that the true value is given as a function of $x$,
$t=t(x;\boldsymbol{\theta})$, which depends on unknown parameters
$\boldsymbol{\theta}=(\theta_1,\ldots,\theta_N)$. The aim of the method of
the least squares is to estimate the parameters $\boldsymbol{\theta}$.  In
addition, the method allows for a simple evaluation of the
goodness-of-fit of the hypnotized function $t(x;\boldsymbol{\theta})$.

Taking the logarithm of the joint probability function and dropping
additive terms that do not depend on the parameters gives the
log-likelihood function,
\bea
\log L(\boldsymbol{\theta}) = -\frac{1}{2}\sum_{i=1}^N
\frac{(m_i-t_i)^2}{\sigma_i^2}\,.
\eea
This is maximized by finding the values of the parameters $\theta$ that
minimize the quantity
\bea
\chi^2(\boldsymbol{\theta})=\sum_{i=1}^N
\frac{(m_i-t_i)^2}{\sigma_i^2}\,,
\label{sce}
\eea
namely the quadratic sum of differences between measured and hypnotized
values, weighted by the inverse of variances. This is the basis of the
{\it method of least squares} (LS), and it it used to define the procedure 
even in cases where individual measurement are not Gaussian, but as 
long as they are independent.

If the measurements are not independent but can be described by an
N-dimensional Gaussian probability distribution function with known
covariance matrix $V$ but unknown mean values, the corresponding
log-likelihood function is obtained from the logarithm of the join
probability function given by eq.~(\ref{gaussdist})\,
\bea
\log L(\boldsymbol{\theta}) = -\frac{1}{2}\sum_{i,j=1}^N
(m_i-t_i(x_i;\boldsymbol{\theta}))V_{ij}^{-1}
(m_j-t_j(x_j;\boldsymbol{\theta}))\,,
\eea
where additive terms not depending on the parameters have been dropped. This is maximized by minimizing the quantity
\bea
\chi^2(\boldsymbol{\theta})=\sum_{i,j=1}^N
(m_i-t_i(x_i;\boldsymbol{\theta}))V_{ij}^{-1}
(m_j-t_j(x_j;\boldsymbol{\theta}))\,
\label{cme}
\eea
which reduces to eq.~(\ref{sce}) if the covariance matrix (and hence
its inverse) are diagonal. The parameter that minimize the $\chi^2$
are called the LS estimators, $\hat\theta_1,\ldots,\hat\theta_m$.
The explicit expression of ${\bf V}$ can be found by taking
\bea
m_i = t_i(\boldsymbol{\theta}) + r_i \sigma_i + \sum_k s_k \Delta_{ik}
\label{midef}
\eea
where $m_i$ is the measurement of data point $i$, $t_i(\boldsymbol{\theta})$ is the
model prediction depending on a set of parameters $\boldsymbol{\theta}$, $\sigma_i$
is the uncorrelated (statistical) error on data point $i$ and
$\Delta_{ik}$ is the correlated (systematic) error from  source
$k$. In eq.~(\ref{midef}), $r_i$ and $s_k$ denote Gaussian random variables
with zero mean and unit variance. These random variables are assumed
to be independent of each other:
\bea
  \lan \Delta r_i \Delta r_j\ran = \lan\Delta s_i \Delta s_j\ran 
= \delta_{ij} \qquad 
  \lan\Delta r_i \Delta s_j\ran = 0\,.
\label{delridelrj}
\eea
From eqs.~(\ref{midef}) and (\ref{delridelrj}) the
covariance matrix ${\bf V}$ of the measurements is given by
\bea
  V_{ij} = \lan\Delta m_i \Delta m_j\ran = 
  \delta_{ij}\sigma_i^2 + \sum_k \Delta_{ik} \Delta_{jk}\,.
\label{vmeasdef}
\eea

Minimizing $\chi^2$ defined by eq.~(\ref{cme}) is impractical
because it involves the inversion of the measurement covariance matrix
(\ref{vmeasdef}) which, in global fits, tends to become very
large. Because the systematic errors of different data sets are in
general uncorrelated (but not always, see~\cite{NMC}) this
matrix takes a block diagonal form and each block could, in principle,
be inverted once and for all.  However, the dimension of these block
matrices can still easily be larger than a few hundred. Furthermore,
if the systematic errors dominate, the covariance matrix might,
numerically, be uncomfortably close to a matrix with the simple
structure $V_{ij} = \Delta_i \Delta_j$, which is singular.

Fortunately, the $\chi^2$ of (\ref{cme}) can be cast in an
alternative form which avoids the inversion of large matrices
\cite{lmethod}:
\bea
  \chi^2 &=& \sum_i \left( \frac{m_i-t_i}{\sigma_i} \right)^2 
  - {\bf B}\;{\bf A}^{-1}\;{\bf B} \nonumber \\
  B_k &=& \sum_i \Delta_{ik} (m_i-t_i)/\sigma_i^2     
\label{cteqchi} \\
\nonumber
  A_{kl} &=&  \delta_{kl}
             + \sum_i \Delta_{ik} \Delta_{il} / \sigma_i^2. 
\eea
The matrix ${\bf A}$ in eq.~(\ref{cteqchi}) has the dimension of the
number of systematic sources only and can be inverted at the
initialization phase of a fitting program once the number of data
points included in the fit (\ie after cuts) is known. An example of
a global QCD fit with error calculations based on the covariance
matrix approach can be found in~\cite{alekhinfit}.

It is remarkable that minimizing eq.~(\ref{cteqchi}) is equivalent to a
fit where {\em both} the parameters ${\bf \theta}$ and  ${\bf s}$ are left
free. In such a fit $\chi^2$ is defined as follows. First, the
effect of the systematic errors is incorporated in the model prediction
\bea
  f_i(\boldsymbol{\theta},{\bf s}) =  t_i(\boldsymbol{\theta}) + \sum_k s_k \Delta_{ik}.
\label{fpsdef}
\eea
Next, $\chi^2$ is defined by
\bea
  \chi^2 = \sum_i \left( \frac{m_i-f_i(\boldsymbol{\theta},{\bf s})}{\sigma_i} 
\right)^2 + \sum_k s_k^2.   
\label{xpsdef}
\eea
The second term in eq.~(\ref{xpsdef}) serves to constrain the fitted
values of ${\bf s}$. The presence of this term is easily understood if
one takes the view that the calibration of each experiment yields a set
of `measurements' $s_k = 0 \pm 1$~\cite{dagostini2}.

Because $f$ is linear in ${\bf s}$ the minimization with respect to the
systematic parameters can be done analytically. It is easy to show, by
solving the equations $\partial \chi^2 / \partial ds_k = 0$, that this 
leads to the $\chi^2$ given by eq.~(\ref{cteqchi}) which, in turn, is
equivalent to eq.~(\ref{cme}), see~\cite{lmethod}. The relation
between the optimal values of ${\bf s}$, the matrix ${\bf A}$ and the
vector ${\bf B}$ of eq.~(\ref{cteqchi}) is
\begin{equation}
  {\bf s} = {\bf A}^{-1} {\bf B}\,.
\label{eq:bestsys}
\end{equation}

\subsubsection*{Remarks}

\begin{itemize}
\item If we have data with correlated systematic uncertainties,
the minimization of eq.~(\ref{sce}) called also the {\it simplest}
$\chi^2$ {\it estimator} (SCE), is an estimator of the $\chi^2$
given by eq.~(\ref{cme}), {\it covariance matrix estimator} (CME).  
It can be shown \cite{alekhinchi} that the value of the error due to
improved statistical precision for the SCE does not necessarily
decrease after adding a new point.
Qualitatively, for SCE the fitted curve tightly follows the
data points and, if these points are shifted due to the systematic
errors fluctuations, the parameter gains appropriate systematic
errors. At the same time, since for the CME the information on data
correlations is explicitly included in $\chi^2$, the correlated
fluctuation of data due to systematic shift does not necessary leads
to the fitted curve shift and parameter deviation gets smaller than
for SCE.
\item If we have strong correlations, the correlation matrix
is almost singular and we may find pathological results. 
Eq.~(\ref{cme}) can be written as
\bea
\chi^2=\sum_{i,j=1}^N\frac{m_i-t_i}{\sigma_i}\rho_{ij}^{-1}
\frac{m_j-t_j}{\sigma_j}\,.
\eea
If we take as an example the case with only two data, we have
\bea
\rho&=&\l(\matrix{1 & 1-\delta \cr 1-\delta & 1}\r) 
\\ \nonumber
\rho^{-1}&=&\frac{1}{2\delta-\delta^2}
\l(\matrix{1 & -1+\delta \cr -1+\delta & 1}\r) 
\,. 
\eea
Thus, we get
\bea
\chi^2&\approx&\frac{1}{2\delta-\delta^2}\l[
\l(\frac{m_1-t_1}{\sigma_1}\r)^2+\l(\frac{m_2-t_2}{\sigma_2}\r)^2\r. 
\nonumber \\
&-&\l.2\frac{(m_1-t_1)(m_2-t_2)}{\sigma_1\sigma_2}\r]\,,
\eea
that can be zero not only if $m_i=t_i$. Notice also that
the covariance matrix has $N$ diagonal terms and $N(N-1)$
off-diagonal ones. If off-diagonal terms are large, their 
effect may be stronger than that of the diagonal terms.
An example of such a pathological result is shown in
Fig.~\ref{fig:shift}.

\begin{figure}[t]
\begin{center}
\epsfig{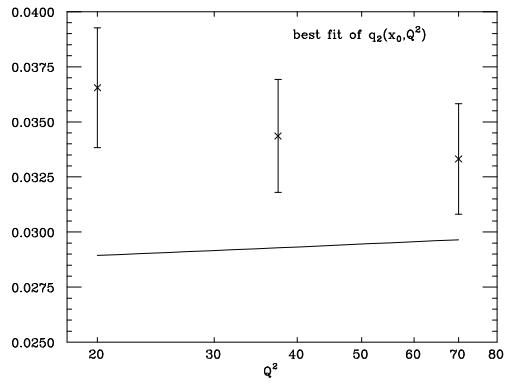}  
\end{center}
\caption{\it Pathological fit of the second truncated moment with $x_0=0.2$}
\label{fig:shift}
\end{figure}

\end{itemize}

\chapter{Numerical routines}

\noindent
Here we reproduce the codes given in \cite{NumRec} that we have used to
generate random numbers. 
The random numbers generator {\tt ran1} is very trustful as it is not 
known any statistical test 
that it fails to pass, except when the number of calls starts to become
on the order of the period $m$, say $>10^8$. For our applications
the maximum number of calls is less than $10^5$.

\begin{center}
\begin{verbatim}
      FUNCTION ran1(idum)
      INTEGER idum,IA,IM,IQ,IR,NTAB,NDIV
      REAL ran1,AM,EPS,RNMX
      PARAMETER (IA=16807,IM=2147483647,
     *AM=1./IM,IQ=127773,IR=2836,
     *NTAB=32,NDIV=1+(IM-1)/NTAB,EPS=1.2e-7,RNMX=1.-EPS)
      INTEGER j,k,iv(NTAB),iy
      SAVE iv,iy
      DATA iv /NTAB*0/, iy /0/
      if (idum.le.0.or.iy.eq.0) then
        idum=max(-idum,1)
        do 11 j=NTAB+8,1,-1
          k=idum/IQ
          idum=IA*(idum-k*IQ)-IR*k
          if (idum.lt.0) idum=idum+IM
          if (j.le.NTAB) iv(j)=idum
 11      continue
        iy=iv(1)
      endif
      k=idum/IQ
      idum=IA*(idum-k*IQ)-IR*k
      if (idum.lt.0) idum=idum+IM
      j=1+iy/NDIV
      iy=iv(j)
      iv(j)=idum
      ran1=min(AM*iy,RNMX)
      return
      END
\end{verbatim}
\end{center}

\noindent
The routine {\tt gasdev} simply returns normally distributed Gaussian
random numbers with zero mean and unit variance, 
using {\tt ran1(idum)} as the source of random numbers.

\begin{center}
\begin{verbatim}
      FUNCTION gasdev(idum)
      INTEGER idum
      REAL gasdev
      INTEGER iset
      REAL fac,gset,rsq,w1,w2,ran1
      SAVE iset,gset
      DATA iset/0/
      if (iset.eq.0) then
1       w1=2.*ran1(idum)-1.
        w2=2.*ran1(idum)-1.
        rsq=w1**2+w2**2
        if(rsq.ge.1..or.rsq.eq.0.)goto 1
        fac=sqrt(-2.*log(rsq)/rsq)
        gset=w1*fac
        gasdev=w2*fac
        iset=1
      else
        gasdev=gset
        iset=0
      endif
      return
      END
\end{verbatim}
\end{center}

\noindent
The {\tt idirty} generator is a quick random number generator
used to {\it somewhat} randomize things. What we need in the training of 
neural networks is to process data from experiments not always in the same
order to avoid a bias on the training (see Chapter 5). 
Here the period is 6075.

\begin{center}
\begin{verbatim}
        jran=mod(jran*106+1283,6075)
        idirty=jlo+((jhi-jlo+1)*jran)/6075
\end{verbatim}
\end{center}

\end{document}